\documentclass[12pt,a4paper,final]{iopart}
\usepackage{iopams}
\usepackage{mathrsfs}
\usepackage{graphicx}
\usepackage{color}
\usepackage[breaklinks=true,colorlinks=true,linkcolor=blue,urlcolor=blue,citecolor=blue]{hyperref}
\usepackage{amsfonts}

\expandafter\let\csname equation*\endcsname\relax
\expandafter\let\csname endequation*\endcsname\relax
\usepackage{amsmath}
\usepackage{amssymb}
\usepackage{bm}
\usepackage{dcolumn}
\usepackage[latin1]{inputenc}
\usepackage{latexsym}
\usepackage{rotating}
\usepackage[all]{hypcap} 
\usepackage{mathrsfs}

\usepackage{footmisc}

\newcommand{\s}{\mathfrak{s}} 
\newcommand{\nv}{\mathcal{V}}

\allowdisplaybreaks

\eqnobysec
\definecolor{red  }{rgb}{1,0,0}
\definecolor{blue }{rgb}{0,0,1}
\definecolor{green}{rgb}{0,1,0}
\definecolor{CiteColor}{rgb}{0,0,0.35}
\definecolor{URLColor}{rgb}{0,0,0.35}
\definecolor{darkgreen}{rgb}{0.2,0.7,0.2}

\eqnobysec
\begin{document}

\title[Post-Newtonian templates for binary black-hole inspirals]
{Post-Newtonian templates for binary black-hole inspirals: the effect of the horizon fluxes and the secular change in the black-hole masses and spins}

\author{Soichiro Isoyama$^{1,~2}$}
\address{
$^1$ The Open University of Japan, Chiba 261-8586, Japan \\
$^2$ International Institute of Physics, Universidade Federal do
Rio Grande do Norte, 59070-405, Natal, Brazil
}

\ead{isoyama@yukawa.kyoto-u.ac.jp}

\author{Hiroyuki Nakano$^{3,~4}$}
\address{
$^3$ Faculty of Law, Ryukoku University, Kyoto 612-8577, Japan  \\
$^4$ Department of Physics, Kyoto University, Kyoto 606-8502, Japan
}

\ead{hinakano@law.ryukoku.ac.jp}

\date{\today}

\begin{abstract}
Black holes (BHs) in an inspiraling compact binary system absorb 
the gravitational-wave (GW) energy and angular-momentum fluxes across 
their event horizons and this leads to the secular change 
in their masses and spins during the inspiral phase. 
The goal of this paper is to present ready-to-use, $3.5$ post-Newtonian (PN) 
template families for spinning, non-precessing, binary BH inspirals 
in quasicircular orbits, 
including the $2.5$PN and $3.5$PN horizon-flux contributions 
as well as the correction due to the secular change 
in the BH masses and spins through $3.5$PN order, respectively, in phase. 
We show that, for binary BHs observable by Advanced LIGO 
with high mass ratios (larger than $\sim 10$) 
and large aligned-spins (larger than $\sim 0.7$),  
the mismatch between the frequency-domain template with and without 
the horizon-flux contribution is typically above the $3\%$ mark.
For (supermassive) binary BHs observed by LISA, 
even a moderate mass-ratios and spins can produce a similar level 
of the mismatch. 
Meanwhile, the mismatch due to the secular time variations of the BH masses 
and spins is well below the $1\%$ mark in both cases, 
hence this is truly negligible.
We also point out that neglecting the cubic-in-spin, point-particle 
phase term at $3.5$PN order 
would deteriorate the effect of BH absorption in the template. 
\end{abstract}


\maketitle

\section{Introduction and Summary}
\label{sec:intro}

\subsection{Goals and motivations: }

The first detection of gravitational waves (GWs),
GW150914 from binary black holes (BBHs)~\cite{Abbott:2016blz,
TheLIGOScientific:2016qqj,TheLIGOScientific:2016wfe,TheLIGOScientific:2016uux}
with the succeeding detection, GW151226~\cite{Abbott:2016nmj}, 
GW170104~\cite{Abbott:2017vtc}, 
GW170814~\cite{Abbott:2017oio}, 
and a candidate event, LVT151012~\cite{TheLIGOScientific:2016pea}, 
recorded by Advanced LIGO detectors~\cite{TheLIGOScientific:2016agk,TheLIGOScientific:2016zmo,Abbott:2016jsd} opened a new window on physics and the Universe. 
To perform such GW astrophysics with very high precision 
in the context of ground-based GW detectors, including 
Advanced LIGO, Advanced Virgo~\cite{Aasi:2013wya} 
and KAGRA~\cite{Somiya:2011np, Aso:2013eba} 
as well as planned space-based GW detectors 
such as LISA~\cite{Audley:2017drz} and (B-)DECIGO~\cite{Seto:2001qf, BDECIGO},
it is now crucial to have extremely accurate predictions of GWs emitted 
from BBHs to maximize the extraction of physical information 
from noisy GW signals through the well-known technique of matched filtering; 
cross correlating the noisy detector output 
with the theoretical GW waveforms for the expected GW signal 
(see, e.g.,~\cite{Allen:2005fk} for the algorithm 
used by LIGO Scientific Collaboration).

The waveforms for BBHs in the early inspiral phase are most accurately 
modeled within post-Newtonian (PN) theory~\cite{Poisson:2014pn} 
and there have been rapid progress to push it 
to high PN orders~\cite{Blanchet:2013haa}. 
For the late inspiral, merger and ringdown 
phases~\cite{TheLIGOScientific:2016pea}, 
the PN models are not applicable and 
it is mandatory to use the numerical-relativity (NR) 
simulations based on the
breakthrough~\cite{Pretorius:2005gq,Campanelli:2005dd,Baker:2005vv} 
(see also~\cite{Aylott:2009ya,Aylott:2009tn,Ajith:2012az,Hinder:2013oqa,
Aasi:2014tra,Lovelace:2016uwp}) 
as well as other analytical treatments combined with 
NR waveforms, including effective-one-body
formalism~\cite{Buonanno:1998gg,Buonanno:2000ef,Damour:2008qf,
Damour:2009kr,Barausse:2009xi,Taracchini:2013rva,Damour:2014sva,Bohe:2016gbl} 
and phenomenological models~\cite{Pan:2007nw,Ajith:2007kx,Ajith:2009bn,
Santamaria:2010yb,Hannam:2013oca,Husa:2015iqa,Khan:2015jqa}. 
The BBH waveform models have been further improved over the years 
and many applications to detection have already followed. 
In the context of testing the dynamical sector of
general relativity (GR)~\cite{TheLIGOScientific:2016src}, 
for instance, GW150914, GW151226 and GW170104
showed no statistical significant evidence 
on deviations from PN coefficients of the GW phase predicted by 
GR~\cite{Abbott:2016nmj,Abbott:2017vtc, TheLIGOScientific:2016src}. 
In~\cite{Abbott:2016apu}, GW150914 was directly compared 
with NR simulations and it was shown that they are mutually consistent.
The rate estimation of BBH mergers~\cite{Abbott:2016nhf},
the BBH formation astrophysics~\cite{TheLIGOScientific:2016htt,
TheLIGOScientific:2016wyq}, and 
the multi-messenger astronomy~\cite{Abbott:2016gcq,Abbott:2016iqz} 
are other achievements of GW astrophysics.

In this paper, our primary focus is the improvement of waveforms 
for BBHs in the early inspiral phase, 
where the change in the orbital frequency over an orbital period 
is much smaller than the orbital frequency itself. 
Given that the gravitational radiation causes 
the orbits of isolated binary systems 
to circularize~\cite{Peters:1963ux,Peters:1964zz},
we will consider only the PN-inspirals in quasicircular orbits 
with masses $m_i$ $(i = 1,\,2)$ and (the magnitude of) spins ${S}_i$ 
that are (anti-)aligned and normal to the orbital plane, 
but they have an arbitrary mass ratio. 
(All throughout, we use geometric units, where $G=c=1$,
with the useful conversion factor 
$1 M_{\odot} = 1.477 \; {\rm{km}} = 4.926 \times 10^{-6} \; {\rm{s}}$.) 
In this adiabatic setup, the GW phase of the dominant harmonic 
is twice the orbital phase~\cite{Blanchet:2013haa}.
The orbital phase $\phi(t)$ in terms of the PN barycentric time $t$ 
can be computed by the center-of-mass binding energy $E(t\,;m_i,S_i)$ and 
the energy flux of the gravitational radiation carried out to infinity 
$F_{\infty}(t\,;m_i,S_i)$; 
the state-of-art of their PN approximations including 
spin effects are reviewed in~\cite{Blanchet:2013haa,Mishra:2016whh} 
(see also section~\ref{sec:PN-formulae}).
Motivated by the Bondi-Sachs mass-loss
formula~\cite{Bondi:1962px,Sachs:1962wk} in full GR, 
the (orbital-averaged) change rate of $E$ is assumed to be related 
with $F_{\infty}$ through the balance equation,
\begin{equation}\label{balance0}
\frac{d E}{dt} = - {F}_{\infty}
\end{equation}
for constant masses $m_i$ and spins $S_i$, 
and this combined with the definition 
$d \phi / d t = \pi f$ for the GW frequency (of the dominant harmonic) $f$ 
provides the equation to obtain the evolution of $\phi(t)$.

When at least one of the two companions in binaries is a BH, 
there are additional contributions to computing 
$\phi(t)$, which are due to the slow increase in the BH mass 
(``tidal heating'') and decrease in the BH spin (``tidal torquing'') 
during the inspiral phase~\cite{Alvi:2001mx} 
in a PN order under consideration: 
in the PN theory, a term of relative $O(v^{2n})$ 
where the orbital velocity $v$ defined 
in terms of the GW frequency $f$ by 
\begin{equation}\label{def-v} 
v \equiv ( \pi m f )^{1/3}\,
\end{equation}
with the total mass of the binary $m \equiv m_1 + m_2$ 
is said to be of $n$th PN order. 
First, these ``heating'' and ``torquing'' are energy and angular-momentum 
fluxes across the BH horizon, which are known as the horizon fluxes 
${\cal F}_{\mathrm H}^{i}(t\,;m_i,S_i)$~\cite{Poisson:2004cw,
Nagar:2011aa,Chatziioannou:2012gq,Taracchini:2013wfa,
Chatziioannou:2016kem} 
(or BH absorption~\cite{Mino:1997bw,Sasaki:2003xr,
Fujita:2014eta}) 
to distinguish them from ${F}_{\infty}$. 
The horizon-flux contributions first appear at $2.5$PN order 
for spinning BHs and $4$PN order
for non-spinning BHs~\cite{Poisson:2004cw,Chatziioannou:2012gq,Poisson:1994yf} 
beyond the leading-order quadrupolar flux.
These contributions modify the right hand side of 
the balance equation~\eqref{balance0} beyond that order.  
Second, the absorption of the horizon fluxes leads to 
\textit{a secular change 
in BH masses $m_i$ and spins $S_i$ during the inspiral phase.}
The timescales for the evolution of $m_i$ and $S_i$ are estimated as 
$T_{m} \equiv {m_i} / {{\dot m}_i} = O(v^{-15})$ and 
$T_{S} \equiv {S_i} / {{\dot S}_i} = O(v^{-12})$ 
[see~\eqref{FHt-SM}]
while the radiation-reaction timescale for the (adiabatic) inspiral is 
$T_{rr} \equiv {v}/{\dot v} = O (v^{-8})$ 
[see~\eqref{dvdtT4-I0}]: 
the overdot stands for the derivative with respect to $t$.
The ratios 
\begin{equation}
\frac{T_{rr}}{T_m} = O(v^{7})\,,
\quad 
\frac{T_{rr}}{T_S} = O(v^{4})\ 
\end{equation}
imply that the BH masses $m_i$ and spins $S_i$ in $E$ and $F_{\infty}$ 
are \textit{no longer secularly constants} during the inspiral phase, 
but they rather slowly evolve as a function of $t$ at $3.5$PN order 
for $m_i$ and $2$PN order for $S_i$: 
we recall that the spin effects to the orbital phase first appear 
at $1.5$PN order~\cite{Blanchet:2013haa}. 
Such a $3.5$PN order contribution therefore alters 
the expressions for $E$ and $F_{\infty}$
\footnote{
The time-dependence of ${\cal F}_{\mathrm H}^{i}$ 
through $m_i(t)$ and $S_i(t)$ starts from at $6$PN order, 
which is negligible compared to the PN corrections 
that we consider in this paper.}.
In fact, this is the same PN order of various higher-order spin effects 
such as the leading cubic-in-spin terms~\cite{Marsat:2014xea}.

In short, the first objective of this work is 
to construct the PN template families 
for BBH quasi-circular inspirals that account for 
the effect of horizon fluxes and the secular time variations 
of the BH masses and spins accumulated in the inspiral phase.  
Built on this, 
the second objective of this work is to quantify 
the importance of corresponding corrections to observe GW signals 
from BBHs by Advanced LIGO and LISA. 
While many results have been obtained along those lines 
in the past~\cite{Santamaria:2010yb,Alvi:2001mx,Fujita:2014eta,Poisson:1994yf,
Tagoshi:1997jy,Hughes:2001jr,Yunes:2009ef,Yunes:2010zj,Hannam:2007wf,
Hannam:2010ec,Maselli:2017cmm}, 
they have considered only the correction 
due to the horizon flux restricted to various special cases 
and the emphasis of these works are not always on the application 
to GW detectors.
We improve these results with all possible effects 
of the BH absorption up to the relative $3.5$PN order  
in the context of arbitrary-mass-ratio BBH inspirals 
by bringing to bear the mindset and tools of GW data analysis.

\subsection{Generation of Post-Newtonian waveforms }

To this end, in effect, we have the following two modifications 
in the method to compute the orbital phase $\phi(t)$ 
in the adiabatic approximation; 
Our discussion in section~\ref{sec:PN-dynamics} provides these details. 
\begin{enumerate}
\item The corrected binding energy and 
energy fluxes carried out to infinity 
\begin{equation}
{\cal E} \equiv {\cal E}(t\,; m_i(t),S_i(t))\,,
\quad
{\cal F}_{\infty} \equiv {\cal F}_{\infty}(t\,; m_i(t),S_i(t))\,.
\end{equation}
They account for the modification of $E$ and $F_{\infty}$ at $3.5$PN order 
due to the \textit{secular change} 
in BH masses $m_i(t)$ and spins $S_i(t)$ during the inspiral phase. 
The explicit $3.5$PN expressions for 
${\cal E}$ and ${\cal F}_{\infty}$ as a function of $v(f)$ 
(though redefined in terms of the initial total mass $m^{\mathrm I}$; 
see~\eqref{def-v2}) 
are displayed in~\eqref{def-vE} and~\eqref{def-vF}, respectively;
\item The postulate of the generalized balance equation 
\begin{equation}\label{balanceH-0}
\left(
\frac{\partial {\cal E}}{\partial t}
\right)_{m,\,S} 
= 
- 
{\cal F}_{\infty} 
- 
\sum_{i =1,2} ( 1 - \Gamma_{\mathrm H}^i )\,{\cal F}_{\mathrm H}^i \,, 
\end{equation}
which equates the change rate in ${\cal E}$ 
to ${\cal F}_{\infty}$ and 
the horizon energy flux ${\cal F}_{\mathrm H}^i$. 
By contrast to~\eqref{balance0}, it is important to recognize that 
the left-hand side expression is \textit{the partial derivative} 
with respect to $t$ as the time variation of 
$m_i(t)$ and $S_i(t)$ is no longer negligible 
when taking the total time derivative in ${\cal E}$. 
This generates the additional BH growth factor 
$\Gamma_{\mathrm H}^i \equiv \Gamma_{\mathrm H}^i(t\,; m_i(t))$. 
The explicit $3.5$PN expressions for 
${\cal F}_{\mathrm H}^i$ and $\Gamma_{\mathrm H}^i$ as a function of $v$ 
(in terms of the initial total mass $m^{\mathrm I}$) 
are displayed in~\eqref{FHt-SM} and~\eqref{def-Gamma}, respectively.
\end{enumerate}

In section~\ref{sec:Taylor}, 
we construct five different PN templates 
for spinning, non-precessing, BBH inspirals in quasicircular orbits,  
making use of the corrected binding energy 
${\cal E}$, the corrected energy flux carried out to infinity 
${\cal F}_{\infty}$ and the horizon flux ${\cal F}_{\mathrm H}^i$ 
combined with the generalized balance equation~\eqref{balanceH-0}. 
Our ready-to-use templates keep 
only the leading PN (``Newtonian'') order in the 
polarized amplitude, 
but $3.5$PN accurate in the phase;
they incorporate all known spin terms up to $3.5$PN order 
(except the unknown spin-spin terms of GW tails at $3.5$PN order) 
as well as all possible contributions due to the BH absorption. 
We view our templates as a direct extension of 
the so-called Taylor template families 
(TaylorT1, T2, T3, T4 and F2) without the BH absorption, 
which are available and implemented 
in the \textit{LALSuite: LSC Algorithm Library Suite}; 
see, e.g.,~\cite{Damour:2000zb,Damour:2002kr,
Arun:2004hn,Buonanno:2009zt,Varma:2013kna} 
for the non-spinning inspirals,
and~\cite{Mishra:2016whh,Arun:2008kb,Wade:2013hoa} 
for the spinning inspirals.  
Also, our templates could readily used for 
comparison with NR simulation for BBHs 
in the high-mass ratio and high-spin regime 
(e.g.,~\cite{Mroue:2013xna,Jani:2016wkt,Healy:2017psd}), 
or for refining more realistic search templates 
such as effective-one-body formalism~\cite{Damour:2014sva,Bohe:2016gbl} 
and phenomenological model~\cite{Khan:2015jqa},  
including inspiral, merger and ringdown phases as well.

The amplitude and phase of the GW signals carry information about parameters 
of BBHs, such as masses and spins as well as their location and distance 
to the GW detectors.
Our templates for BBHs therefore provide a natural starting point 
to investigate the importance of BH absorption to their measurability.
Here, we adopt the frequency-domain model TaylorF2 
as our illustrative example, and we postpone the comparison 
of different template families to the future task
\footnote{
Such study for PN templates without BH absorption were 
investigated in, e.g.,~\cite{Buonanno:2009zt,Varma:2013kna,Nitz:2013mxa}. 
}; 
the details of TaylorF2 with BH absorption are provided 
in section~\ref{subsec:F2}.
Since we consider BBHs with (anti-)aligned spins, 
there is no modulation of the amplitude due to the precession. 
In this case, the phasing of GWs is much more 
important than its amplitude for detector applications. 
Using the standard ``stationary phase approximation'', 
the Fourier representation of waveforms is given by~\cite{Buonanno:2009zt}
\begin{equation}\label{h-F2}
{\tilde h}(f) 
\equiv 
{\cal A} f^{-7/6} e^{i \Psi^{\mathrm {F2}}(f) }\,, 
\end{equation}
where $f$ is the GW frequency,
the frequency-domain amplitude is expressed as
${\cal A} \propto {\cal M}^{5/6} Q({\mathrm {angles}})/ D_L$ 
with the (initial value of) chirp mass 
${\cal M} := (m_1 m_2)^{3/5} / m^{1/5}$, 
a function of all the relevant angles $Q({\mathrm {angles}})$ 
(position of the binary, orientation of the GW detector etc.) 
and the luminosity distance $D_L$ between the inspiraling BBH 
and an observer. 
We will calculate the frequency-domain phase $\Psi^{\mathrm {F2}}(f)$ 
in~\eqref{F2-inf} and~\eqref{F2-H} up to $3.5$PN order, 
and the resulting expression has the structure of 
\footnote{
It should be noted that the frequency-domain phase 
$\Psi^{\mathrm {F2}}_{3.5{\mathrm {PN}}}(v)$ in~\eqref{phi-F2-0} 
is not valid if the velocity $v$ is larger than a certain value of ``pole'' 
$v_{\mathrm {pole}}$ because the PN energy flux ${F}_{\infty}$, 
a basic input for TaylorF2, becomes \textit{negative} 
when $v \gtrsim v_{\mathrm {pole}}$ 
for a broad range of the BBH parameters [see section~\ref{subsec:EandF}]. 
The TaylorF2 hence has to be terminated before reaching the ``pole''. 
The precise value of $v_{\mathrm {pole}}$ depends on the BBH
parameters, and we find that it always satisfies 
$v_{\mathrm {pole}} \gtrsim 0.7$ to our examination. 
This indicates that the existence of $v_{\mathrm {pole}}$
is mostly irrelevant when dealing with BBHs in the early inspiral phase, 
but this issue should be borne in mind when one
implements~\eqref{phi-F2-0} for various applications 
(see, e.g.,~\fref{fig:LISA}).
}
\begin{align}\label{phi-F2-0}
\Psi^{\mathrm {F2}}_{3.5{\mathrm {PN}}}(v(f)\,; m, \nu, \chi_i)
&=   
2 \pi f t_c - \Psi_c 
+
\Psi^{\mathrm {F2}}_{\infty}(v\,; m, \nu, \chi_i)
\cr 
& \quad 
+
\frac{3}{128 \nu}
\left\{
1 + 3 \ln \left( \frac{v}{v_{\mathrm {reg}}} \right) 
\right\}
\Psi^{\mathrm {F2}}_{\mathrm {Flux},5}(m, \nu, \chi_i) \cr
& \quad
+
\frac{3 v^2}{128 \nu}
\left\{
\Psi^{\mathrm {F2}}_{\mathrm {Flux},7}(m, \nu, \chi_i)
+
\nu\, \Psi^{\mathrm {F2}}_{\mathrm {BH},7}(m, \nu, \chi_i)
\right\} \,, 
\end{align}
in which the total mass $m$, the symmetric mass ratio 
$\nu \equiv (m_1 m_2) / m^2$ 
and the dimensionless spin parameter $\chi_i \equiv S_i / m_i^2$ 
are given by their \textit{initial values} that the waveform begins. 
In the above expression, the total mass $m$ in the velocity $v$ 
[recall~\eqref{def-v}] is now replaced to its initial value, 
and the constants $\Psi_c, t_c$ and $v_{\mathrm {reg}}$ 
can be chosen arbitrary. 
The point-particle phasing function $\Psi^{\mathrm {F2}}_{\infty}$ 
accounts for the non-spinning, the spin-orbit, the quadratic-in-spin 
and the cubic-in-spin contributions ignoring the BH
absorption~\cite{Mishra:2016whh,Arun:2008kb}. 
On the other hand, $\Psi^{\mathrm {F2}}_{\mathrm {Flux},5}$ 
and $\Psi^{\mathrm {F2}}_{\mathrm {Flux},7}$ 
denote the $2.5$PN leading order (LO) contribution~\cite{Maselli:2017cmm} 
and the $3.5$PN next-to-leading order (NLO) contribution  
to the GW phase due to the horizon flux, respectively. 
The remained phasing function $\Psi^{\mathrm {F2}}_{\mathrm {BH},7}$ 
represents the $3.5$PN correction to the GW phase that is generated by 
the LO secular change in BH masses and spins during the inspiral phase, 
and it is suppressed by the prefactor of the mass ratio, 
$0 \leq \nu \leq 1/4$.

\subsection{Results 1: the error in GW cycles}

A useful estimator to characterize the effects 
of $\Psi^{\mathrm {F2}}_{\mathrm {Flux},5}$, 
$\Psi^{\mathrm {F2}}_{\mathrm {Flux},7}$ 
and 
$\Psi^{\mathrm {F2}}_{\mathrm {BH},7}$ in the phase~\eqref{phi-F2-0} 
on the waveforms is the total number of GW cycles $N$ accumulated 
within a given frequency band of detectors. 
This is defined in terms of the frequency-domain phase $\Psi(f)$ by 
\begin{equation}\label{def-N}
N \equiv 
\frac{1}{2 \pi} \int_{f_{\mathrm {min}}}^{f_{\mathrm {max}}} 
f \left(\frac{d^2 \Psi (f)}{d f^2} \right) d f\,.
\end{equation}
The substitution of~\eqref{phi-F2-0} into~\eqref{def-N} gives 
the relative number of GW cycles $\Delta N$ contributed 
by each term in $\Psi^{\mathrm {F2}}_{3.5{\mathrm {PN}}}$ 
and accumulated within the frequency band 
$f \in [f_{\mathrm {min}},\,f_{\mathrm {max}}]$. 
We generally consider that the contribution is likely to be 
\textit{negligible} if it is less than one radian. 
\begin{figure}[tbp]
\begin{tabular}{cc}  
\begin{minipage}[t]{.45\hsize}
  \centering
  \includegraphics[clip, width=\columnwidth]{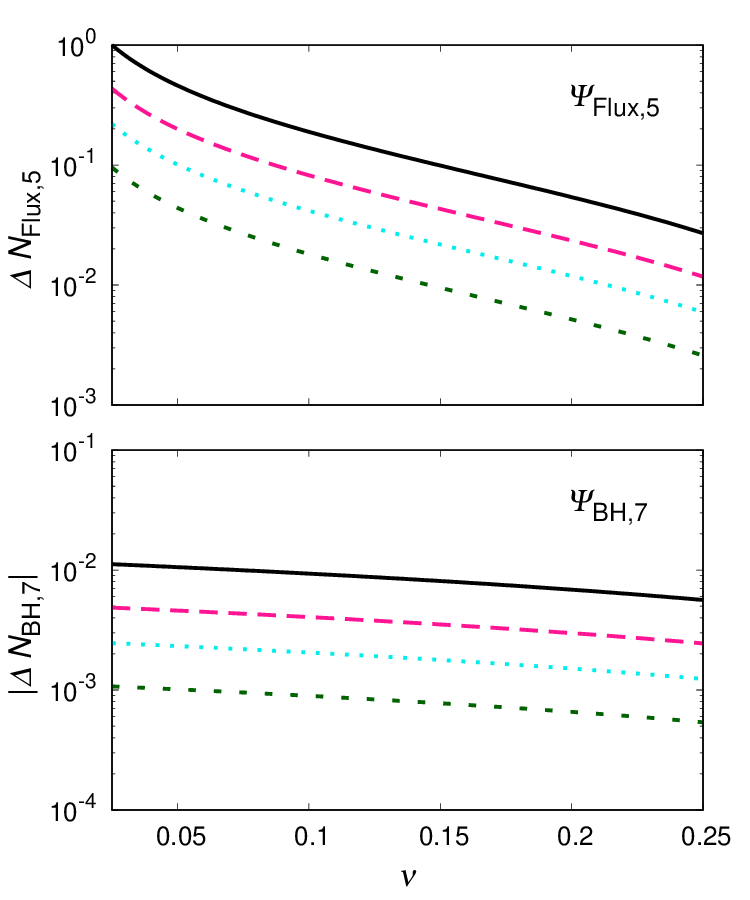}
\end{minipage}
\qquad 
\begin{minipage}[t]{.45\hsize}
  \centering
  \includegraphics[clip, width=\columnwidth]{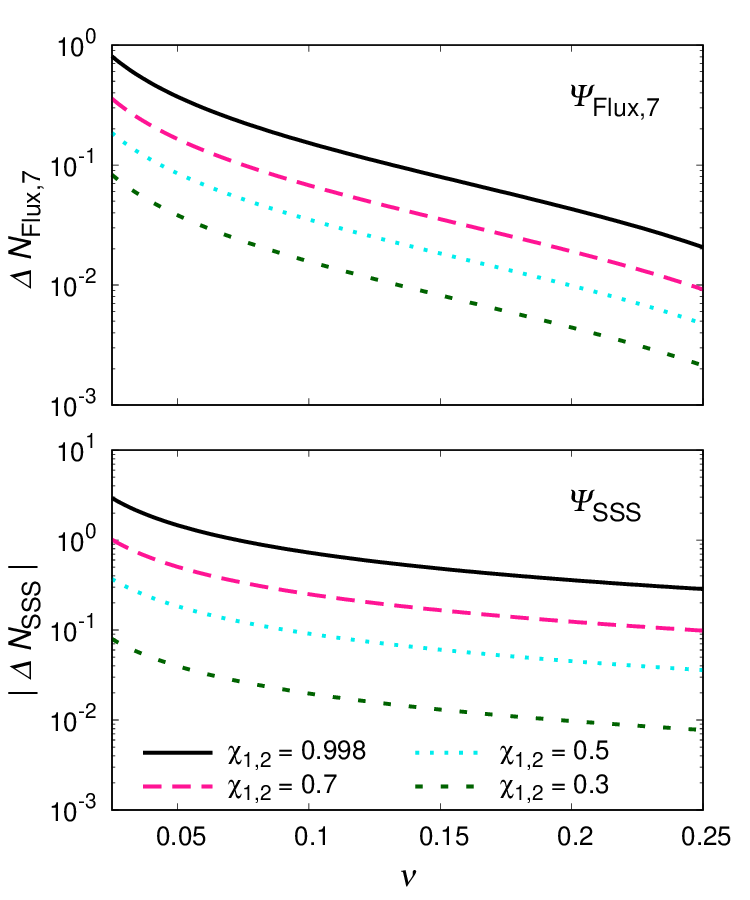}
  \end{minipage}
\end{tabular}
\caption{The relative number of GW cycles $\Delta N$ 
accumulated in ground based detectors,
LIGO/VIRGO/KAGRA frequency band $m f \in [0.0035,\,0.018]$ 
as a function of the initial symmetric mass ratio $\nu$ for different values 
of the initial aligned spins $\chi_1 = \chi_2$, 
for the contribution of the LO horizon-flux term 
$\Psi^{\mathrm {F2}}_{\mathrm {Flux},5}$ (Top left), 
the NLO horizon-flux term 
$\Psi^{\mathrm {F2}}_{\mathrm {Flux},7}$ (Top right), 
the LO term due to the secular change in BH intrinsic parameters
$\Psi^{\mathrm {F2}}_{\mathrm {BH},7}$ (Bottom left) 
and (for comparison) the LO cubic-in-spin term without the BH absorption 
$\Psi^{\mathrm {F2}}_{\mathrm {SSS}}$ (Bottom right). 
A nearly extremely spinning BBH with $\chi_{1,2} = 0.998$ is in 
the Novikov-Thorne limit for BHs spun up by accretion~\cite{Thorne:1974ve}. 
For $\Delta N_{\mathrm {BH},7}$ and $\Delta N_{\mathrm {SSS}}$, 
their absolute values 
$|\Delta N_{\mathrm {BH},7}|$ and $|\Delta N_{\mathrm {SSS}}|$ are plotted 
because they become negative in this parameter region.}
\label{fig:LIGO}
\end{figure}

Figure~\ref{fig:LIGO} shows the contributions of 
$\Psi^{\mathrm {F2}}_{\mathrm {Flux},5}$,
$\Psi^{\mathrm {F2}}_{\mathrm {Flux},7}$ and 
$\Psi^{\mathrm {F2}}_{\mathrm {BH},7}$ 
(including their prefactors in~\eqref{phi-F2-0})
to $\Delta N_{\mathrm {Flux},5}$, $\Delta N_{\mathrm {Flux},7}$ 
and $\Delta N_{\mathrm {BH},7}$, respectively, 
as a function of the initial value of the mass ratio $\nu$ 
with different initial values of the aligned spins 
$\chi_1 = \chi_2$ accumulated 
within the GW frequency $m f \in [0.0035,\,0.018]$ 
in terms of the initial total mass $m$. 
The choice of our frequency band comes from the fact that 
this agrees with the inspiral portion of 
the ``PhenomD'' model~\cite{Khan:2015jqa} and thus 
it is the most relevant band for plausible BBH parameters measured 
by ground-based GW detector such as Advanced LIGO, Advanced Virgo and KAGRA.
For comparison, we also show the same results for 
the cubic-in-spin pieces 
$(3 v^2 \Psi^{\mathrm {F2}}_{\mathrm {SSS}}) / (128 \nu)$ 
in the point-particle phase $\Psi^{\mathrm {F2}}_{\infty}(v)$, 
which generates $\Delta N_{\mathrm {SSS}}$ 
at $3.5$PN order [see~\eqref{F2-inf}]. 
In this case, we find that individual contributions 
$\Delta N_{\mathrm {Flux},5},\Delta N_{\mathrm {Flux},7}$ 
and $\Delta N_{\mathrm {BH},7}$ are all negligible.
They are always smaller than $|\Delta N_{\mathrm {SSS}}|$ 
and, in particular, 
the value of $\Delta N_{\mathrm {BH},7}$ is highly suppressed 
due to the prefactor $\nu$ for $\Psi^{\mathrm {F2}}_{\mathrm {BH},7}$ 
in~\eqref{phi-F2-0}; we have $\Delta N_{\mathrm {BH},7} \sim \nu^{0}$ 
while others scale as 
$\Delta N_{\mathrm {Flux},5} \sim \Delta N_{\mathrm {Flux},7} 
\sim \Delta N_{\mathrm {SSS}} \sim \nu^{-1}$. 
We also note that the magnitude of 
$\Delta N_{\mathrm {Flux},5},\Delta N_{\mathrm {Flux},7}, 
\Delta N_{\mathrm {BH},7}$ and $\Delta N_{\mathrm {SSS}}$  
become smaller 
for BBHs with the same magnitude of spins anti-aligned 
with the orbital angular momentum.
These results are consistent with the previous study 
by Alvi~\cite{Alvi:2001mx}, 
where for BBHs with the total mass $m$ ranging from 
$5.0 M_\odot$ to $50.0 M_\odot$ and aligned spins $\chi_{1,2} = 0.998$, 
only negligible contribution of $\Delta N_{\mathrm {Flux},5}$ is observed.

However, we find that the sum of $\Delta N_{\mathrm {Flux},5}$ 
and $\Delta N_{\mathrm {Flux},7}$ are marginally non-negligible 
for BBHs with high-mass ratio $\nu \lesssim 0.05$
and high spins $ \chi_{1,2} \gtrsim 0.90$. 
In~\fref{fig:LIGO}, we see that NLO ($3.5$PN) horizon-flux contribution 
$\Delta N_{\mathrm {Flux},7}$ can be as much as 
LO ($2.5$PN) horizon-flux contribution $\Delta N_{\mathrm {Flux},5}$. 
The origin of these comparable contributions can be easily
understood from the fact that for the given $\nu$ and $\chi_{1,2}$ 
the NLO phase coefficient $\Psi^{\mathrm {F2}}_{\mathrm {Flux},7}$ 
in~\eqref{phi-F2-0} [or~\eqref{F2-H}] is $O(10)$ larger than 
the LO phase coefficient $\Psi^{\mathrm {F2}}_{\mathrm {Flux},5}$. 
Because of this, the corresponding NLO horizon-flux term 
in the integrand~\eqref{def-N} also becomes larger than 
the LO horizon-flux term when $m f \gtrsim  0.0110$ in our case. 
In fact, the sum of $\Delta N_{\mathrm {Flux},5}$ and 
$\Delta N_{\mathrm {Flux},7}$ are almost the same as 
the cubic-in-spin contribution $\Delta N_{\mathrm {SSS}}$ 
and they could compensate each other; recall that 
$\Delta N_{\mathrm {SSS}}$ is negative while 
$\Delta N_{\mathrm {Flux},5}$ and $\Delta N_{\mathrm {Flux},7}$ are positive.
Therefore, by contrast to the prior belief~\cite{Alvi:2001mx}, 
the horizon-flux contributions to the number of GW cycles $N$ 
could be marginally non-negligible 
when we account for \textit{both} the LO and NLO terms.

\begin{figure}[tbp]
\begin{tabular}{cc}  
\begin{minipage}[t]{.45\hsize}
  \centering
  \includegraphics[clip, width=\columnwidth]{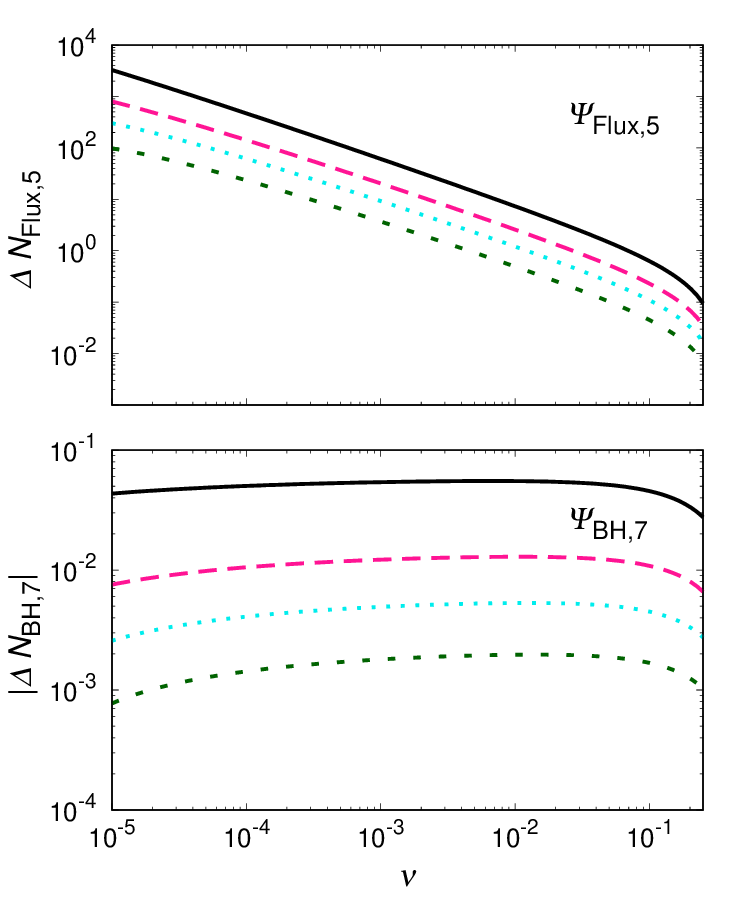}
\end{minipage}
\qquad 
\begin{minipage}[t]{.45\hsize}
  \centering
  \includegraphics[clip, width=\columnwidth]{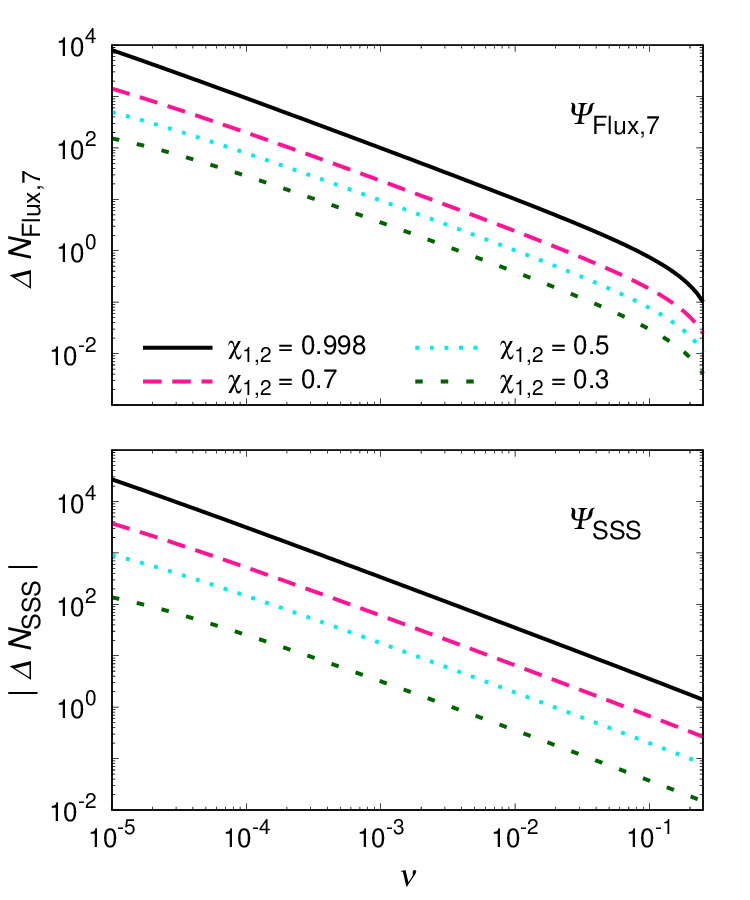}
  \end{minipage}
\end{tabular}
\caption{The relative number of GW cycles $\Delta N$ 
accumulated in a space based detector, LISA frequency band 
$m f \in [2.0 \times 10^{-4} \, \nu^{-3/8},\,  m f_{\mathrm {ISCO}}]$ 
as a function of the initial symmetric mass ratio $\nu$ for different values 
of the initial aligned spins $\chi_1 = \chi_2$. 
For $\chi_{1,2} = 0.998$, the upper cutoff of the frequency 
is $m f_{\mathrm {pole}} \sim 0.109$, which is smaller than 
the ISCO frequency $m f_{\mathrm {ISCO}} \sim 0.134$, 
to validate TaylorF2 model. 
Labels are the same as for~\fref{fig:LIGO}, 
and we plot the absolute value of $\Delta N_{\mathrm {BH},7}$ 
and $\Delta N_{\mathrm {SSS}}$ as they become negative 
in the given parameter region.}
\label{fig:LISA}
\end{figure}

Figure~\ref{fig:LISA} is similar to~\fref{fig:LIGO} 
except that the frequency band is now chosen as 
$m f \in [2.0 \times 10^{-4} \nu^{-3/8},\, m f_{\mathrm {ISCO}}]$,  
where $f_{\mathrm {ISCO}}$ is twice the frequency 
of the innermost stable circular orbit (ISCO) for Kerr geometry 
with mass $m$ and aligned equal-spin $\chi_{1,2}$~\cite{Bardeen:1972fi}, 
namely, 
\begin{equation}\label{def-ISCO}
\pi m f_{\mathrm {ISCO}} \equiv 
\{ ( 3 + Z_2 - [(3 - Z_1) (3 + Z_1 + 2 Z_2)]^{1/2} )^{3/2} 
+ \chi_{1,2}\}^{-1}
\end{equation}
with 
$
Z_1 \equiv 1 + (1 - \chi_{1,2}^2)^{1/3} 
[(1 + \chi_{1,2})^{1/3} + (1 - \chi_{1,2})^{1/3} ] 
$
and 
$
Z_2 \equiv (3 \chi_{1,2}^2 + Z_1^2)^{1/2}
$.
Roughly speaking, the choice of this frequency band is motivated 
by the one year observation of BBHs with the initial total mass 
$m \sim O(10^{6}) M_{\odot}$ before reaching ISCO~\cite{Berti:2004bd}, 
and this covers plausible BBH parameters for LISA.  
In this case, the ech horizon-flux contribution 
$\Delta N_{\mathrm {Flux},5}$ and $\Delta N_{\mathrm {Flux},7}$ 
become non-negligible for BBHs with high-mass ratio $\nu \lesssim 0.01$ 
and moderate aligned-spins $\chi_{1,2} \gtrsim 0.50$. 
They rapidly grow as 
$\Delta N_{\mathrm {Flux},5} \sim \nu^{-1} \ln(\nu)$ 
and 
$\Delta N_{\mathrm {Flux},7} \sim \Delta N_{\mathrm {SSS}}  
\sim \nu^{-5/4}$ as $\nu$ decreases, 
and their values become as large as $O(10^{3}) \sim O(10^{4})$ 
when $\nu \sim 10^{-5}$, depending on the values of $\chi_{1,2}$. 
While our results for high-mass-ratio inspirals 
$(\nu \lesssim 10^{-3})$ are only indicative 
because the PN approximation is not so accurate 
for these BBHs~\cite{Yunes:2008tw,Zhang:2011vha,Sago:2016xsp,Fujita:2017wjq}, 
these results are basically consistent with previous results made 
by many authors, which showed that for quasicircular, 
extreme mass-ratio BBH inspirals with $\nu \sim 10^{-6}$ and 
nearly extremal spins the horizon-flux effects significantly 
increases the duration of inspiral
phase~\cite{Fujita:2014eta,Tagoshi:1997jy,Hughes:2001jr,Yunes:2009ef}.

Meanwhile,~\fref{fig:LISA} shows that $\Delta N_{\mathrm {BH},7}$ 
is negligible even for the LISA-type detector. 
One would question this result because the scaling 
$\Delta N_{\mathrm {BH},7} \sim \nu^{-1/4}$ is expected 
given the prefactor $\nu$ for $\Psi^{\mathrm {F2}}_{\mathrm {BH},7}$ 
in~\eqref{phi-F2-0}, 
and it could be pronounced when $\nu$ is sufficiently small. 
However, the explicit calculation shows that its coefficient that 
depends on the spins is at most $O(10^{-3})$ even when $\chi_{1,2} = 0.998$. 
Given the range of mass ratio that we consider here, 
the term $\Delta N_{\mathrm {BH},7} \sim \nu^{-1/4}$ therefore 
does not dominate compared to other $\nu$-dependent terms 
in $\Delta N_{\mathrm {BH},7}$, which have the positive powers in $\nu$.

\subsection{Results 2: the mismatch for Advanced LIGO and LISA}
\label{subsec:result2}

While the GW detectors are sensitive to the evolution of the GW phase 
of BBH inspirals, the relative number of GW cycles 
$\Delta N_{\mathrm {Flux},5}$, $\Delta N_{\mathrm {Flux},7}$ 
and $\Delta N_{\mathrm {BH},7}$ accumulated in a detector's 
frequency-band is not a robust estimator 
for the amount of information contained in each phase correction 
$\Psi^{\mathrm {F2}}_{\mathrm {Flux},5}, 
\Psi^{\mathrm {F2}}_{\mathrm {Flux},7}$ 
and $\Psi^{\mathrm {F2}}_{\mathrm {BH},7}$. 
For the measurement of BBHs by Advanced LIGO and LISA, 
such information becomes manifest only when aided 
by the matched filtering~\cite{Allen:2005fk}.

In section~\ref{sec:match}, we compute an optimized cross-correlation 
(usually called \textit{match}~\cite{Owen:1995tm,Owen:1998dk}) 
between the two TaylorF2 waveforms with and without each phase correction 
$\Psi^{\mathrm {F2}}_{\mathrm {Flux},5}, 
\Psi^{\mathrm {F2}}_{\mathrm {Flux},7}$ 
and $\Psi^{\mathrm {F2}}_{\mathrm {BH},7}$ due to BH absorption 
as a measure of template imperfection; 
the definition of the match is detailed below in section~\ref{subsec:MF}. 
The match is weighted by the detector noise spectrum 
that we hope to observe the GW signal with, 
and can quantify the \textit{faithfulness}~\cite{Damour:1997ub} 
of our PN template in observing GW signals of BBHs by Advanced LIGO and LISA. 
A match of unity means that the template is a very precise representation 
of the target GW signal. A value less than unity means that the
template reproduces the signal only imperfectly and hence it is unfaithful. 
We below consider that the \textit{mismatch} ($\equiv$ 1 - match) 
due to template imperfection is \textit{significant} 
if it is larger than $3\%$~\cite{Damour:1997ub}. 

We compute the match by taking the target GW signal 
to be the TaylorF2 waveforms in~\eqref{h-F2} with the complete $3.5$PN phase 
$\Psi^{\mathrm {F2}}_{3.5{\mathrm {PN}}}$ in~\eqref{phi-F2-0}, 
and by taking five different templates to be the same TaylorF2 waveforms 
as the target signal except each template neglects 
one of the following phase contributions: 
(1) the LO horizon-flux term 
$\Psi^{\mathrm {F2}}_{\mathrm {Flux},5}$; 
(2) the NLO horizon-flux term 
$\Psi^{\mathrm {F2}}_{\mathrm {Flux},7}$; 
(3) the LO term due to the secular change 
in BH mass and spins $\Psi^{\mathrm {F2}}_{\mathrm {BH},7}$; 
(4) all phase terms due to BH absorption 
$\Psi^{\mathrm {F2}}_{\mathrm {H,all}}
=
\{
\Psi^{\mathrm {F2}}_{\mathrm {Flux},5},\, 
\Psi^{\mathrm {F2}}_{\mathrm {Flux},7},\,
\Psi^{\mathrm {F2}}_{\mathrm {BH},7}
\}$;
(5) (for comparison) the LO cubic-in-spin term 
$\Psi^{\mathrm {F2}}_{\mathrm {SSS}}$ in the non-absorption, 
point-particle phase term $\Psi^{\mathrm {F2}}_{\infty}$ 
[recall~\eqref{phi-F2-0}]. 
The match for Advanced LIGO accounts for the noise curve of 
its ``zero-detuning, high-power'' 
configuration~\cite{Harry:2010zz} [see~\eqref{S-LIGO}], 
which is the design goal of Advanced LIGO, and we consider the frequencies 
in the interval $m f \in [0.0035,\,0.018]$, 
using the same setup in figure~\ref{fig:LIGO}. 
At the same time, the match for LISA takes into account 
its latest noise curve~\cite{Babak:2017tow}, 
which includes the improvement successfully demonstrated 
by LISA Pathfinder~\cite{Armano:2016bkm}~[see~\eqref{S-LISA}], 
and we chose the frequency range  
$m f \in [2.0 \times 10^{-4} \, \nu^{-3/8},\, m f_{\mathrm {ISCO}}]$ 
with twice the frequency of the ISCO for Kerr geometry 
$f_{\mathrm {ISCO}}$ [recall~\eqref{def-ISCO}], 
in order to echo to the setup in figure~\ref{fig:LISA}. 
The rationale for our choice of the frequency range is that 
we are interested in analyzing the (dis)agreement of TaylorF2 
with and without the correction due to BH absorption, 
and our interval can provide a baseline for fair comparison 
between such two different TaylorF2;  
recall that the different choice for the frequency range 
affects each template in the same way. 
A full discussion about mismatch calculation 
is presented in section~\ref{subsec:match}. 

\begin{figure}[tbp]
\begin{tabular}{cc}  
\begin{minipage}[t]{.45\hsize}
  \centering
  \includegraphics[clip, width=\columnwidth]{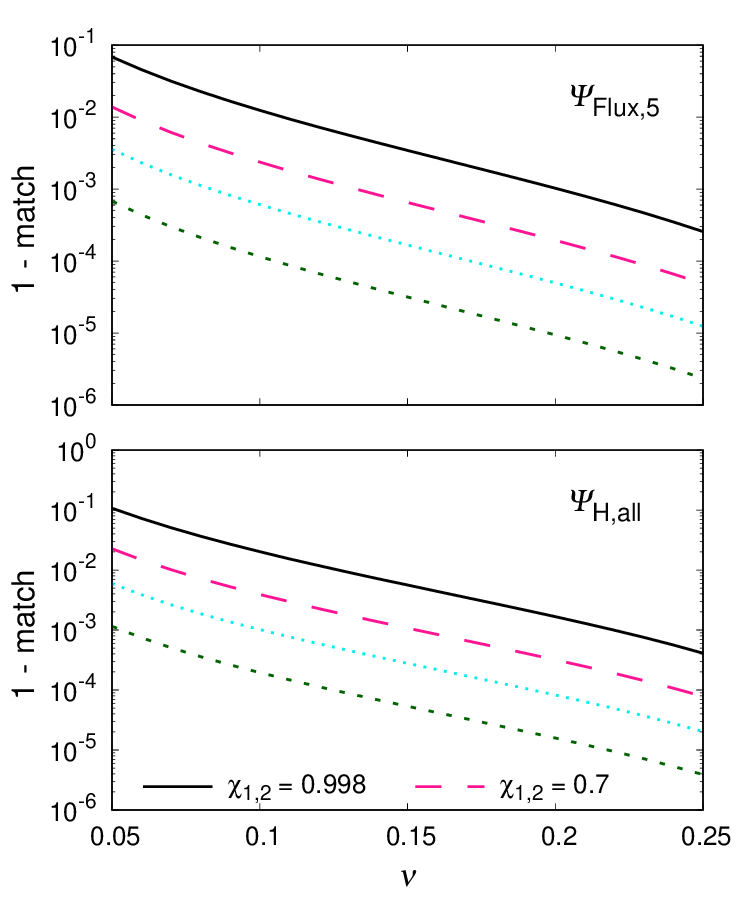}
\end{minipage}
\qquad 
\begin{minipage}[t]{.45\hsize}
  \centering
  \includegraphics[clip, width=\columnwidth]{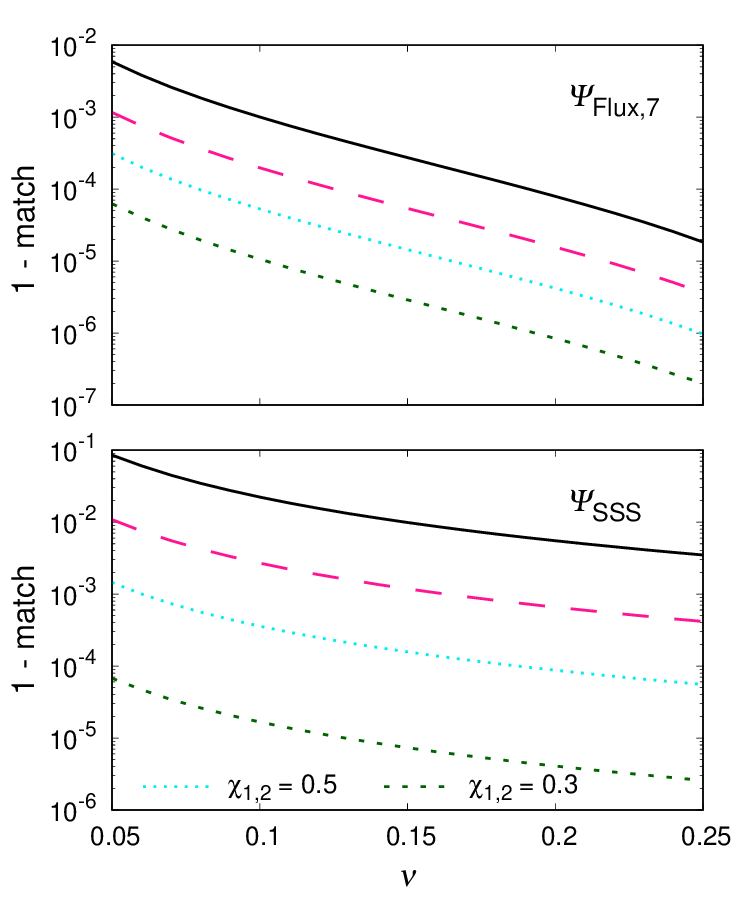}
  \end{minipage}
\end{tabular}
\caption{The mismatch ($ \equiv 1 - {\mathrm {match}}$) 
between the two TaylorF2 templates with and without 
each phase correction due to BH absorption 
accumulated in the Advanced LIGO frequency band $m f \in [0.0035,\,0.018]$, 
where the initial total mass chosen to be $m = 60.0 M_{\odot}$. 
The results are plotted as a function of 
the symmetric mass ratio $\nu$ for different values 
of the initial aligned-spins $\chi_1 = \chi_2$, 
and they are grouped into four panels according to 
what is neglected in the TaylorF2 phase in~\eqref{phi-F2-0}; 
\textit{Top left:} the neglect of the LO horizon-flux term 
$\Psi^{\mathrm {F2}}_{\mathrm {Flux},5}$. 
\textit{Top right:} the neglect of the NLO horizon-flux term 
$\Psi^{\mathrm {F2}}_{\mathrm {Flux},7}$. 
\textit{Bottom left:} the neglect of all phase terms due to BH absorption 
$\Psi^{\mathrm {F2}}_{\mathrm {H,all}}$, 
including all horizon-flux terms 
$\Psi^{\mathrm {F2}}_{\mathrm {Flux},5}$ and
$\Psi^{\mathrm {F2}}_{\mathrm {Flux},7}$ 
as well as the LO term due to the secular change in BH intrinsic parameters 
$\Psi^{\mathrm {F2}}_{\mathrm {BH},7}$. 
\textit{Bottom right:} (for comparison) 
the neglect of the LO cubic-in-spin term  
$\Psi^{\mathrm {F2}}_{\mathrm {SSS}}$ in the non-absorption, 
point-particle phase term $\Psi^{\mathrm {F2}}_{\infty}$. } 
\label{fig:LIGOcs}
\end{figure}

For BBHs observable by Advanced LIGO, 
we summarize the mismatch for each imperfect template 
with our complete BH-absorption TaylorF2 in figure~\ref{fig:LIGOcs}, 
assuming that the initial total mass of the BBH is $m = 60.0 M_{\odot}$ 
(this corresponds to $f \in [11.8,\, 60.9]$~Hz) for different values of 
initial aligned-spins. 
Because the mismatch due to the neglect of 
$\Psi^{\mathrm {F2}}_{\mathrm {BH},7}$ is always below 
the $10^{-7}$ mark even with nearly extremal aligned-spins 
$\chi_{1,2} = 0.998$, the corresponding mismatch is not displayed here.

Overall, the mismatch in figure~\ref{fig:LIGOcs} follows 
the similar trend for the relative contribution to GW cycles $\Delta N$ 
in figure~\ref{fig:LIGO} and supports the broader conclusion 
that we can draw from it; 
the effects BH absorption on GWs are significant 
for high-mass-ratio, high-aligned-spin BBHs. 
More specifically, not including the phase term 
$\Psi^{\mathrm {F2}}_{\mathrm {H,all}}$ introduces 
the significant mismatch when the BBH is in high-mass ratio regime 
$\nu \lesssim 0.083$ with nearly extremal aligned-spins 
$\chi_{1,2} \sim 0.098$ 
as well as at very high-mass ratio region $\nu \lesssim 0.05$ 
with large aligned-spins $\chi_{1,2} \gtrsim 0.70$. 
Looking at top two panels, we particularly see that 
the inclusion of $\Psi^{\mathrm {F2}}_{\mathrm {Flux},5}$ is crucial 
as this dominates the mismatch due to the neglect of 
$\Psi^{\mathrm {F2}}_{\mathrm {H,all}}$; 
by contrast to $\Delta N_{\mathrm {Flux},7}$ in figure~\ref{fig:LIGO}
the mismatch due to neglecting $\Psi^{\mathrm {F2}}_{\mathrm {Flux},7}$ 
never becomes significant for BBHs considered here.  
In the bottom two panels, we also see that 
the mismatch due to neglecting $\Psi^{\mathrm {F2}}_{\mathrm {SSS}}$ is 
as significant as that of $\Psi^{\mathrm {F2}}_{\mathrm {H,all}}$; 
recall that $\Psi^{\mathrm {F2}}_{\mathrm {H,all}}$ consists of  
both linear-in-spin and cubic-in-spin terms [see~\eqref{F2-H}]. 
This suggests that one would also need to include 
$\Psi^{\mathrm {F2}}_{\mathrm {SSS}}$ 
if we wish to fully exploit information about BH absorption 
in $\Psi^{\mathrm {F2}}_{\mathrm {H,all}}$ 
by measuring BBHs by Advanced LIGO.

We emphasize that the mismatch in figure~\ref{fig:LIGOcs} 
is \textit{only indicative};  
the resulting mismatch depends on the upper and lower cutoff 
frequencies that we consider here. 
Their interpretation in the context of actual GW search is thus delicate. 
For instance, if we instead take the frequency interval 
$m f \in [0.0035,\, m f_{\mathrm {ISCO/pole}}]$, 
the mismatch plotted in figure~\ref{fig:LIGOcs} is 
\textit{increased} by a factor of $O(10)$. 
In this case, the mismatch due to neglecting 
$\Psi^{\mathrm {F2}}_{\mathrm {H,all}}$ can be above the $3\%$ mark 
even for the high-mass-ratio BBH $(\nu \lesssim 0.10)$ 
with moderate aligned-spins ($\chi_{1,2} \gtrsim 0.50)$ 
as well as the almost equal-mass ratio BBH $(\nu \lesssim 0.20)$ 
with near extremal aligned-spins $(\chi_{1,2} \sim 0.998)$. 
Another example is the frequency interval $m f \in [10.0m,\, 845.0]$ 
considered by Alvi~\cite{Alvi:2001mx}. 
For BBHs with the initial total mass $ m= \{5.0, 20.0, 50.0\} M_{\odot}$ 
and the initial aligned-spins $\chi_{1,2} = 0.998$, 
Alvi showed for such BBHs with symmetric mass ratio $\nu \gtrsim 0.16$
(corresponding to $m_2 /m_1 \leq 4 $) 
that $\Delta N_{\mathrm {Flux},5}$ accumulated in his frequency range 
is far less than one radian; see Table IV of~\cite{Alvi:2001mx}. 
Focusing on his configurations $m= 20(50) M_{\odot}$ 
with the mass ratio $\nu = \{0.25, 0.22, 0.16\}$ 
(corresponding to his choice $m_2 /m_1 = \{1, 2, 4\}$), however, 
we find that the neglect of $\Psi^{\mathrm {F2}}_{\mathrm {Flux},5}$ 
accumulated in the same frequency range produces the mismatch
$0.74\% (0.37\%),\, 1.7\% (0.87\%)$ and $7.1\% (3.8\%)$, respectively. 
While these values have no direct implication to an actual GW search for BBHs, 
we feel that more investigation would be needed to assess 
if the corrections $\Psi^{\mathrm {F2}}_{\mathrm {Flux},5}$ 
as well as $\Psi^{\mathrm {F2}}_{\mathrm {H,all}}$ in a realistic template 
are truly too small to be observed 
in current ground-based detectors, 
including Advanced LIGO/Virgo and KAGRA.

\begin{figure}[tbp]
\begin{tabular}{cc}  
\begin{minipage}[t]{.45\hsize}
  \centering
  \includegraphics[clip, width=\columnwidth]{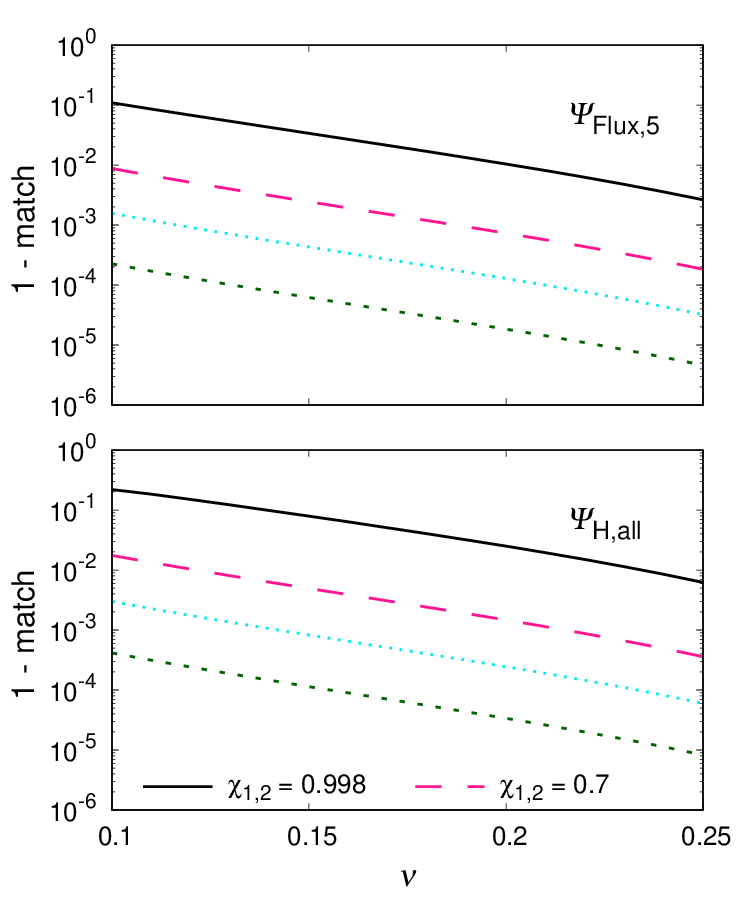}
\end{minipage}
\qquad 
\begin{minipage}[t]{.45\hsize}
  \centering
  \includegraphics[clip, width=\columnwidth]{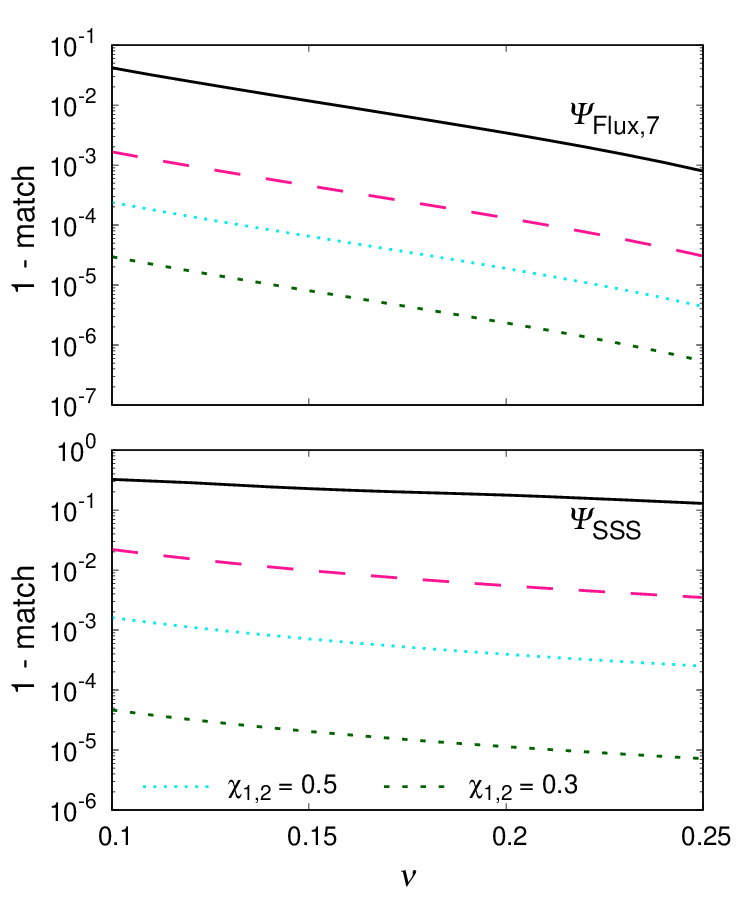}
  \end{minipage}
\end{tabular}
\caption{The mismatch ($ \equiv 1 - {\mathrm {match}}$) 
between the two TaylorF2 templates with and without 
each phase correction due to BH absorption accumulated 
in a space based detector, LISA frequency band 
$m f \in [2.0 \times 10^{-4} \, \nu^{-3/8},\, m f_{\mathrm {ISCO}}]$, 
where the initial total mass is chosen to be $m = 10^{6} M_{\odot}$. 
For $\chi_{1,2} = 0.998$, the upper cutoff of the frequency 
is $m f_{\mathrm {pole}} \sim 0.109$. 
The label and grouping of the panels are the same as 
for figure.~\ref{fig:LIGOcs}.}
\label{fig:LISAcs}
\end{figure}

Figure~\ref{fig:LISAcs} is similar to figure~\ref{fig:LIGOcs} 
except that it summarizes the mismatch of each imperfect template 
for supermassive BBHs observable by LISA, 
assuming that the initial total mass of the BBHs is 
$m = 10^{6} M_{\odot}$ (corresponding, e.g., $f \lesssim 0.022$Hz 
for nearly extremal aligned-spins ${\chi}_{1,2} = 0.998$). 
Because our TaylorF2 is not expected to be reliable 
for very high mass-ratio configuration ($\nu \lesssim O(10^{-2})$), 
we only consider supermassive BBHs with low mass-ratio 
in the range $\nu \in [0.10,\,0.25]$
%
\footnote{Although our analysis is not expected to be valid for 
BBHs in high mass-ratio regime, 
we point out that all mismatch becomes above the $O(10)\%$ mark 
irrespective to the value of $\chi_{1,2}$: 
except for that due to the neglect of $\Psi^{\mathrm {F2}}_{\mathrm {BH},7}$, 
which is always below the $O(10^{-3})\%$ mark. 
This is expected from figure~\ref{fig:LISA} 
because the large phase difference between two templates 
could easily produce the significant mismatch.}.
%
As was the case for BBHs observable by Advanced LIGO, 
figure~\ref{fig:LISAcs} supports the broader conclusion 
that we can draw from the figure~\ref{fig:LISA}, 
which showed  the relative contribution to GW cycles $\Delta N$; 
the mismatch due to the neglect $\Psi^{\mathrm {F2}}_{\mathrm {H,all}}$ is 
significant when supermassive BBHs is 
in almost equal-mass regime $\nu \lesssim 0.19$ 
with nearly extremal aligned-spins ${\chi}_{1,2} \gtrsim 0.998$ 
as well as in the moderate mass-ratio regime $\nu \lesssim 0.10$ 
with large aligned-spins $\chi_{1,2} \gtrsim 0.70$.  
For the case of LISA, we also see that the mismatch from neglecting 
$\Psi^{\mathrm {F2}}_{\mathrm {Flux},7}$ 
can become significant if the BBH is at large mass ratio $\nu \sim 0.1$ 
with nearly extremal aligned-spins ${\chi}_{1,2} \gtrsim 0.998$. 
This suggests that the NLO contribution 
$\Psi^{\mathrm {F2}}_{\mathrm {Flux},7}$ 
is likely to be as important as LO contribution 
$\Psi^{\mathrm {F2}}_{\mathrm {Flux},5}$ 
for measuring high-mass-ratio, high-aligned-spin supermassive BBHs. 
However, the mismatch due to the neglect of 
$\Psi^{\mathrm {F2}}_{\mathrm {BH},7}$ never becomes significant 
for all supermasiive BBHs that we consider here.
The resulting mismatch is always below $10^{-5}$ mark, 
and we therefore do not plot it in figure~\ref{fig:LISAcs}.

It is interesting to observe that for supermassive BBHs 
in the large-spin regime ($\chi_{1,2} \gtrsim 0.70$), 
the mismatch from neglecting $\Psi^{\mathrm {F2}}_{\mathrm {SSS}}$ 
depends on the mass ratio only weakly. 
It becomes therefore significant over the full range of mass ratios. 
In addition, $\Psi^{\mathrm {F2}}_{\mathrm {SSS}}$ produces  
the larger mismatch than that from $\Psi^{\mathrm {F2}}_{\mathrm {H,all}}$ 
unless supermasiive BBHs have relatively small aligned-spins 
$\chi_{1,2} \lesssim 0.50$.  
This prompts us to suggest that the contribution 
$\Psi^{\mathrm {F2}}_{\mathrm {SSS}}$ to the PN template should not be 
neglected for supermassive BBHs to measure the BH-absorption effects by LISA.

Before concluding our paper, 
once more, we should point out that the mismatch presented  
in figure~\ref{fig:LISAcs} is only indicative 
as its value is quite sensitive to our choice of the frequency range. 
If the upper cutoff frequency is instead chosen as twice of the frequency 
of the ISCO of the Schwarzschild space time 
$m f_{\mathrm {ISCO, Sch}} (\sim 0.022) 
\ll m f_{\mathrm {pole}} (\sim 0.109) $ [recall~\eqref{def-ISCO}], 
all mismatch plotted in figure.~\ref{fig:LISAcs} 
is \textit{decreased} by a factor of $O(10)$. 
We find, nevertheless, that our results give a useful idea 
about the impact of BH absorption in the context of LISA, 
and we expect this helps future modeling efforts 
for GWs from suppermassive BBHs.

\vspace{5mm}

\textit{In summary}, we found the following four main results: 
\begin{enumerate}
\item 
For the case of BBHs observable by Advanced LIGO, 
not including the LO horizon-flux phase term 
$\Psi^{\mathrm {F2}}_{\mathrm {Flux},5}$ 
will typically cause a significant phasing error in PN templates 
if BBHs are in high-mass ratio $(\nu \lesssim 0.10)$ 
and high aligned-spins $({\chi}_{1,2} \gtrsim 0.70)$ regime;
\item 
For the case of supermassive BBHs observable by LISA,
the inclusion of both $\Psi^{\mathrm {F2}}_{\mathrm {Flux},5}$ and  
$\Psi^{\mathrm {F2}}_{\mathrm {Flux},7}$ in PN templates will be mandatory. 
Almost equal-mass supermassive BBHs $(\nu \lesssim 0.20)$ 
with nearly extremal aligned-spins $({\chi}_{1,2} \sim 0.998)$ 
can produce significant phasing errors if they are neglected, 
and they become more evident for the lower-spin BBHs 
with decreasing mass-ratio;
\item
The phasing error due to the LO secular change 
in BH masses and spins $\Psi^{\mathrm {F2}}_{\mathrm {BH},7}$ 
is truly negligible for all BBHs measurable by Advanced LIGO and LISA; 
\item
The $3.5$PN cubic-in-spin phase term $\Psi^{\mathrm {F2}}_{\mathrm {SSS}}$ 
in the non-absorption, point-particle phase term 
$\Psi^{\mathrm {F2}}_{\infty}$ [recall~\eqref{phi-F2-0}] 
causes the significant phasing error 
as much as $\Psi^{\mathrm {F2}}_{\mathrm {H,all}}$, 
thus should not be neglected when one includes 
the correction due to BH absorption to PN templates. 
\end{enumerate}

\vspace{5mm}

\textit{A note on the limitation of this work---}
Our work is intended as a proof-of-principle, 
hence there are various limitations, 
which are summarized as follows.  

First, our templates did not account for the precession effects in BBHs; 
despite its importance, for example, in the detection of BBHs 
by Advanced LIGO~\cite{Bustillo:2016gid,Williamson:2017evr},
the (PN) computation of the horizon flux 
for the spin-precessing BBHs is beyond the current state of the art, 
even in the test spinning-particle limit. 
Second, we considered only the dominant $(2,2)$ 
spin-weighted-spherical-harmonic mode 
of GW waveforms, and neglected the contribution from their higher harmonics. 
For the case of BBH with high-mass-ratio and high-spins, 
these modes can be significant during the last stages of
inspiral~\cite{Arun:2008kb} and the corresponding systematic errors 
to the templates due to neglecting such higher modes become 
larger~\cite{Varma:2014jxa,Varma:2016dnf}.
While the main goal of this paper motivates us to limit 
the scope to the dominant $(2,2)$ mode, 
our work could be improved by the corrections from these higher modes, 
for instance, making use of the method recently proposed 
in~\cite{London:2017bcn}. 
Third, although there have been highly accurate predictions for inspiral, 
merger and ringdown gravitational waveforms for BBHs, 
we did not use the effective-one-body waveforms~\cite{Bohe:2016gbl} 
or the phenomenological (so-called PhenomD) waveforms~\cite{Khan:2015jqa} 
as our reference GW signals. 
As seen in figures~\ref{fig:LIGOcs} and~\ref{fig:LISAcs}, 
the corrections to the template due to BH absorption become 
most significant in high-mass-ratio regime with high aligned-spins, 
but very few NR simulation is currently available there.
These models therefore are not well calibrated for such ``extreme'' BBHs.
Given the spirit of this work, 
we rather focus on only the (early) inspiral phase, sticking 
to use the simple TaylarF2 model in the GW frequency range 
$mf \leq 0.018$ for Advanced LIGO and $f \leq f_{\mathrm {ISCO}}$ for LISA. 
In these cases, the resulting matches would be difficult 
to interpret as they have no immediate application in actual 
GW searches for BBHs, but they provide a conceptually clean setup to study 
the impact of BH absorption on the template. 

Despite such limitations, we feel that our findings in this paper would 
motivate further explorations of GW data-analysis tasks 
for spinning, non-precessing BBHs. 
The impact of BH absorption for parameter-estimate predictions 
for LIGO-type and LISA-type detectors could be investigated in future work. 
The issue of the corresponding systematic biases 
in (PN) template families due to the effect of BH absorption 
would be also a relevant extension of our work; 
systematic errors in GW observations 
was already discussed in the context of neutron-star
binaries~\cite{Yagi:2013baa,Yagi:2013sva,Favata:2013rwa} 
as well as BBHs~\cite{Hinder:2013oqa,Kumar:2016dhh,Abbott:2016wiq}.

\vspace{5mm}

In the remainder of the paper, we detail the results presented above. 
In section~\ref{sec:PN-formulae},
we summarize the PN expressions for the center-of-mass binding energy,
the GW energy flux to infinity and the horizon energy fluxes. 
Then, the secular change in BH masses and spins is calculated.
In section~\ref{sec:PN-dynamics},
we present the corrected binding energy and GW energy flux 
as well as the generalized balance equation
that include the appropriate effects of the secular change
in the BH mass and spin during the inspiral phase. 
The PN template families from the generalized balance equation are
given in section~\ref{sec:Taylor}. 
Finally in section~\ref{sec:match} 
we calculate the match between frequency-domain PN templates 
with and without the effect of BH absorption. 
The mismatch for other BBH configurations that were not covered 
in this section are displayed in section~\ref{subsec:match-mas}.

\section{The PN approximants}
\label{sec:PN-formulae}

For the convenience of the reader, 
we in this section recapitulate the explicit $3.5$PN expressions 
for the center-of-mass binding energy $E$,  
the GW energy flux carried out to infinity ${F}_{\infty}$ 
and the horizon energy (and angular momentum) fluxes 
${\cal F}_{\mathrm {H}}^i\,(i = 1,\,2)$ for the pinning, non-precessing 
quasicircular BBH with constant masses $m_i$ and 
constant-in-magnitude spin vectors ${\bf S}_i$, 
beyond their LO Newtonian terms.
We shall import many relevant results from the review~\cite{Blanchet:2013haa} 
and references~\cite{Chatziioannou:2016kem,Marsat:2014xea,Bohe:2015ana} 
as well as literatures cited in these references. 
We then compute the LO PN expressions 
for the secular change in the BH mass and spin 
during the inspiral phase, which are at $3.5$PN and $2$PN order, 
respectively. 

Following the notation used in~\cite{Blanchet:2013haa}, 
we define the projected value of the spin vectors ${\bf{S}}_i$ 
along the unit normal ${\boldsymbol \ell}$ to the orbital plane 
by ${S}_i \equiv {\bf{S}}_i \cdot {\boldsymbol {\ell}}$, 
and introduce two combinations of them: 
\begin{equation}\label{def-spin}
S_{\ell} \equiv {S_1} + {S_2}\,,
\quad
\Sigma_{\ell}
\equiv \frac{{S_2}}{X_2} - \frac{{S_1}}{X_1}\,,
\end{equation}
where $X_i \equiv m_i/m$.
We here also introduce the dimensionless spin parameter $\chi_i$ by 
\begin{equation}\label{def-chi}
{\chi}_i \equiv \frac{S_i}{m_i^2}\,.
\end{equation}
In our notation, the spin parameter takes $-1 < {\chi}_i < 1$; 
its positive (negative) value corresponds to the aligned (anti-aligned) 
configuration with respect to the orbital angular momentum of the binary. 
Assuming $m_1 < m_2$, \eqref{def-spin} and~\eqref{def-chi} are related by 
\begin{align}\label{S-to-chi}
S_{\ell} &=
\frac{m^2}{4} (1 + \Delta)^2 \chi_1 + \frac{m^2}{4} (1 - \Delta)^2 \chi_2 \,, 
\cr
\Sigma_{\ell} &= 
- \frac{m^2}{2} (1 + \Delta) \chi_1 + 
\frac{m^2}{2} (1 - \Delta) \chi_2 \,,
\end{align}
which can be inverted to give 
\begin{equation}\label{chi-to-S}
\chi_1 = 
\frac{2 \{ (1 + \Delta) S_{\ell} - 2 \nu \Sigma_{\ell} \}}
{m^2 (1 + \Delta)^2}\,, 
\quad
\chi_2 = 
\frac{2 \{ (1 - \Delta) S_{\ell} + 2 \nu \Sigma_{\ell} \}}
{m^2 (1 - \Delta)^2}\,,
\end{equation}
where 
\begin{equation}\label{def-Delta}
\Delta \equiv - \sqrt{1 - 4 \nu}\,,
\quad
\nu \equiv \frac{m_1 m_2}{m^2} = X_1 X_2\,.
\end{equation}

\subsection{The binding energy 
and the energy flux emitted to the infinity}
\label{subsec:EandF}

Schematically, the $3.5$PN binding energy is expressed as 
\begin{equation}\label{def-E}
E \equiv 
-\frac{m \nu}{2} v^2 
\left(
E_{\mathrm {NS}}
+
\frac{v^3}{m^2} E_{\mathrm {SO}}
+
\frac{v^4}{m^4} E_{\mathrm {SS}} 
+
\frac{v^7}{m^6} E_{\mathrm {SSS}} 
+ 
O(v^8)
\right)\,,
\end{equation}
where $E_{\mathrm {NS}}, E_{\mathrm {SO}}, 
E_{\mathrm {SS}}$ and $E_{\mathrm {SSS}}$ denote the non-spinning, 
spin-orbit (SO, linear-in-spin), spin-spin (SS, quadratic-in-spin),
and spin-spin-spin (SSS, cubic-in-spin) contributions to $E$ 
and all depend on the parameters 
$m,\nu,S_{\ell}$ and $\Sigma_{\ell}$ as a function of $v$.
Their explicit expressions are provided 
in~(232) and (415) of~\cite{Blanchet:2013haa} 
for $E_{\mathrm {NS}}$ 
\footnote{
The $4$PN expression for $E_{\mathrm {NS}}$ is recently 
computed in~\cite{Damour:2014jta,Damour:2016abl,Bernard:2016wrg}. 
}
and $E_{\mathrm {SO}}$ respectively, 
in~(3.33) of~\cite{Bohe:2015ana} for $E_{\mathrm {SS}}$ 
and in~(6.17) of~\cite{Marsat:2014xea} for $E_{\mathrm {SSS}}$. 
The expressions for $E_{\mathrm {SS}}$ and $E_{\mathrm {SSS}}$ include 
constants $\kappa_{\pm}$ and $\lambda_{\pm}$ that characterize 
the deformation of a small object in a binary 
due to its own spin angular-momentum, 
and they take $\kappa_+ = \lambda_+ = 2$ and $\kappa_- = \lambda_- = 0$ 
for a BBH~\cite{Marsat:2014xea,Bohe:2015ana}. 
The expressions are then given by 
\begin{align}\label{E-compt}
E_{\mathrm {NS}}
&= 
1+ \left( - \frac{3}{4} - \frac{1}{12 }{\nu} \right) v^{2} 
+ \left( -\frac{27}{8} + \frac{19}{8} \nu 
-\frac{1}{24} \nu^2 \right) {v}^{4} \cr 
& \quad + \left\{ -\frac {675}{64}+ 
\left( \frac{34445}{576}-\frac{205\,{\pi}^{2}}{96} \right) \nu 
-\frac {155}{96}\nu^2 -\frac {35}{5184} \nu^3 \right\} {v}^{6} \,, \cr
E_{\mathrm {SO}}
&= 
\left(
\frac{14}{3}\,S_{\ell} + 2 \Delta \Sigma_{\ell} \right)
+ \left\{ \left( 11 - \frac {61}{9} \nu \right) S_{\ell} 
+ \left( 3 - \frac{10}{3}\nu \right) \Delta\,\Sigma_{\ell} \right\} v^{2} \cr 
& \quad + 
\left\{  
\left( {\frac {135}{4}}- \frac {367}{4}\nu + \frac {29}{12} \nu^2 \right) 
S_{\ell} 
+ \left( \frac {27}{4} -39\,{\nu} + \frac{5}{4}\,{\nu}^{2} \right) 
\Delta\,\Sigma_{\ell} \right\} {v}^{4}\,, \cr
E_{\mathrm {SS}}
&= 
-4 S^{2}_{\ell} - 4 \Delta \Sigma_{\ell}S_{\ell} 
- (1 - 4 \nu) \Sigma_{\ell}^{2} \cr 
& \quad + 
\left\{ 
 \left( -\frac {25}{9}+ \frac{10}{3}\,\nu \right) S_{\ell}^{2} 
 + \left( \frac{10}{3} + \frac{10}{3}\,{\nu} \right) 
 \Delta \Sigma_{\ell} S_{\ell}  
 + \left( \frac{5}{2}- \frac{15}{2}\nu -\frac{10}{3} \nu^2 \right) 
 \Sigma_{\ell}^{2} \right\} v^2 \,, \cr
E_{\mathrm {SSS}}
&= 
-8 S_{\ell}^{3} - 16 \Delta \Sigma_{\ell} {S}_{\ell}^{2} 
-10 (1 - 4 \nu) \Sigma_{\ell}^{2} S_{\ell} 
- 2 (1 - 4 \nu) \Delta \Sigma_{\ell}^{3} \,.
\end{align} 
We note that $E$ is \textit{complete} up to the relative 3.5PN order. 

Similarly, the $3.5$PN energy flux associated 
with gravitational radiation carried out to infinity is written as
\begin{equation}\label{def-Finf}
{F}_{\infty} \equiv 
\frac{32}{5} \nu^2 v^{10} 
\left(
{F}_{\mathrm {NS}}
+
\frac{v^3}{m^2} {F}_{\mathrm {SO}}
+
\frac{v^4}{m^4} {F}_{\mathrm {SS}} 
+
\frac{v^7}{m^6} {F}_{\mathrm {SSS}}
+
O(v^8)
\right)\,,
\end{equation}
where ${F}_{\mathrm {NS}}, {F}_{\mathrm {SO}}, 
{F}_{\mathrm {SS}}$ and ${F}_{\mathrm {SSS}}$ 
denote the non-spinning, SO, SS, 
and SSS parts of ${F}_{\infty}$ 
and all depend on the parameters 
$m,\nu,S_{\ell}$ and $\Sigma_{\ell}$ as a function of $v$.
Their explicit expressions are provided 
in~(314) and (414) of~\cite{Blanchet:2013haa} 
for ${F}_{\mathrm {NS}}$ and ${F}_{\mathrm {SO}}$ 
\footnote{
The $4$PN expression for ${F}_{\mathrm {SO}}$ is also available 
in~\cite{Marsat:2013caa}.}
respectively,
in~(4.14) of~\cite{Bohe:2015ana} for ${F}_{\mathrm {SS}}$ 
and in~(6.19) of~\cite{Marsat:2014xea} for ${F}_{\mathrm {SSS}}$.
Once again setting $\kappa_\pm$ and $\lambda_\pm$ 
in ${F}_{\mathrm {SS}}$ and ${F}_{\mathrm {SSS}}$ 
as $\kappa_+ = \lambda_+ = 2$ and $\kappa_- = \lambda_- = 0$,  
their expressions for a BBH read 
\begin{align}\label{Finf-compt}
{F}_{\mathrm {NS}}
&= 
1 + \left( -\frac {1247}{336}- \frac {35}{12} \nu \right) v^2
+4\,\pi \,{v}^{3}
+ \left( 
-\frac{44711}{9072}+ \frac{9271}{504} \nu + \frac{65}{18} \nu^2 
\right) {v}^{4} \cr
& \quad +
\left( -\frac {8191}{672}- \frac {583}{24}\nu \right) \pi \,{v}^{5} 
+
\left\{ \frac {6643739519}{69854400} + \frac{16}{3} {\pi}^{2}
-\frac {1712}{105} \,\gamma_{\rm E}
-\frac {856}{105}\,\ln  \left( 16\,v^2 \right)  \right. \cr
& \quad+ 
\left. 
\left( -\frac {134543}{7776} + \frac {41\,}{48} {\pi}^{2}\right) \nu 
- \frac{94403}{3024} \nu^2 -\frac{775}{324} \nu^3 \right\} {v}^{6} \cr 
& \quad+ 
\left( 
- \frac{16285}{504} + \frac{214745}{1728} \nu + \frac{193385}{3024} \nu^2 
\right) \pi \,{v}^{7}\,, \cr
{F}_{\mathrm {SO}}
&= 
-4\,S_{\ell} - \frac{5}{4} \Delta \,\Sigma_{\ell} 
+ \left\{  
\left( -\frac{9}{2} + \frac{272}{9} \nu \right) S_{\ell} 
+ \left( -\frac{13}{16} + \frac{43}{4} \nu \right) \Delta \,\Sigma_{\ell} 
\right\} v^2 \cr 
& \quad +
\left( -16 S_{\ell} - \frac {31}{6}\,\Delta \Sigma_{\ell} \right) \pi \,v^3 \cr
& \quad + 
\left\{  
\left( \frac {476645}{6804} + \frac {6172}{189} \nu 
- \frac{2810}{27} \nu^2 \right) S_{\ell} 
+ \left( \frac {9535}{336} + \frac {1849}{126} \nu 
-\frac{1501}{36}\nu^2 \right) \Delta \,\Sigma_{\ell} \right\} {v}^{4}\,, \cr
{F}_{\mathrm {SS}}
&= 
8 {S}_{\ell}^{2} + 8 \Delta \Sigma_{\ell} S_{\ell} 
+ \left( \frac {33}{16}- 8\nu \right) {\Sigma}^{2}_{\ell} 
- \left\{  
\left( {\frac {3839}{252}} + 43 \nu \right) {S}^{2}_{\ell} 
+ \left( \frac {1375}{56} + 43 \nu \right) \Delta \Sigma_{\ell} S_{\ell} 
\right.
\cr
& \quad \left.
+ \left( \frac {227}{28} - \frac {3481}{168} \nu - 43 {\nu}^{2} \right) 
{\Sigma}^2_{\ell} \right\} v^2 + O(v^3) \,, \cr
{F}_{\mathrm {SSS}}
&= 
-\frac{16}{3} {S}_\ell^{3} + \frac{2}{3} \Delta \Sigma_\ell S_\ell^2 
+ \left( \frac{9}{2} - \frac{56}{3} \nu \right) {\Sigma}_{\ell}^{2} S_{\ell} 
+ \left( \frac{35}{24} -6 \nu \right) \Delta {\Sigma}_{\ell}^{3}\,,
\end{align} 
where $\gamma_{\rm E} = 0.577 \dots$ are Euler constant. 
As we indicated in ${F}_{\mathrm {SS}}$ with the term of $O(v^3)$, 
${F}_{\infty}$ is \textit{incomplete} because 
its $1.5$PN terms due to the SS tail contribution, 
which affects ${F}_{\infty}$ at the $3.5$PN order, 
are still unknown~\cite{Bohe:2015ana}, 
except those in the test-particle limit
$\nu \to 0$~\cite{Tanaka:1996ht,Messina:2017yjg,FS:2017pp}. 
These terms have yet to be computed in the future. 

In addition, we note that ${F}_{\infty}$ has 
a pole at $v = v_{\mathrm {pole}}$ and become even \textit{negative} 
when $v > v_{\mathrm {pole}}$. 
This unphysical behavior was first pointed out 
in the test-particle limit~\cite{Tagoshi:1996gh,Isoyama:2012bx}, 
and the same issue happens for the finite-mass case. 
Fortunately, our examination suggests $v_{\mathrm {pole}} \gtrsim 0.70$ 
for a broad range of the BBH parameters, 
where the PN expansion will lose
accuracy~\cite{Yunes:2008tw,Zhang:2011vha,Sago:2016xsp}, 
and it should be taken over by the result 
from NR simulations. 
Moreover, if we compare the value of $v_{\mathrm {pole}}$ 
to the nominal value of the ISCO of the Kerr metric [recall~\eqref{def-ISCO}]
with mass $m$ and \textit{(dimensionless) effective spin} 
$\chi_{\mathrm {eff}}$ adopted in the phenomenological (``Phenom'')
model~\cite{Ajith:2009bn,Santamaria:2010yb,Khan:2015jqa} 
\begin{equation}\label{def-chi-eff}
\chi_{\mathrm {eff}}
\equiv
\frac{(1 + \Delta)}{2} \chi_1 + \frac{(1 - \Delta)}{2} \chi_2 \,, 
\end{equation}
we have $v_{\mathrm {ISCO}} \gtrsim 0.70$ only 
when $\chi_{\mathrm {eff}} \gtrsim 0.98$. 
These results assure that the pole in ${F}_{\infty}$ is not 
a serious obstacle in modeling BBHs during the inspiral phase
except each individual BH has the nearly extremal spins,  
reaching the Novikov-Thorne limit $\chi_{1,2} = 0.998$ 
for BHs spun up by accretion~\cite{Thorne:1974ve}.

\subsection{The horizon fluxes}
\label{subsec:H-flux}

Reference~\cite{Chatziioannou:2016kem} provides the ready-to-use 
formulas of the horizon energy and angular-momentum fluxes 
for a BH in a quasicircular BBH, 
up to $1.5$PN order beyond the LO horizon fluxes, 
which are at $2.5$PN order beyond the quadrupolar fluxes. 
While their expressions at the relative $1.5$PN order 
do not recover the expressions 
in the test-particle limit $\nu \to 0$~\cite{Fujita:2014eta,Tagoshi:1997jy} 
\footnote{
Strictly speaking, the test-particle limit in this sentence refers 
to the case where a point particle is assumed to be non-spinning. 
For the case of the horizon fluxes 
emitted from a \textit{spinning test-particle} 
on the circular equatorial orbit in Kerr spacetime, 
the current state of the art is a numerical work by Han~\cite{Han:2010tp} 
as well as an analytical work by Sago and Fujita~\cite{FS:2017pp} 
to $6$PN order beyond the quadrupolar fluxes, 
although Han's result is controversial~\cite{Harms:2015ixa}. 
},
for the purpose of our analysis at the $3.5$PN accuracy level, 
we only need them up to the relative $1$PN order 
that do agree with the test-mass results. 
Importing the results in~(42) and (43) of~\cite{Chatziioannou:2016kem}, 
the horizon energy and angular-momentum fluxes 
for the spinning, non-precessing, quasicircular BBH are defined by 
\begin{equation}\label{def-FH}
{\cal F}_{\mathrm {H}}^i 
\equiv 
\left \langle \frac{d m_i}{d t} \right \rangle 
= 
\Omega_{\mathrm {tidal}} 
(\Omega_{\mathrm {H}} - \Omega_{\mathrm {tidal}})\, C_{v,i}\,, 
\end{equation} 
and 
\begin{equation}\label{FHt-SM0}
\left \langle \frac{d |S_i|}{d t} \right \rangle
\equiv
\Omega_{\mathrm {tidal}}^{-1}
\left \langle \frac{d m_i}{d t} \right \rangle\,, 
\end{equation}
respectively; the angular-bracket operation 
in~\eqref{def-FH} and~\eqref{FHt-SM0} indicates 
the long-term average~\cite{Poisson:2004cw}.  
Here, $t$ is the PN barycentric time,  
the angular velocity of the tidal field $\Omega_{\mathrm {tidal}} $ is 
\begin{equation}\label{Omega-t}
\Omega_{\mathrm {tidal}} 
= 
\epsilon_i \frac{v^3}{m} \left(1 - \nu v^2 + O(v^3) \right)\,,
\end{equation}
with $\epsilon_i = +1 \,(-1)$ if the orbital and spin angular momentum
of the unperturbed BH are aligned (anti-aligned), 
the angular velocity of the unperturbed Kerr BH is 
\begin{equation}\label{Omega-H}
\Omega_{\mathrm H} 
= 
\frac{|\chi_i|}{2 m_i (1 + \sqrt{1 - \chi_i^2})}\,,
\end{equation}
and [recall that $X_i \equiv m_i/m$.]
\begin{align}
C_{v,i}
&\equiv 
-\frac{16}{5} m_i^2  X_i^2 \nu^2 ( 1 + \sqrt{1 - {\chi_i}^2} ) v^{12} \cr 
& \quad \times 
\left\{
1 + 3 \chi_i^2 
+ \frac{1}{4}
\left(
3 (2 + \chi_i^2) + 2 X_i  (2 + 3 X_i) (1 + 3 \chi_i^2)
\right) v^2 + O(v^{3})
\right\}\,. 
\end{align}

For the purpose of computing a change in the BH mass and spin, 
it is more useful to write the horizon fluxes 
in terms of the velocity $v$, which is a coordinate-invariant parameter. 
Once again importing the results in~(46) and (47) 
of~\cite{Chatziioannou:2016kem}, they are given by 
\begin{equation}\label{FHX-SM0}
\left \langle \frac{d m_i}{d v} \right \rangle 
=
\Omega_{\mathrm {tidal}} \left \langle \frac{d |S_i|}{d v} \right \rangle 
=
\Omega_{\mathrm {tidal}} (\Omega_{\mathrm {H}} - \Omega_{\mathrm {tidal}}) \,
C^{'}_{v,i}\,,
\end{equation}
where 
\begin{align}
C^{'}_{v,i} 
&\equiv 
-\frac{1}{2}  m_i^3 X_i \nu (1 + \sqrt{1 - {\chi_i}^2} ) v^{3} \cr 
& \quad \times 
\left\{
1 + 3 \chi_i^2 
+ 
\left(
\frac{1}{336} (1247 + 2481 \chi_i^2) 
+ \frac{5}{4} (3 - X_i) X_i (1 + 3 \chi_i^2)
\right) v^2 + O(v^{3}) 
\right\}\,. \cr 
& \quad  
\end{align}

It should be noted that~\eqref{def-FH} and~\eqref{FHX-SM0} 
are displayed in factorized-resumed forms 
because of the factor $\Omega_{\mathrm H} -\Omega_{\mathrm {tidal}}$. 
As a result, these expressions include uncontrolled $1.5$PN 
remainders of $O(v^3)$, which are not allowed to keep in our analysis. 
To avoid contamination with such uncontrolled higher PN-order terms, 
we substitute~\eqref{Omega-t} and~\eqref{Omega-H}
into~\eqref{def-FH} and~\eqref{FHX-SM0} 
and then re-expand them in the power of $v$. 
The resulting power series is then explicitly truncated 
at the relative $1$PN order beyond the LO horizon fluxes, 
which gives 
\begin{align}
\label{FHt-SM}
\left \langle \frac{d m_i}{d t} \right \rangle
&= 
-\frac{8}{5} X_i^3 \nu^2 \s_i v^{15} 
-\frac{2}{5} X_i^3 \nu^2 
\left\{
3 \chi_i ( 2 + \chi_i^2)  - 4 \nu \s_i + 4 X_i \s_i + 6 X_i^2 \s_i
\right\} v^{17}  + O(v^{18}) \,, \\
\label{FHX-SM}
\left \langle \frac{d m_i}{d v} \right \rangle
&= 
-\frac{1}{4} X_i^3 m \nu \s_i v^{6} \cr 
& \quad 
-\frac{1}{16} X_i^3 m \nu 
\left\{
\chi_i \left(
\frac{1247}{84} + \frac{827}{28} \chi_i^2
\right) - 4 \nu \s_i + 15 X_i \s_i -5 X_i^2 \s_i
\right\} v^{8} + O(v^{9}) \,, \cr 
& 
\end{align}
where 
\begin{equation}\label{def-fraks}
\s_i \equiv \chi_i(1 + 3 \chi_i^2)\,.
\end{equation}
These PN expressions are manifestly at $3.5$PN order 
beyond the LO quadrupolar piece 
of the PN energy flux carried to the infinity. 
[For example, compare~\eqref{FHt-SM} to the PN energy flux 
to infinity in~\eqref{def-Finf}, where the LO PN-term is at $O(v^{10})$.]
In the rest of our analysis, we use only these fully expanded forms 
as the horizon energy fluxes 
and similarly for the horizon angular-momentum fluxes.

\subsection{
Mass and spin evolution of a spinning black hole 
in the quasicircular BBH}
\label{subsec:secularH}

The flux formulas in~\eqref{FHX-SM} can be solved iteratively 
to give the secular changes in $m_i$ and $S_i$ during the inspiral phase 
as a function of $v$. 
When we compute the LO solutions of the secular changes, 
the quantities $m$, $\nu$ and $\chi_i$ that 
appear in the right-hand-side expressions of~\eqref{FHX-SM} 
are taken to be constants and hence we can integrate them immediately. 
Making use of the relation $\epsilon_i |S_i| = S_i$, 
the LO solutions in terms of the parameters 
$m$, $\nu$, $S_{\ell}$ and $\Sigma_{\ell}$ 
in~\eqref{def-E} and~\eqref{def-Finf} are given by 
\begin{align}\label{evolve-mSv}
m (v) &= m^{\mathrm I} + \delta m(v)\,,
\quad
\nu (v) = \nu^{\mathrm I} + \delta \nu(v)\,, \cr 
S_{\ell} (v) &= S_{\ell}^{\mathrm I} + \delta S_{\ell}(v)\,,
\quad 
\Sigma_{\ell} (v) = 
\Sigma_{\ell}^{\mathrm I} + \delta \Sigma_{\ell} (v)\,.
\end{align}
Here, the quantities $m^{\mathrm I}$,
$\nu^{\mathrm I}$, $S_{\ell}^{\mathrm I}$ 
and $\Sigma_{\ell}^{\mathrm I}$ are the initial values 
of $m$, $\nu$, $S_{\ell}$ and $\Sigma_{\ell}$, respectively. 
The secular changes $\delta m(v)$,
$\delta \mu(v)$, $\delta S_{\ell}(v)$ 
and $ \delta \Sigma_{\ell} (v)$ are given by
\begin{align}\label{def-Dm}
\delta m (v) 
&=
\frac{1}{56}
m^{\mathrm I} 
\left(
C_1^m \s^{\mathrm I}_1 + C_2^m \s^{\mathrm I}_2 
\right) v^7 +  O(v^9)\,, \cr 
\delta \nu (v) 
&=
\frac{1}{56} \nu^{\mathrm I} 
\left(
C_1^\nu \s^{\mathrm I}_1 + C_2^\nu \s^{\mathrm I}_2
\right)v^7 +  O(v^9)\,, \\
\label{def-DS}
\delta S_{\ell} (v) 
&=
\frac{1}{32} 
(m^{\mathrm I})^2 
\left(
C_1^S \s^{\mathrm I}_1 + C_2^S \s^{\mathrm I}_2
\right) v^4 +  O(v^6)\,, \cr
\delta \Sigma_{\ell} (v) 
&=
\frac{1}{32} 
(m^{\mathrm I})^2  
\left(
C_1^\Sigma \s^{\mathrm I}_1 + C_2^\Sigma \s^{\mathrm I}_2 
\right) v^4 +  O(v^6)\,, 
\end{align}
with coefficients
\begin{align}\label{C-mS}
C_1^m &\equiv
-(1 + \Delta^{\mathrm I} )  \nu^{\mathrm I}
+(3 + \Delta^{\mathrm I} ) (\nu^{\mathrm I} )^2 \,,
\quad 
C_2^m \equiv
-(1 - \Delta^{\mathrm I}) \nu^{\mathrm I}
+(3 - \Delta^{\mathrm I}) (\nu^{\mathrm I} )^2 \,, \cr
C_1^\nu &\equiv 
   (1 + \Delta^{\mathrm I} ) \nu^{\mathrm I}
-2 (2 + \Delta^{\mathrm I} ) (\nu^{\mathrm I} )^2 \,,
\quad 
C_2^\nu \equiv
    (1 - \Delta^{\mathrm I}) \nu^{\mathrm I}
- 2 (2 - \Delta^{\mathrm I} ) (\nu^{\mathrm I} )^2  \,, \cr
C_1^S &\equiv 
- (1 + \Delta^{\mathrm I} ) \nu^{\mathrm I}
+ (3 + \Delta^{\mathrm I} ) (\nu^{\mathrm I} )^2   \,,
\quad 
C_2^S \equiv
- (1 - \Delta^{\mathrm I}) \nu^{\mathrm I}
+ (3 - \Delta^{\mathrm I}) (\nu^{\mathrm I} )^2 \,, \cr
C_1^\Sigma &\equiv 
(1 + \Delta^{\mathrm I}) \nu^{\mathrm I} 
-2 (\nu^{\mathrm I} )^2 \,,
\quad 
C_2^\Sigma \equiv
-( 1 - \Delta^{\mathrm I} ) \nu^{\mathrm I} 
+ 2 (\nu^{\mathrm I} )^2  \,.
\end{align}
Here, $\s^{\mathrm I}_i$ and $\Delta^{\mathrm I}$ 
are the initial values of parameters $\s_i$ and $\Delta$, respectively; 
recall~\eqref{def-Delta} and~\eqref{def-fraks}.
We note that~\eqref{def-Dm} and~\eqref{def-DS} 
are at $3.5$PN and $2$PN orders, respectively, 
and both vanish in the test-particle limit $\nu \to 0$. 

For example, in the case of the equal-mass aligned-spin BBH 
with $\chi_{1,2} = 0.994$ and $\nu = 0.250$, we have 
\begin{equation*}
\frac{\delta m(v = 0.350)}{m^{\mathrm I} } 
\approx
-5.66 \times 10^{-6}\,,
\quad 
\frac{\delta S(v = 0.350)}{ (m^{\mathrm I})^2 } 
\approx
-2.31 \times 10^{-4}\,,
\end{equation*}
and $\delta \nu(v) = \delta \Sigma(v) = 0$. 
Interestingly, the current NR simulation for BBHs is matured enough 
to measure such order of the change in mass and spin of each individual BH
at late time in the inspiral (although depending on the numerical resolution 
and simulation parameters~\cite{Scheel:2014ina,Carlos}). 
Therefore, the inclusion of such secular effects would be useful 
for a future comparison between simulation and PN models. 

\section{The adiabatic approximation 
with the black-hole absorption effect}
\label{sec:PN-dynamics}

In this section, we consider how the PN binding energy $E$ in~\eqref{def-E}, 
the PN energy flux to infinity ${F}_{\infty}$ in~\eqref{def-Finf} 
and the balance equation in~\eqref{balance0} are altered 
due to the horizon flux ${\cal F}_{\mathrm H}^{i}$ in~\eqref{FHt-SM} 
for each BH in a BBH and the corresponding secular change 
in its mass and spin~\eqref{evolve-mSv}.
They will be the basic inputs for modeling the BBH inspirals 
in the adiabatic approximation, including the effects of BH absorption.

\subsection{The corrected PN binding energy and energy flux}
\label{subsec:EF-H}

The PN method to deduce the expressions for 
$E$ in~\eqref{def-E} and ${F}_{\infty}$ in~\eqref{def-Finf} are based 
on binary systems of spinning point particles 
with \textit{constant} masses and spins, 
not on those of extended bodies (or tidally perturbed BHs)
with time-dependent masses $m_i(t)$ and spins $S_i(t)$ 
as a function of PN barycentric time $t$. 
At the same time, however, we recall that LO multipoles of the PN metric 
around each spinning particle for such $E$ and ${F}_{\infty}$ are chosen 
so that they coincide with the expressions for an isolated Kerr
BH~\cite{Marsat:2014xea,Bohe:2015ana}. 
Motivated by the above fact, it seems then natural to assume that 
the PN binding energy ${\cal E}$ and 
the PN GW energy fluxes ${\cal F}_{\infty}$ of a BBH 
with the time-dependent mass $m_i(t)$ and spin $S_i(t)$ of 
each individual BH are obtained through~\eqref{def-E} and~\eqref{def-Finf} 
with a simple substitution,
\begin{equation}\label{t-mS}
(m,\,\nu,\,S_{\ell},\,\Sigma_{\ell}) 
\to 
(m(t),\,\nu(t),\,S_{\ell}(t),\,\Sigma_{\ell}(t))\,.
\end{equation}
We will content ourselves with this assumption
\footnote{This assumption might be rigorously proved 
if we would start from a formulation 
for the PN two-body problem where the small body are directly modeled 
as an extended object~\cite{Futamase:2007zz}.} 
in this paper.

In practice, it is more convenient to adopt the velocity $v$ rather than $t$ 
because the explicit expressions for ${E}$ and ${F}_{\infty}$ 
as well as the secular changes in $m_i(t)$ and $S_i(t)$ in~\eqref{evolve-mSv} 
are all given as functions of the velocity $v(t)$. 
The subtle point here is the time-dependence of $v$ 
through the secularly evolving total mass $m(t)$ [recall~\eqref{def-v}]. 
We clarify this by introducing a convenient ``velocity'' parameter
\begin{equation}\label{def-nv}
\nv \equiv ( \pi m(t) f)^{1/3}\,,
\end{equation}
while we redefine the original velocity $v$ in terms of 
the initial value of the total mass $m^{\mathrm I}$ by 
\begin{equation}\label{def-v2}
v \equiv ( \pi m^{\mathrm I} f)^{1/3}\,.
\end{equation}
They are mutually related to each other through 
\begin{equation}\label{v-nv}
\frac{\nv}{v} = 
1 + \frac{1}{3} \frac{\delta m(v)}{m^{\mathrm I}} + O(v^{8})\,,
\end{equation}
and the difference is of order $3.5$PN; recall~\eqref{evolve-mSv}. 
The substitution~\eqref{t-mS} thus implies the additional insertion 
$v \to {\nv}$ for ${E}(v)$ and ${F}_{\infty}(v)$ 
in addition to $m(v)$, $\nu(v)$, $S_{\ell}(v)$ and $\Sigma_{\ell}(v)$.

Keeping this in mind, the steps required to compute 
the PN expressions for the corrected binding energy ${\cal E}(v)$ 
and the energy flux ${\cal F}_{\infty}(v)$ with BH absorption 
are as follows. 
We first compute ${\cal E}(\nv)$ and ${\cal F}_{\infty}(\nv)$, 
making use of the substitution~\eqref{t-mS} as well as $v \to {\nv}$ 
into~\eqref{def-E} and~\eqref{def-Finf}. 
It should be noted that the difference in the parameterizations 
for time-dependent BH masses and spins are all negligible 
up to the relative $5.5$PN order; recall from~\eqref{v-nv} that 
$m (\nv) = m(v) + O(v^{14})\,,
\nu (\nv) = \nu(v) + O(v^{14})\,,  
S_{\ell} (\nv) = S_{\ell}(v) + O(v^{11})
$
and
$ 
\Sigma_{\ell} (\nv) = \Sigma_{\ell}(v) + O(v^{11}) 
$.
Next, we re-expand the resulting expression in the power of $v$, 
making use of~\eqref{v-nv}.
After a simple algebra, we find that the explicit $3.5$PN expressions 
for ${\cal E}(v)$ and ${\cal F}_{\infty}(v)$ are given by 
\begin{align}
\label{def-vE}
{\cal E} (v) 
&=
E^{\mathrm I}(v) + \delta {\cal E}(v) \,, \\
\label{def-vF}
{\cal F}_{\infty}(v) 
&= 
{F}_{\infty}^{\mathrm I}(v) + \delta {\cal F}_{\infty}(v) \,.
\end{align}
Here, the non-absorption (point-particle) terms 
$E^{\mathrm I}$ and ${F}_{\infty}^{\mathrm I}$ 
are defined in terms of the \textit{initial values} of 
BH masses and spins by~\eqref{def-E} and~\eqref{def-Finf} 
with the substitution 
$(m,\,\nu,\,S_{\ell},\,\Sigma_{\ell}) 
\to 
(m^{\mathrm I} ,\,\nu^{\mathrm I} 
,\,S_{\ell}^{\mathrm I} ,\,\Sigma_{\ell}^{\mathrm I})$, respectively. 
At the same time, $\delta {\cal E}$ and $\delta {\cal F}_{\infty}$ 
describe the corrections 
\textit{due to the secular change in BH masses and spins}: 
\begin{align}\label{eval-DeltaE}
\delta {\cal E}
&= 
-\frac{m^{\mathrm I} \nu^{\mathrm I}}{2}
\left(
C_1^E\,\s^{\mathrm I}_1 + C_2^E\, \s^{\mathrm I}_2
\right) v^9 + O(v^{11})\,, \\ 
\delta {\cal F}_{\infty}
&= 
\frac{32}{5}( \nu^{\mathrm I} )^{2}
\left(
C_1^F \, \s^{\mathrm I}_1 + C_2^F \, \s^{\mathrm I}_2
\right) v^{17} + O(v^{19}) \,,
\end{align}
with the initial value of $\s^{\mathrm I}_i$ in~\eqref{def-fraks} 
and coefficients 
\begin{align}\label{C-E}
C_1^E &\equiv
-\frac{2}{21}(1 + \Delta^{\mathrm I}) \nu^{\mathrm I}
+ 
\left( 
\frac{23}{112} + \frac{5}{336} \Delta^{\mathrm I}
\right) (\nu^{\mathrm I})^{2} \,, \cr 
C_2^E &\equiv
-\frac{2}{21}(1 - \Delta^{\mathrm I} ) \nu^{\mathrm I} 
+ 
\left( 
\frac{23}{112} - \frac{5}{336} \Delta^{\mathrm I}
\right) (\nu^{\mathrm I})^{2} \,, \\
C_1^F &\equiv
\frac{167}{2688}(1 + \Delta^{\mathrm I}) \nu^{\mathrm I}
-
\left( 
\frac{41}{224} + \frac{79}{1344} \Delta^{\mathrm I}
\right) (\nu^{\mathrm I})^{2} \,, \cr 
C_2^F &\equiv
\frac{167}{2688}(1 - \Delta^{\mathrm I} ) \nu^{\mathrm I} 
- 
\left( 
\frac{41}{224} - \frac{79}{1344} \Delta^{\mathrm I}
\right) (\nu^{\mathrm I})^{2} \,. 
\end{align}
The corrections $\delta {\cal E}$ and $\delta {\cal F}_{\infty}$ 
are at $3.5$PN order beyond their LO Newtonian terms. 
They come from the Newtonian ($0$PN) terms and 
the LO ($1.5$PN) SO terms 
in~\eqref{def-E} and~\eqref{def-Finf}, 
which couple with $(\delta m (v) ,\, \delta \nu (v) )$ 
at $3.5$PN order and 
$(\delta S_{\ell} (v) ,\, \delta \Sigma_{\ell} (v) )$ 
at $2$PN order, respectively [recall~\eqref{evolve-mSv}].  
Particularly, we observe that $C_i^{E}$ and $C_i^{F}$ vanish 
in the test-particle limit $\nu^{\mathrm I} \to 0$. 
This is expected results from~\eqref{def-Dm} and~\eqref{def-DS} 
that as well vanish in this limit.

The PN expressions in~\eqref{def-vE} and~\eqref{def-vF} 
are particularly useful for practical application because they only involve 
$m^{\mathrm I},\,\nu^{\mathrm I},\,S_{\ell}^{\mathrm I}$ 
and $\Sigma_{\ell}^{\mathrm I}$, all of which are constants.

\subsection{The generalized balance equation}
\label{subsec:balanceH}

We next generalize the PN balance equation in~\eqref{balance0}
to relate the corrected $3.5$PN binding energy ${\cal E}$ 
to the corrected $3.5$PN energy flux ${\cal F}_{\infty}$,  
incorporating the horizon flux ${\cal F}_{\mathrm H}^{i}$. 
Our main objective with this subsection is to fully clarify 
the assumptions that were (implicitly) made for the PN balance equation 
with the horizon flux in the literature, 
and to show how they are naturally generalized for additionally 
including the secular change in BH masses and spins 
accumulated in the inspiral phase. 
The following is patterned after a similar discussion
produced by Le Tiec, Blanchet and Whiting~\cite{LeTiec:2011ab}.

A starting point for our analysis is the Bondi-Sachs mass-loss formula 
in full GR~\cite{Bondi:1962px,Sachs:1962wk}: 
\begin{equation}\label{mass-loss1}
\frac{d M_{\mathrm {B}}(U)}{dU} = - {\cal F}_{\infty}(U)\,,
\end{equation}
where $M_{\mathrm B}(U)$ is the Bondi mass of the system 
at a null retarded-time coordinate $U \equiv T-R$ associated 
with an asymptotically Bondi-type coordinate system 
$\{T,R\}$, and 
${\cal F}_{\infty}
\equiv
\int_{{\mathscr I}^+} |{\cal N}|^2 d \Omega$ 
is the (exact) GW energy flux 
given by the surface integral at future null infinity 
${\mathscr I}^+$ of the News function ${\cal N}$. 
Applying~\eqref{mass-loss1} to the case of a gravitationally bound isolated 
system such as a BBH, in principle, the generalized PN balance law 
for ${\cal E}$, ${\cal F}_{\infty}$ and ${\cal F}_{\mathrm H}^{i}$ 
should be derived through the implementation of~\eqref{mass-loss1} 
in the PN theory. 

However, such derivation is quite nontrivial because 
the Bondi mass $M_{\mathrm B}(U)$ is not \textit{a priori} 
guaranteed to be related with the corrected PN binding energy ${\cal E}$.  
In fact, these two notions of mass (or energy) is conceptually different: 
for asymptotically flat spacetimes 
$M_{\mathrm {B}}(U)$ is defined in the full GR 
as a surface integral at future null infinity 
while ${\cal E}$ (or rather $E$ for spinning point-particle binaries) 
is defined by one of the ten Noether charges 
associated with the Poincar\'{e} group symmetries of the specific background 
Minkowski metric, which involves the near-zone PN metric 
produced by the conservative part of the orbital dynamics of a BBH only 
(discarding the dissipative radiation-reaction effect).
Clearly, neither the background Minkowski spacetime or 
the clear distinction between the conservative and dissipative parts 
of the orbital dynamics does not exist in full GR. 

Our aim in this section is not to provide a rigorous proof 
of such identification $M_{\mathrm {B}}(U)$ to ${\cal E}$, 
following from first principles in GR. 
Instead, motivated by the similarity between 
the PN balance formula~\eqref{balance0} 
and the exact mass-loss formula~\eqref{mass-loss1} 
\footnote{The PN balance equation~\eqref{balance0} is proved 
up to the relative $1.5$PN order for generic gravitationally bound isolated 
matter source~\cite{Blanchet:1996vx}.
},
we rather \textit{postulate} that there exists a spacelike hypersurface 
$t = {\mathrm {const}.}$ in terms of the PN barycentric time $t$ such that 
\begin{equation}\label{def-PNMB}
M_{\mathrm {B}}(U) 
= 
{\cal E}(t) + \sum_{i = 1,\,2} m_i(t)\,,
\end{equation}
where $m_i(t)$ is the Christodoulou mass of each tidally perturbed BH 
in a BBH defined in terms of its apparent horizon 
(see, e.g.,~\cite{Chu:2009md} 
for its precise definition in the context of NR simulation). 
It should be emphasized that the identification~\eqref{def-PNMB} 
is not always unique because there is no unique way in relating 
the outgoing null coordinate $U$ in an asymptotically Bondi-type coordinate 
system to a time-coordinate in the near-zone of the PN source. 
Despite that, the recent comparison of the binding energy for a BBH 
between the PN theory and NR simulations suggests that
the identification~\eqref{def-PNMB} might be sound and
natural~\cite{Damour:2011fu,LeTiec:2011dp,Nagar:2015xqa}. 
Henceforth, we will thus admit the validity of~\eqref{def-PNMB}
to the relative $3.5$PN order.

Based on the above observation, 
the generalized balance equation is now obtained simply by
inserting ~\eqref{def-PNMB} into~\eqref{mass-loss1}. 
The (orbital-averaged) result reads 
\begin{equation}\label{balance1}
\frac{d {\cal E}}{dt} 
= 
- {\cal F}_{\infty} 
- \sum_{i = 1,\,2} {\cal F}_{\mathrm {H}}^i \,, 
\end{equation}
where ${\cal F}_{\mathrm {H}}^i$ is the horizon energy flux in~\eqref{def-FH}.
This is just the standard balance law used in the past, 
accounting for
${\cal F}_{\mathrm {H}}^i$~\cite{Alvi:2001mx,Fujita:2014eta,Tagoshi:1997jy,
Hughes:2001jr,Yunes:2009ef,Maselli:2017cmm}.
In addition to its physically obvious character, 
this indeed recovers~\eqref{balance0} when ${\cal F}_{\mathrm {H}}^i$ 
are absent and the mass and spin of each BH in a BBH are constant. 
Recall~\eqref{def-FH} that ${\cal F}_{\mathrm {H}}^i$ 
starts from $2.5$PN order beyond the LO quadrupolar piece
of ${\cal F}_{\infty}$ for a spinning BH. 
They therefore affects the right hand side of~\eqref{balance1} 
at that accuracy level. 
At the same time, the total time-derivative of ${\cal E}$ 
is evaluated as (taking the average over a orbital period) 
\begin{equation}\label{dEdt0}
\frac{d {\cal E}}{dt}
= 
\left(
\frac{\partial {\cal E}}{\partial t}
\right)_{m,\,S} 
+
\sum_{i = 1,\,2}
\left\langle
\frac{d m_i}{d t}
\right\rangle
\left\{
\left( \frac{\partial {\cal E}}{\partial m_i} \right)_{v,\,S}
+
\frac{1}{\Omega_{\mathrm {tidal}}}\,
\left( \frac{\partial {\cal E}}{\partial S_i} \right)_{v,\,m}
\right\}\,,
\end{equation} 
where we used~\eqref{FHt-SM0}; 
recall that ${\cal E}(t) 
= 
{\cal E}(v(t)\,; m_i(v(t)),\, S_i(v(t)))$ 
and 
$({\partial {\cal E}}/{\partial t})_{m,\,S}
=
(d v/ dt) ({\partial {\cal E}}/{\partial v})_{m,\,S}
$. 
Equations~\eqref{FHt-SM} and~\eqref{def-vE} indicate 
that $\langle {d m_i}/{d t} \rangle = O(v^{15})$ 
and  
$({\partial {\cal E}}/{\partial m_i} )_{v,\,S} 
=
(1 / \Omega_{\mathrm {tidal}})
({\partial {\cal E}}/{\partial S_i} )_{v,\,m} = O(v^2)$, 
which means that \textit{$m_i(t)$ and $S_i(t)$ in ${\cal E}(t)$
separately affects the left hand side of~\eqref{balance1}} 
at the $3.5$PN accuracy level, in addition to 
the contribution from ${\cal F}_{\mathrm {H}}^i$ to its right hand side. 

For the construction of GW models, 
it would be more convenient to rewrite~\eqref{balance1}, 
making use of~\eqref{def-FH} and~\eqref{dEdt0} 
together with the expressions in~\eqref{def-vE} and~\eqref{def-vF}. 
A simple calculation gives 
\begin{equation}\label{balanceH}
\left(
\frac{\partial {\cal E}}{\partial t}
\right)_{m,\,S} 
= 
- {\cal F}_{\mathrm {eff}}(v)\,, 
\end{equation}
where we define \textit{the effective flux} by 
\begin{equation}\label{def-Eff}
{\cal F}_{\mathrm {eff}}(v)
\equiv 
{\cal F}_{\infty}(v)
+
\sum_{i =1,2} ( 1 - \Gamma_{\mathrm H}^i(v) ) 
{\cal F}_{\mathrm H}^i (v) \,, 
\end{equation}
with the BH's growth factor 
\begin{equation}\label{def-Gamma}
\Gamma_{\mathrm H}^1
\equiv 
\left(
\frac{3}{4} - \frac{3}{4} \Delta^{\mathrm I} - \frac{1}{6} \nu^{\mathrm I} 
\right) 
v^2 + O(v^3)\,, 
\quad
\Gamma_{\mathrm H}^2
\equiv
\left(
\frac{3}{4} + \frac{3}{4} \Delta^{\mathrm I} - \frac{1}{6} \nu^{\mathrm I} 
\right)
v^2 + O(v^3)\,.
\end{equation} 
Notice that the combination 
$\Gamma_{\mathrm H}^i {\cal F}_{\mathrm H}^i$ 
is once again at $3.5$PN order beyond their LO Newtonian terms, 
and vanishes in the test-particle limit $\nu \to 0$ 
[recall \eqref{def-Delta} and~\eqref{FHt-SM}].

The expression in~\eqref{balanceH} is the same as what was given
in~\eqref{balanceH-0}. 
Once again, this is practically useful 
because it involves only the \textit{partial} derivative with respect to $t$, 
keeping $m_i(t)$ and $S_i(t)$ fixed. 
Furthermore, its explicit dependence on $m_i(t)$ and $S_i(t)$ 
only appears through their initial values, that is, 
$m^{\mathrm I},\,\nu^{\mathrm I},\,S_{\ell}^{\mathrm I}$ 
and $\Sigma_{\ell}^{\mathrm I}$. 
In this sense, 
the generalized balance equation~\eqref{balanceH} is a simple superseding of 
the original balance equation in~\eqref{balance0} with the substitution 
$(E,\,F_{\infty}) \to ({\cal E},\,{\cal F}_{\mathrm {eff}})$ 
when we wish to account for all effects of BH absorption.

\section{The PN template families with the effects of BH absorption}
\label{sec:Taylor}

Using the generalized balance equation presented in~\eqref{balanceH}, 
we in this section construct a family of ready-to-use PN templates 
for a spinning, non-precessing, quasicircular BBH
for all mass scales, including the effect of both the horizon flux 
and the secular change in the BH mass and spin accumulated in 
the inspiral phase. 

The part of our PN templates without the effects of BH absorption 
(non-absorption, point-particle part) incorporates 
all $3.5$PN corrections currently available in the literature, 
that is, we include the non-spinning, SO, 
SS and SSS terms up to the relative $3.5$PN order 
beyond the Newtonian order. 
In addition, the BH absorption part of templates incorporates 
the contribution from the horizon energy flux ${\cal F}_{\mathrm {H}}^i$ 
in~\eqref{def-FH} up to $3.5$PN order beyond the LO quadrupolar flux 
and that from the LO ($3.5$PN) secular change in the BH mass and spin 
in~\eqref{evolve-mSv}.
This provides the entirely consistent $3.5$PN templates for BBH inspirals 
with the effects of BH absorption: 
except the unknown SS pieces of the GW tails at $3.5$PN order 
in ${\cal F}_{\infty}$ that have yet to be computed 
\footnote{ 
The horizon flux ${\cal F}_{\mathrm {H}}^i$ in~\eqref{def-FH} 
and the secular change in the BH mass and spin 
in~\eqref{evolve-mSv} involve only the SO and SSS terms. 
This is consistent with the spin effects considered 
in the non-absorption, point-particle part of our $3.5$PN templates. 
}.

In the adiabatic approximation, the master equation of the model 
is the evolution equation for the orbital phase $\phi(t)$. 
Together with the definition $d \phi / d t = \pi f$ 
for the GW frequency (of the dominant harmonic) $f$, 
the generalized balance equation in~\eqref{balanceH} can be 
used to give 
\begin{align}
\label{Kepler}
\frac{d \phi}{d t} &= \frac{v^3}{m(v)} \,, \\ 
\label{dvdt}
\frac{d v}{d t} 
&=
-
\frac{{\cal F}_{\mathrm {eff}}(v)}
{( {\partial {\cal E}} / {\partial v} )_{m,\,S}}\,,
\end{align}
where the corrected binding energy ${\cal E}$ 
and the effective flux ${\cal F}_{\mathrm {eff}}$ 
are given in~\eqref{def-vE} and~\eqref{def-Eff}, respectively. 
Notice that our set of differential equations~\eqref{Kepler}
and~\eqref{dvdt} is designed 
so that it explicitly depends only on the initial values of masses 
and spins of the BBH systems.

Once we obtain the solutions $v(t)$ and $\phi(t)$, 
they can be then used to construct the strain 
of the so-called restricted waveforms $h(t)$
(for the dominant $(2,2)$ mode of the spin-weighted spherical harmonic index), 
for which we write 
\begin{equation}\label{t-strain}
h(t) = F_+ h_+(t) + F_\times F_\times (t)\,,
\end{equation}
with antenna pattern functions of the detector $F_+$ and $F_\times$ 
as well as the plus and cross polarizations 
\begin{align}\label{t-polarization}
h_+(t)  &= -\frac{2 m(t) \nu(t)}{D_L}\, ( m(t) \omega(t))^{2/3} \,
(1 + \cos^2 \Theta)\cos 2 \phi(t)\,, \cr
h_\times(t) &= -\frac{2 m(t) \nu(t)} {D_L}\, ( m(t) \omega(t))^{2/3} \, 
2 \cos \Theta \sin 2 \phi(t)\,,
\end{align}
where $D_L$ is the luminosity distance between the inspiraling BBH system 
and an observer, $\Theta$ is the inclination angle between 
the direction of the GW propagation and the orbital angular momentum, 
and $\omega = \pi f$ is the circular-orbit frequency of the BBH.

To ease the comparison between our templates and those 
without BH absorption available 
in literature~\cite{Mishra:2016whh,Damour:2000zb,Damour:2002kr,
Arun:2004hn,Buonanno:2009zt,Varma:2013kna,
Arun:2008kb,Wade:2013hoa},
we below follow the naming convention of \cite{Buonanno:2009zt} 
(with the exception of TaylorEt~\cite{Gopakumar:2007jz,Gopakumar:2007vh,
Bose:2008ix}, which we do not discuss in this paper). 
We shall provide explicit $3.5$PN expressions 
for the spin-dependent terms in the non-absorption part 
and for full BH absorption part of the template; 
the complete expressions for the spin-independent terms 
in the non-absorption part can be found 
in~\cite{Buonanno:2009zt} up to $3.5$PN order, 
and~\cite{Varma:2013kna} up to $22$PN order 
in the test-particle limit $\nu \to 0$.

Our presentation in this section is largely patterned after 
Buonanno et al.~\cite{Buonanno:2009zt} 
and Ajith~\cite{Ajith:2011ec}. 
For improved readability, 
we will thereafter drop the indices `I' 
for the initial values of quantities 
and use the symmetric and anti-symmetric combination of 
a spin parameter $\chi_i$, namely, 
\begin{equation}\label{chi-as}
\chi_s \equiv \frac{1}{2} (\chi_1 + \chi_2)\,,
\quad
\chi_a \equiv \frac{1}{2} (\chi_1 - \chi_2)\,.
\end{equation} 
They are straightforwardly converted to $S_{\ell}$ 
and $\Sigma_{\ell}$ through~\eqref{chi-to-S} 
(and vice versa via~\eqref{S-to-chi}).

\subsection{TaylorT1}
\label{subsec:T1}

We define the TaylorT1 approximant for the orbital phase 
$\phi^{\mathrm {T1}}(t)$ by the solution of the set of differential equations 
in~\eqref{Kepler} and~\eqref{dvdt}, 
leaving the PN expressions for ${\cal E}$ and ${\cal F}_{\mathrm {eff}}$ 
as they appear in these equations as a ratio of polynomials. 
The solution $\phi^{\mathrm {T1}}(t)$ 
can be obtained by numerically solving~\eqref{Kepler} and~\eqref{dvdt} 
with respect to $v$.

We usually chose $v^{\mathrm {T1}} = v_0$ 
with the total mass $m$ at $t = 0$ as initial conditions, 
and set up the initial phase $\phi^{\mathrm {T1}} = \phi_0$ 
to be either $0$ or $\pi/2$. 
Also, the waveform should be terminated before $v^{\mathrm {T1}}$ reaches 
its nominal value of $v_{\mathrm {ISCO}}$, 
which may be the ISCO of the Kerr metric
with the final value of the total mass $m(v_{\mathrm {ISCO}})$ 
and effective spin $\chi_{\mathrm {eff}}(v_{\mathrm {ISCO}})$ 
[recall~\eqref{def-chi-eff}], 
or the pole $v_{\mathrm {pole}} \sim 0.70$ 
in the PN energy flux $F_{\infty}$ 
if $v_{\mathrm {ISCO}}$ is larger than $v_{\mathrm {pole}}$. 

In this paper, we do not compute the numerical solutions 
$v^{\mathrm {T1}}(t)$ and $\phi^{\mathrm {T1}}(t)$,
but it could be straightforwardly obtained without any technical obstacle. 
A resummation method may be also useful~\cite{Isoyama:2012bx,
Pan:2010hz,Nagar:2016ayt} to avoid $v_{\mathrm {pole}}$ 
and to improve the convergence as well as 
accuracy of $F_{\infty}$.

\subsection{TaylorT4}
\label{subsec:T4}

TaylorT4 model without BH absorption 
is originally proposed in~\cite{Buonanno:2002fy}. 
Built on this, we define TaylorT4 approximant 
of the orbital phase $\phi^{\mathrm {T4}}(t)$
with BH absorption by expanding the ratio of polynomials 
${{\cal F}_{\mathrm {eff}}}
/{( {\partial {\cal E}} / {\partial v} )_{m,\,S}}$ 
in~\eqref{dvdt} to a consistent PN order, 
which is $3.5$PN order in our calculation, 
and then numerically solving~\eqref{Kepler} 
together with the obtained PN approximant of $v^{\mathrm {T4}}(t)$ as input.

We divide, for convenience, ${d v^{\mathrm {T4}}}/{d t}$ into 
\begin{equation}\label{dvdt-T4}
\frac{d v^{\mathrm {T4}}}{d t} 
= 
\frac{d v^{\mathrm {T4}}_{\infty}}{d t} 
+
\frac{d v^{\mathrm {T4}}_{\mathrm H}}{d t}\,.
\end{equation}
The no-absorption term ${d v^{\mathrm {T4}}_{\infty}}/{d t}$ is 
defined by 
\begin{equation}\label{dvdtI-T4}
\frac{d v^{\mathrm {T4}}_{\infty}}{d t} 
= 
-
\frac{{F}_{\infty}(v)}{({d E} / {d v} )}\,,
\end{equation}
after expanding the ratio of polynomials 
${F}_{\infty} / {({d E} / {d v} )}$ up to 3.5PN order 
[recall~\eqref{def-E} and~\eqref{def-Finf}]. 
The result has the following structure 
\begin{equation}\label{dvdtT4-I0}
\frac{d v^{\mathrm {T4}}_{\infty}}{d t} 
=
\frac{32}{5} \frac{\nu}{m} v^9 
\left(
{\dot v}^{\mathrm {T4}}_{\mathrm {NS}}
+
{v^3} {\dot v}^{\mathrm {T4}}_{\mathrm {SO}}
+
{v^4} {\dot v}^{\mathrm {T4}}_{\mathrm {SS}}
+
{v^7} {\dot v}^{\mathrm {T4}}_{\mathrm {SSS}}
+ 
O(v^8)
\right)\,,
\end{equation}
where ${\dot v}^{\mathrm {T4}}_{\mathrm {NS}},
{\dot v}^{\mathrm {T4}}_{\mathrm {SO}}, 
{\dot v}^{\mathrm {T4}}_{\mathrm {SS}}$ 
and 
${\dot v}^{\mathrm {T4}}_{\mathrm {SSS}}$ 
denote the non-spinning, SO, SS 
and SSS contributions. 
The full expression for ${\dot v}^{\mathrm {T4}}_{\mathrm {NS}}$ 
is given in~(3.6) of~\cite{Buonanno:2009zt}, 
and the other expressions are listed as 
\begin{align}\label{T4-compt0}
{\dot v}^{\mathrm {T4}}_{\mathrm {SO}}
&= 
\left\{  
\left( 
-{\frac {130325}{756}} + {\frac {1575529}{2592}} \nu 
-{\frac {341753}{864}} \nu^2 
+{\frac {10819}{216}} \nu^3 
\right) \chi_s  \right. \cr 
& \quad + \left. 
\left( 
\frac{130325}{756} + {\frac {796069 }{2016}} \nu -{\frac {100019}{864}} \nu^2 
\right) \Delta \chi_a 
\right\} {v}^{4} \cr 
& \quad +  
\left(  
\left( - {\frac {75}{2}} + {\frac {74}{3}} \nu \right) \chi_s
-{\frac {75}{2}} \Delta \chi_a
\right) \pi {v}^{3} \cr 
& \quad +
\left\{  
\left( 
-{\frac {31319}{1008}} + {\frac {22975}{252}} \nu - \frac{79}{3} \nu^2  
\right) \chi_s 
+ \left( -{\frac {31319}{1008}} + {\frac {1159}{24}} \nu \right) \Delta \chi_a 
\right\} {v}^{2} \cr 
& \quad + 
\left( -{\frac {113}{12}} + {\frac {19}{3}} \nu \right) \chi_s 
-{\frac {113}{12}} \Delta \chi_a \,, \cr
{\dot v}^{\mathrm {T4}}_{\mathrm {SS}}
&= 
12 \pi 
\left( 
\chi_s^2  + 2 \Delta \chi_a \chi_s + \left( 1 - 4 \nu \right) \chi_a^2
\right) {v}^{3} 
+ \left\{
\left( 
\frac{53353}{672} - \frac {16231}{96} \nu + \frac{1163}{24} \nu^2 
\right) \chi_s^2  \right. \cr 
& \quad + \left.
\left( 
\frac{53353}{336} -\frac {3165}{16} \nu 
\right) \Delta \chi_a \chi_s 
+ \left( 
\frac{53353}{672} - \frac {77575}{224} \nu + 86 \nu^2 
 \right) \chi_a^2
\right\} {v}^{2} \cr 
& \quad + 
\left( \frac {81}{16} - \frac{1}{4} \nu \right) \chi_s^2 
+ \frac{81}{8} \Delta \chi_a \chi_s
+ \left( \frac{81}{16} -20\,{\nu}  \right) \chi_a^2 \,, \cr
{\dot v}^{\mathrm {T4}}_{\mathrm {SSS}} 
&= 
\left( 
-{\frac {1559}{24}} +{\frac {519}{8}} \nu -\frac{3}{2} \nu^2 
\right) \chi_s^3
+ \left( 
-{\frac {1559}{8}} +{\frac {1531}{12}}  \nu 
\right) \Delta \chi_a \chi_s^2 \cr 
& \quad 
+ \left( 
-{\frac {1559}{8}} + {\frac {20161}{24}} \nu -{\frac {748}{3}} \nu^2 
\right) \chi_a^2 \chi_s
+\left( 
-{\frac {1559}{24}} + {\frac {773}{3}} \nu
\right) \Delta \chi_a^3\,.
\end{align} 
Our expression for ${\dot v}^{\mathrm {T4}}_{\mathrm {SO}}$ 
recovers that in appendix A of~\cite{Nitz:2013mxa} 
after correcting differences in the notation.

Similarly, we write for the BH absorption term 
${d v^{\mathrm {T4}}_{\mathrm {H}}}/{d t}$ as 
\begin{align}\label{dXdt-T4H}
\frac{d v^{\mathrm {T4}}_{\mathrm {H}}}{d t} 
= 
\frac{32}{5} \frac{\nu}{m}  v^{14} 
\left\{
{\dot v}^{\mathrm {T4}}_{{\mathrm {Flux}},5}
+
v^2 
\left(
{\dot v}^{\mathrm {T4}}_{{\mathrm {Flux}},7}
+
\nu\, {\dot v}^{\mathrm {T4}}_{{\mathrm {BH}},7}
\right)
+ 
O(v^3)
\right\}\,.
\end{align}
Above, ${\dot v}^{\mathrm {T4}}_{{\mathrm {Flux}},5}$ 
and ${\dot v}^{\mathrm {T4}}_{{\mathrm {Flux}},7}$ 
only accounts for the LO ($2.5$PN) and NLO ($3.5$PN) 
horizon-flux contributions to 
${d v^{\mathrm {T4}}_{\mathrm {H}}}/{d t} $, respectively, 
with the substitution 
$\delta m = \delta \nu = \delta \chi_s = \delta \chi_a 
= \Gamma^{i}_{\mathrm {H}} = 0$ 
[recall \eqref{chi-to-S},~\eqref{evolve-mSv} 
and~\eqref{def-Gamma}]. 
On the other hand, ${\dot v}^{\mathrm {T4}}_{{\mathrm {BH}},7}$ of 
order $3.5$PN corresponds to the residual effect of the LO secular change 
in the BH mass and spin.
We note that ${\dot v}^{\mathrm {T4}}_{{\mathrm {BH}},7}$ is suppressed 
by the prefactor of the mass ratio, $ 0.0 \leq \nu \leq 0.25$. 
Their explicit expressions are summarized as 
\begin{align}\label{T4-comptH}
{\dot v}^{\mathrm {T4}}_{\mathrm {Flux},5} 
&= 
\left( -\frac{1}{4} + \frac{3}{4} \nu \right) \chi_s (1 + 3 \chi_s^2) 
+
\left( -\frac{9}{4} + \frac{9}{4} \nu  \right) 
\Delta \chi_a \chi_s^2 \cr 
& \quad 
+ 
\left( -\frac{9}{4} + {\frac {27}{4}} \nu \right) \chi_a^2 \chi_s 
+ \left( -\frac{1}{4} + \frac{1}{4}\nu \right) \Delta \chi_a (1 + 3 \chi_a^2) 
\,, \cr
{\dot v}^{\mathrm {T4}}_{\mathrm {Flux},7} 
&= 
\left( 
-{\frac {51}{16}} + {\frac {211}{16}} \nu -9\,{{\nu}}^{2}
\right) \chi_s^3 
+ 
\left( -{\frac {153}{16}} + {\frac {327}{16}} \nu - {\frac {21}{4}} \nu^2 
\right) \Delta \chi_a \chi_s^2 \cr 
& \quad 
+ \left\{ 
\left( 
-{\frac {153}{16}} + {\frac {633}{16}} \nu - 27 \nu^2
\right) \chi_a^2 
-{\frac {11}{8}}  + \frac{16}{3} \nu - 3 \nu^2 
\right\} \chi_s \cr 
& \quad 
+ \left( 
-{\frac {51}{16}} + {\frac {109}{16}} \nu - \frac{7}{4} \nu^2 
 \right) \Delta \chi_a^3 
+ \left( 
-{\frac {11}{8}} +{\frac {31}{12}} \nu -{\frac {7}{12}} \nu^2 
\right) \Delta \chi_a \,, \cr 
{\dot v}^{\mathrm {T4}}_{\mathrm {BH},7} 
&= 
\left( {\frac {781}{448}} - {\frac {33}{8}} \nu \right) 
\chi_s (1 + 3 \chi_s^2) 
+ 
\left( 
{\frac {7029}{448}} -{\frac {1287}{224}} \nu 
\right) \Delta \chi_a \chi_s^2  \cr 
& \quad 
+
\left( 
{\frac {7029}{448}} - {\frac {297}{8}} \nu 
\right) \chi_a^2 \chi_s 
+
\left( 
{\frac {781}{448}} - {\frac {143}{224}} \nu 
\right) 
\Delta \chi_a (1 + 3 \chi_a^2) \,.
\end{align}

Inserting the numerical solution of~\eqref{dvdt-T4} into~\eqref{Kepler}, 
we then obtain TaylorT4 approximation 
of the orbital phase $\phi^{\mathrm {T4}}(t)$. 
The initial and terminating conditions for TaylorT4 
can be set up the same as those in the case of TaylorT1. 
We emphasize that the formula~\eqref{dvdt-T4} is not valid 
beyond $v^{\mathrm {T4}}(t) \gtrsim v_{\mathrm {pole}}$ 
although its right hand side is a regular function of $v$.

\subsection{TaylorT2}
\label{subsec:T2}

TaylorT2 approximant is based on 
the equivalent differential forms of~\eqref{Kepler} and~\eqref{dvdt}, 
which are now expressed in terms of $v$ as 
\begin{equation}\label{T2-t}
\frac{d \phi}{d v} 
= 
\frac{v^3}{m(v)} \frac{d t}{d v}\,,
\quad 
\frac{d t}{d v} 
= 
-\frac{( {\partial {\cal E}} / {\partial v} )_{m,\,S}}
{{\cal F}_{\mathrm {eff}}(v)}\,. 
\end{equation}
The right hand side of each expression is re-expanded 
as a single Taylor expansion in the power of $v$ 
and truncated at $3.5$PN order in our calculation.
The above differential equations are then integrated analytically 
to give the closed form solutions 
$\phi^{\mathrm {T2}}(v)$ and $t^{\mathrm {T2}}(v)$.

Following section~\ref{subsec:T4}, we write for the full solution by
\begin{align}\label{phi-T2}
\phi^{\mathrm {T2}}(v) 
&= 
\phi^{\mathrm {T2}}_{\mathrm {ref}}
+
\phi^{\mathrm {T2}}_{\infty}(v) 
+
\phi^{\mathrm {T2}}_{\mathrm H}(v)\,, \\
\label{t-T2}
t^{\mathrm {T2}}(v) 
&= 
t^{\mathrm {T2}}_{\mathrm {ref}}
+
t^{\mathrm {T2}}_{\infty}(v) 
+
t^{\mathrm {T2}}_{\mathrm H}(v)\,,
\end{align}
where $t^{\mathrm {T2}}_{\mathrm {ref}}$ 
and
$\phi^{\mathrm {T2}}_{\mathrm {ref}}$
are integration constants. 
The non-absorption part of the solutions 
$\phi^{\mathrm {T2}}_{\infty}(v)$ and $t^{\mathrm {T2}}_{\infty}(v)$ 
is derived by the set of differential equations 
\begin{equation}\label{T2-ODE-inf}
\frac{d \phi^{\mathrm {T2}}_{\infty}}{d v} 
= 
\frac{v^3}{m} \frac{d t^{\mathrm {T2}}_{\infty}}{d v}\,,
\quad
\frac{d t^{\mathrm {T2}}_{\infty}}{d v} 
=
-\frac{( d {E} / d v )}{ F_{\infty}(v) }\,,
\end{equation}
after expanding the right side of each expressions 
up to the relative $3.5$PN order 
[recall that $m = m^{\mathrm {I}}$ in~\eqref{T2-ODE-inf}]. 
They may have the following general structure: 
\begin{align}\label{T2-inf}
\phi^{\mathrm {T2}}_{\infty} 
&= 
-\frac{1}{32 \nu v^5}
\left(
\phi^{\mathrm {T2}}_{\mathrm {NS}}
+
{v^3} \phi^{\mathrm {T2}}_{\mathrm {SO}}
+
{v^4} \phi^{\mathrm {T2}}_{\mathrm {SS}}
+
{v^7} \phi^{\mathrm {T2}}_{\mathrm {SSS}}
+ 
O(v^8)
\right)\,, \cr
t^{\mathrm {T2}}_{\infty}
&= 
-\frac{5 m}{256 \nu v^8}
\left(
t^{\mathrm {T2}}_{\mathrm {NS}}
+
{v^3} t^{\mathrm {T2}}_{\mathrm {SO}}
+
{v^4} t^{\mathrm {T2}}_{\mathrm {SS}}
+
{v^7} t^{\mathrm {T2}}_{\mathrm {SSS}}
+ 
O(v^8)
\right)\,.
\end{align}
Equations (3.8a) and (3.8b) in~\cite{Buonanno:2009zt} provide 
the full $3.5$PN expressions for the non-spinning terms 
$\phi^{\mathrm {T2}}_{\mathrm {NS}}$ 
and $t^{\mathrm {T2}}_{\mathrm {NS}}$, respectively. 
The results for the SO, SS and 
SSS contributions to the solutions 
read 
\begin{align}\label{T2-phi0}
\phi^{\mathrm {T2}}_{\mathrm {SO}}
&= 
\left\{  
\left( 
-{\frac {25150083775}{24385536}} 
+{\frac {10566655595}{6096384}} \nu 
-{\frac {1042165}{24192}} \nu^2 
+{\frac {5345}{288}} \nu^3 
\right) \chi_s \right. \cr 
& \quad \left. 
+ \left( 
-{\frac {25150083775}{24385536}}
+{\frac {26804935}{48384}} \nu 
-{\frac {1985}{384}} \nu^2 
 \right) \Delta \chi_a  \right\} {v}^{4} \cr 
 &\quad 
+ \pi \left\{  
\left( {\frac {1135}{6}}- 130 \nu   \right) \chi_s 
+{\frac {1135}{6}} \Delta \chi_a 
\right\} {v}^{3} \cr 
& \quad 
+ \left\{ 
\left( 
-{\frac {732985}{2016}} + {\frac {6065 }{18}} \nu + {\frac {85}{2}} \nu^2 
\right) \chi_s
+ \left( -{\frac {732985}{2016}} -{\frac {35}{2}} \nu  \right) \Delta \chi_a
 \right \} \ln \left( \frac{v}{v_{\mathrm {reg}}} \right){v}^{2} \cr 
& \quad 
+ \left( {\frac {565}{24}} - {\frac {95}{6}} \nu \right) \chi_s 
+ {\frac {565}{24}} \Delta \chi_a \,, \cr
\phi^{\mathrm {T2}}_{\mathrm {SS}}
&=  
\pi 
\left\{  
\left( -{\frac {285}{4}} + 5\,{\nu}  \right) \chi_s^2 
-{\frac {285}{2}} \Delta \chi_s \chi_s 
+ \left( -{\frac {285}{4}} + 280\,{\nu}  \right) \chi_a^2 
\right\} {v}^{3} \cr
& \quad 
+ \left\{ \left( 
{\frac {75515}{1152}} 
-{\frac {232415}{2016}} \nu 
+{\frac {1255}{36}} \nu^2 
\right) \chi_s^2 
+\left( {\frac {75515}{576}} -{\frac {8225}{72}} \nu \right) 
\Delta \chi_a \chi_s \right. \cr 
&\quad \left. 
+ 
\left( {\frac {75515}{1152}} -{\frac {263245}{1008}} - 120 \nu^2 \right) 
\chi_a^2 
\right\} {v}^{2} \cr 
& \quad 
+ \left( -{\frac {405}{16}} + \frac{5}{4} \nu \right) \chi_s^2 
-{\frac {405}{8}} \Delta \chi_a \chi_s 
+ \left( -{\frac {405}{16}} + 100\,\nu \right) \chi_a^2  \,, \cr
\phi^{\mathrm {T2}}_{\mathrm {SSS}}
&= 
\left( 
{\frac {14585}{192}} -{\frac {475}{48}} \nu + {\frac {25}{6}} \nu^2 
\right) \chi_s^3
+ \left( 
{\frac {14585}{64}} - {\frac {215}{16}} \nu 
\right) 
\Delta \chi_a \chi_s^2 \cr 
& \quad 
+ \left( 
{\frac {14585}{64}} - {\frac {3635}{4}} \nu + 10 \nu^2 
\right) 
\chi_a^2 \chi_s 
+
\left( 
{\frac {14585}{192}}-{\frac {595}{2}} \nu 
\right) 
\Delta \chi_a^3 \,, 
\end{align} 
and 
\begin{align}\label{T2-t0}
t^{\mathrm {T2}}_{\mathrm {SO}}
&= 
\left\{
\left( 
{\frac {5030016755}{1524096}} 
-{\frac {2113331119}{381024}} \nu 
+{\frac {208433}{1512}} \nu^2 
-{\frac {1069}{18}} \nu^3 
\right) \chi_s \right. \cr 
& \quad \left. 
+ \left( 
{\frac {5030016755}{1524096}} 
- {\frac {5360987}{3024}} \nu 
+ {\frac {397}{24}} \nu^2 
\right) \Delta \chi_a \right\} {v}^{4} \cr 
& \quad 
+ 
\pi \left\{  
\left( -{\frac {454}{3}}  + 104\,{\nu} \right) \chi_s 
-{\frac {454}{3}} \Delta \chi_a \right\} {v}^{3}  \cr 
& \quad 
+ \left\{  
\left(
{\frac {146597}{756}} -{\frac {4852}{27}} \nu -{\frac {68}{3}} \nu^2 
\right) \chi_s 
+ \left( 
{\frac {146597}{756}} + {\frac {28}{3}} \nu 
\right) \Delta \chi_a 
\right\} {v}^{2} \cr 
& \quad 
+ \left( {\frac {226}{15}}-{\frac {152}{15}} \nu \right) \chi_s 
+{\frac {226}{15}} \Delta \chi_a
 \,, \cr
t^{\mathrm {T2}}_{\mathrm {SS}}
&= 
4 \pi 
\left\{
\left( 57 -4 \nu  \right) \chi_s^2 
+ 114 \Delta \chi_a \chi_s
+ \left( 57 - 224 \, \nu \right) \chi_a^2  
\right\} {v}^{3} \cr 
& \quad 
+ \left\{
\left( 
-{\frac {15103}{288}} +{\frac {46483}{504}} \nu -{\frac {251}{9}} \nu^2
\right) \chi_s^2 
+ \left(
-{\frac {15103}{144}} +{\frac {1645}{18}} \nu 
 \right)\Delta \chi_s \chi_s  \right. \cr 
& \quad \left. 
+ \left( 
-{\frac {15103}{288}} + {\frac {52649}{252}} \nu + 96\,{{\nu}}^{2}
\right) \chi_a^2
 \right\} {v}^{2} \cr 
& \quad 
+ \left( 
-{\frac {81}{8}} + \frac{1}{2} \nu \right) \chi_s^2 
-{\frac {81}{4}} \Delta \chi_s \chi_s 
+\left( -{\frac {81}{8}} +40\,{\nu} \right) \chi_a^2 \,, \cr
t^{\mathrm {T2}}_{\mathrm {SSS}}
&= 
\left( 
-{\frac {2917}{12}} + {\frac {95}{3}} \nu -{\frac {40}{3}} \nu^2
 \right) \chi_s^3
+
\left( 
-{\frac {2917}{4}} + 43\,{\nu}
\right) \Delta \chi_a \chi_s^2 \cr 
& \quad
+ \left( 
-{\frac {2917}{4}} + 2908 \nu - 32 \nu^2 
\right) \chi_a^2 \chi_s 
+ \left( 
-{\frac {2917}{12}} + 952\,{\nu}
\right) \Delta \chi_a^3 \,.
\end{align} 
Here the ``regulator'' $v_{\mathrm {reg}}$ for the log terms 
in $\phi^{\mathrm {T2}}_{\mathrm {SO}}$ can be chosen 
either the value at ISCO of the Kerr metric $v_{\mathrm {ISCO}}$, 
which may have the final total mass $m(v_{\mathrm {ISCO}})$ 
and effective spin $\chi_{\mathrm {eff}}(v_{\mathrm {ISCO}})$ 
[recall~\eqref{def-chi-eff}],
or that of the pole $v_{\mathrm {pole}}$ in the PN energy flux $F_{\infty}$. 
Our expression for $\phi^{\mathrm {T2}}_{\infty}$ 
recovers~(3.4) of~\cite{Ajith:2011ec} with non-precessing spins 
up to $2.5$PN order 
(as spin terms in this reference is truncated at $2.5$PN order).

Meanwhile, the BH absorption part of the solutions 
$\phi^{\mathrm {T2}}_{\mathrm H}(v)$ 
and $t^{\mathrm {T2}}_{\mathrm H}(v)$ may be expressed as 
\begin{align}\label{T2-H}
\phi^{\mathrm {T2}}_{\mathrm {H}} 
&= 
-\frac{1}{32 \nu }
\left\{
\ln \left( \frac{v}{v_{\mathrm {reg}}} \right) 
\phi^{\mathrm {T2}}_{\mathrm {Flux},5}
+
{v^2} 
\left( 
\phi^{\mathrm {T2}}_{\mathrm {Flux},7} 
+
\nu \, \phi^{\mathrm {T2}}_{\mathrm {BH},7}
\right)
+
O(v^3)
\right\}\,, \cr
t^{\mathrm {T2}}_{\mathrm {H}}
&= 
-\frac{5 m}{256 \nu v^3}
\left\{
t^{\mathrm {T2}}_{\mathrm {Flux},5}
+
{v^2} 
\left( 
t^{\mathrm {T2}}_{\mathrm {Flux},7} 
+
\nu \, t^{\mathrm {T2}}_{\mathrm {BH},7}
\right)
+
O(v^3)
\right)\,,
\end{align}
where $\phi^{\mathrm {T2}}_{\mathrm {Flux},5}$ and 
$t^{\mathrm {T2}}_{\mathrm {Flux},5}$ 
as well as 
$\phi^{\mathrm {T2}}_{\mathrm {Flux},7}$ and 
$t^{\mathrm {T2}}_{\mathrm {Flux},7}$
denote the LO ($2.5$PN) and NLO ($3.5$PN) 
contributions only from the horizon energy flux 
with the substitution 
$\delta m = \delta \nu = \delta \chi_s = \delta \chi_a 
= \Gamma^{i}_{\mathrm {H}} = 0$, respectively, 
and $\phi^{\mathrm {T2}}_{\mathrm {BH},7}$ and 
$t^{\mathrm {T2}}_{\mathrm {BH},7}$ 
are the corrections due to the LO secular change 
in the BH mass and spin, which are suppressed by 
the mass ratio $\nu$. 
They read
\begin{align}\label{T2-phiH}
\phi^{\mathrm {T2}}_{\mathrm {Flux},5}
&= 
\left( -{\frac {5}{4}}+{\frac {15}{4}} \nu \right) 
\chi_s (1 + 3 \chi_s^2)
+ \left( -{\frac {45}{4}} + {\frac {45}{4}} \nu \right) 
\Delta \chi_a \chi_s^2 \cr 
& \quad 
+ \left( -{\frac {45}{4}} + {\frac {135}{4}} \nu \right) 
\chi_a^2 \chi_s 
+ \left( -\frac{5}{4} + \frac{5}{4} \nu  \right) 
\Delta \chi_a (1 + 3 \chi_a^2) \,, \cr
\phi^{\mathrm {T2}}_{\mathrm {Flux},7}
&= 
\left( 
- {\frac {7285}{448}} + {\frac {21295}{448}} \nu + {\frac {135}{16}} \nu^2 
\right) \chi_s^3
+ \left( 
-{\frac {21855}{448}} + {\frac {20175}{448}} \nu + {\frac {285}{16}} \nu^2 
\right) \Delta \chi_a \chi_s^2 \cr 
& \quad 
+ \left\{  
\left( 
-{\frac {21855}{448}} + {\frac {63885}{448}} \nu + {\frac {405}{16}} \nu^2 
\right) \chi_a^2 
-{\frac {8335}{1344}} + {\frac {24445}{1344}} \nu + {\frac {45}{16}} \nu^2 
\right\} \chi_s \cr 
& \quad 
+ \left( 
- {\frac {7285}{448}} + {\frac {6725}{448}} \nu + {\frac {95}{16}} \nu^2 
\right) \Delta \chi_a^3  
+ \left( 
-{\frac {8335}{1344}} + {\frac {7775}{1344}} \nu + {\frac {95}{48}} \nu^2 
\right) \Delta \chi_a \,, \cr 
\phi^{\mathrm {T2}}_{\mathrm {BH},7}
&=  
\left(
{\frac {8195}{2688}}-{\frac {715}{112}} \nu 
\right) \chi_s (1 + 3 \chi_s^2) 
+
\left( 
{\frac {24585}{896}} - {\frac {165}{64}} \nu 
\right) \Delta \chi_a \chi_s^2 \cr
& \quad 
+ 
\left( 
{\frac {24585}{896}} - {\frac {6435}{112}} \nu 
\right) \chi_a^2 \chi_s 
+
\left( 
{\frac {8195}{2688}} - {\frac {55}{192}} \nu 
\right) \Delta \chi_a (1 + 3 \chi_a^2 ) \,,
\end{align} 
and 
\begin{align}\label{T2-tH}
t^{\mathrm {T2}}_{\mathrm {Flux},5}
&= 
 \left( \frac{2}{3} -2 \nu  \right) \chi_s (1 + 3 \chi_s^2)
+ 6 \left( 1 - \nu \right) \Delta \chi_a \chi_s^2 \cr 
& \quad 
+ 6 \left( 1 - 3 \nu \right) \chi_a^2 \chi_s 
+ \left( \frac{2}{3} - \frac{2}{3} \nu \right) 
\Delta \chi_a (1 + 3 \chi_a^2) \,, \cr
t^{\mathrm {T2}}_{\mathrm {Flux},7}
&= 
\left(
{\frac {1457}{28}} - {\frac {4259}{28}} \nu - 27\,{{\nu}}^{2}
\right) \chi_s^3
+ \left( 
{\frac {4371}{28}} - {\frac {4035}{28}} \nu - 57\,{{\nu}}^{2}
\right) \Delta \chi_a \chi_s^2 \cr
& \quad 
+ \left\{  
\left( 
{\frac {4371}{28}} - {\frac {12777}{28}} \nu - 81\,{{\nu}}^{2}
\right) \chi_a^2 
+ {\frac {1667}{84}} -{\frac {4889}{84}} \nu - 9\,{{\nu}}^{2} 
\right\} \chi_s \cr 
& \quad 
+ \left( 
{\frac {1457}{28}} - {\frac {1345}{28}} \nu - 19 \nu^2 
\right) \Delta \chi_a^3
+ \left( 
{\frac {1667}{84}} - {\frac {1555}{84}} \nu - {\frac {19}{3}} \nu^2 
\right) \Delta \chi_a \,, \cr
t^{\mathrm {T2}}_{\mathrm {BH},7}
&= 
\left( 
-{\frac {1831}{168}} + {\frac {167}{7}} \nu 
\right) \chi_s (1 + 3 \chi_s^2)
+
\left( 
- {\frac {5493}{56}} + {\frac {519}{28}} \nu 
\right) 
\Delta \chi_a \chi_s^2 \cr 
& \quad
+ 
\left( 
-{\frac {5493}{56}} +{\frac {1503}{7}} \nu 
\right) \chi_a^2 \chi_s 
+ \left( 
-{\frac {1831}{168}} + {\frac {173}{84}} \nu 
\right) \Delta \chi_a ( 1 + 3 \chi_a^2 )\,. 
\end{align} 
We note that $t^{\mathrm {T2}}_{\mathrm {ref}}$ in~\eqref{t-T2} 
has to be chosen to satisfy $t^{\mathrm {T2}}(v_0) = 0$ 
with the initial condition $v = v_0$ 
while $\phi^{\mathrm {T2}}_{\mathrm {ref}}$ is arbitrary, 
typically taken as either $0$ or $\pi/2$. 
Also, both of our solutions in~\eqref{phi-T2} and~\eqref{t-T2} 
are valid only when $v < v_{\mathrm {pole}}$ 
and they should not be extended all the way 
to $v_{\mathrm {ISCO}}$ if $ v_{\mathrm {pole}} \leq v_{\mathrm {ISCO}}$ 
for given spins $\chi_{a,s}$.

\subsection{TaylorT3}
\label{subsec:T3}

TaylorT3 approximant is the ``inverse'' of TaylorT2~\cite{Blanchet:1996pi}. 
That is, TaylorT2 expression $t^{\mathrm {T2}}(v)$ 
in~\eqref{t-T2} is explicitly inverted to obtain 
TaylorT3 expression $v^{\mathrm {T3}}(\theta)$, 
where we define the dimensionless time variable 
\begin{equation}\label{def-theta}
\theta 
\equiv 
\left(
\frac{\nu}{5m} (t^{\mathrm {T2}}_{\mathrm {ref}} - t)
\right)^{-1/8}\,.
\end{equation}
Then, $v^{\mathrm {T3}}(\theta)$ is used to obtain 
an explicit TaylorT3-representation of the orbital phase 
$\phi^{\mathrm {T3}}(\theta) \equiv 
\phi^{\mathrm {T2}} (v^{\mathrm {T3}}(\theta))$. 
The above procedure yields TaylorT3 approximants: 
\begin{align}\label{phi-T3}
\phi^{\mathrm {T3}}(\theta) 
&= 
\phi^{\mathrm {T3}}_{\mathrm {ref}}
+
\phi^{\mathrm {T3}}_{\infty}(\theta) 
+
\phi^{\mathrm {T3}}_{\mathrm H}(\theta)\,, \\
\label{f-T3}
F^{\mathrm {T3}}(\theta) 
&= 
F^{\mathrm {T3}}_{\infty}(\theta) 
+
F^{\mathrm {T3}}_{\mathrm H}(\theta)\,,
\end{align}
where 
$\phi^{\mathrm {T3}}_{\mathrm {ref}}$ 
is an integration constant and $F^{\mathrm {T3}}$ 
is the GW frequency of the dominant $(2,2)$ 
spin-weighted-spherical harmonic mode 
[recall~\eqref{def-v2}]. 

The non-absorption part of the solution $F^{\mathrm {T3}}_{\infty}(\theta)$ 
is computed by inverting the corresponding TaylorT2 solution 
$t^{\mathrm {T2}}_{\infty}(v)$. 
This is then fed into $\phi^{\mathrm {T2}}_{\infty}(F(v))$ to obtain 
the non-absorption part of the orbital phase 
$\phi^{\mathrm {T3}}_{\infty}(\theta) = 
\phi^{\mathrm {T2}}_{\infty} (F^{\mathrm {T3}}_{\infty}(\theta))$. 
Like the expressions in~\eqref{T2-inf}, 
their general structures are 
\begin{align}\label{T3-inf}
\phi^{\mathrm {T3}}_{\infty} 
&= 
-\frac{1}{\nu \theta^5}
\left(
\phi^{\mathrm {T3}}_{\mathrm {NS}}
+
{\theta^3} \phi^{\mathrm {T3}}_{\mathrm {SO}}
+
{\theta^4} \phi^{\mathrm {T3}}_{\mathrm {SS}}
+
{\theta^7} \phi^{\mathrm {T3}}_{\mathrm {SSS}}
+ 
O(\theta^8)
\right)\,, \cr
F^{\mathrm {T3}}_{\infty}
&= 
\frac{\theta^3}{8 \pi m }
\left(
F^{\mathrm {T3}}_{\mathrm {NS}}
+
{\theta^3} F^{\mathrm {T3}}_{\mathrm {SO}}
+
{\theta^4} F^{\mathrm {T3}}_{\mathrm {SS}}
+
{\theta^7} F^{\mathrm {T3}}_{\mathrm {SSS}}
+ 
O(\theta^8)
\right)\,.
\end{align}
Equations (3.9a) and (3.9b) in~\cite{Buonanno:2009zt} provide 
the full $3.5$PN expressions for the non-spinning terms 
$\phi^{\mathrm {T3}}_{\mathrm {NS}}$ 
and $F^{\mathrm {T3}}_{\mathrm {NS}}$, respectively. 
The explicit expressions for the spin contributions in~\eqref{T3-inf} 
take
\begin{align}\label{T3-phi0}
\phi^{\mathrm {T3}}_{\mathrm {SO}}
&= 
\left\{  
\left( 
-{\frac {6579635551}{260112384}}
+{\frac {1496368361}{37158912}} \nu 
+{\frac {840149}{3096576}} \nu^2
+{\frac {12029}{18432}} \nu^3
\right) \chi_s \right. \cr 
& \quad \left. 
+ \left( 
-{\frac {6579635551}{260112384}}
+ {\frac {143169605}{12386304}} \nu
-{\frac {2591}{9216}} \nu^2 
\right) \Delta \chi_a
\right\} {\theta}^{4} \cr 
& \quad 
+ \pi \left\{  
\left( {\frac {6127 }{1280}}
-{\frac {1051 }{320}} \nu \right) \chi_s 
+{\frac {6127}{1280}} \Delta \chi_a \right) {\theta}^{3} \cr 
& \quad 
+
\left\{
\left( 
-{\frac {732985}{64512}}
+{\frac {6065}{576}} \nu 
+{\frac {85}{64}} \nu^2 
 \right) \chi_s 
+ \left( 
-{\frac{732985}{64512}}
-{\frac {35}{64}} \nu 
\right) \Delta \chi_a 
 \right\} 
 \ln \left(\frac{\theta}{\theta_{\mathrm {reg}}}\right) {\theta}^{2} \cr 
& \quad  
+ \left( {\frac {113}{64}}-{\frac {19}{16}} \nu \right) \chi_s 
+{\frac {113}{64}} \Delta \chi_a \,, \cr
\phi^{\mathrm {T3}}_{\mathrm {SS}}
&= 
\pi 
\left\{  
\left( 
-{\frac {3663}{2048}}+{\frac {63}{512}} \nu \right) \chi_s^2 
-{\frac {3663}{1024}} \Delta \chi_a \chi_s 
+ \left( 
-{\frac {3663}{2048}} +{\frac {225}{32}} \nu \right) \chi_a^2 
\right\} {\theta}^{3} \cr 
& \quad 
+
\left\{  
\left( 
{\frac {16928263}{13762560}}
-{\frac {288487}{143360}} \nu 
+{\frac {76471}{122880}} \nu^2 
\right) \chi_s^2
+ \left( 
{\frac {16928263}{6881280}}
-{\frac {453767}{245760}} \nu 
\right) \Delta \chi_a \chi_s \right. \cr 
& \quad \left. 
+ \left( 
{\frac {16928263}{13762560}}
-{\frac {2336759}{491520}} \nu 
-{\frac {1715}{512}} \nu^2 
 \right) \chi_a^2
\right\} {\theta}^{2} \cr 
& \quad 
+ \left( 
-{\frac {1215}{1024}} + {\frac {15}{256}} \nu \right) \chi_s^2
-{\frac {1215}{512}} \Delta \chi_a \chi_s 
+ \left( -{\frac {1215}{1024}}  + {\frac {75}{16}} \nu \right) \chi_a^2
\,, \cr
\phi^{\mathrm {T3}}_{\mathrm {SSS}}
&= 
\left( 
{\frac {67493}{32768}} - {\frac {111}{256}} \nu + {\frac {219}{2048}} \nu^2 
 \right) \chi_s^3  
+ \left( 
{\frac {202479}{32768}} - {\frac {5771}{8192}} \nu 
\right) \Delta \chi_a \chi_s^2 \cr 
& \quad 
+ \left( 
{\frac {202479}{32768}} - {\frac {203365}{8192}} \nu + {\frac {125}{128}} \nu^2
 \right) \chi_a^2 \chi_s 
+ \left( 
{\frac {67493}{32768}} - {\frac {4135}{512}} \nu 
\right) \Delta \chi_a^3\,
\end{align} 
and 
\begin{align}\label{T3-F0}
F^{\mathrm {T3}}_{\mathrm {SO}}
&= 
\left\{  
\left( 
{\frac {6579635551}{650280960}}
-{\frac {1496368361}{92897280}} \nu 
-{\frac {840149}{7741440}} \nu^2 
-{\frac {12029}{46080}} \nu^3
\right) \chi_s \right. \cr 
& \quad \left. 
+ \left( 
{\frac {6579635551}{650280960}}
-{\frac {28633921}{6193152}} \nu 
+{\frac {2591}{23040}} \nu^2
\right) \Delta \chi_a  
\right\} {\theta}^{4} \cr 
& \quad 
+ \pi \left\{  
\left( 
-{\frac {6127}{6400}} + {\frac {1051}{1600}} \nu 
\right) \chi_s 
-{\frac {6127}{6400}} \Delta \chi_a 
\right\} {\theta}^{3} \cr 
& \quad 
+\left\{
\left( 
{\frac {146597}{64512}}
-{\frac {1213}{576}} \nu 
-{\frac {17}{64}} \nu^2 
\right) \chi_s 
+ \left( 
{\frac {146597}{64512}} + {\frac {7}{64}} \nu 
\right) \Delta \chi_a  \right\} {\theta}^{2} \cr 
& \quad 
+ \left( {\frac {113}{160}}-{\frac {19}{40}} \nu \right) \chi_a
+{\frac {113}{160}} \Delta \chi_a \,,\cr
F^{\mathrm {T3}}_{\mathrm {SS}}
&=  
\pi \left\{  
\left( {\frac {3663}{5120}}-{\frac {63}{1280}} \nu \right) \chi_s^2 
+{\frac {3663}{2560}} \Delta \chi_a \chi_s
+ \left( {\frac {3663}{5120}}-{\frac {45}{16}} \nu \right) \chi_a^2 
\right\} {\theta}^{3} \cr 
& \quad 
+ \left\{
\left( 
-{\frac {16928263}{68812800}}
+{\frac {288487}{716800}} \nu
-{\frac {76471}{614400}} \nu^2 
\right) \chi_s^2 \right. \cr 
& \quad \left. 
+ \left( 
- {\frac {16928263}{34406400}} 
+ {\frac {453767}{1228800}} \nu 
\right) \Delta \chi_a \chi_s \right. \cr 
& \quad \left. 
+ \left( 
-{\frac {16928263}{68812800}} 
+{\frac {2336759}{2457600}} \nu 
+{\frac {343}{512}} \nu^2 
\right) \chi_a^2
 \right\} {\theta}^{2} \cr
& \quad 
+ \left( -{\frac {243}{1024}} + {\frac {3}{256}} \nu \right) \chi_s^2
- {\frac {243}{512}}\Delta \chi_a \chi_s  
+ \left( 
-{\frac {243}{1024}} + {\frac {15}{16}} \nu 
 \right) \chi_a^2
\,, \cr
F^{\mathrm {T3}}_{\mathrm {SSS}}
&= 
\left( 
-{\frac {67493}{81920}} + {\frac {111}{640}} \nu - {\frac {219}{5120}} \nu^2 
 \right) \chi_s^3 
+ \left( 
-{\frac {202479}{81920}} + {\frac {5771}{20480}} \nu 
\right) \Delta \chi_a \chi_s^2 \cr 
& \quad 
+ \left( 
- {\frac {202479}{81920}} + {\frac {40673}{4096}} \nu - {\frac {25}{64}} \nu^2 
 \right) \chi_a^2 \chi_s 
+ \left( 
-{\frac {67493}{81920}} + {\frac {827}{256}} \nu 
\right) \Delta \chi_a^3\,.
\end{align} 
The ``regulator'' $\theta_{\mathrm {reg}}$ for the log terms 
in $\phi^{\mathrm {T3}}_{\mathrm {SO}}$ may be chosen either the value at ISCO 
$\theta^{\mathrm {T3}}_{\mathrm {ISCO}}$ 
of the Kerr metric with the final total mass $m(\theta_{\mathrm {ISCO}})$ 
and effective spin $\chi_{\mathrm {eff}}(\theta_{\mathrm {ISCO}})$,  
[recall~\eqref{def-chi-eff}]
or that of the pole $\theta_{\mathrm {pole}}$ 
in the PN energy flux $F_{\infty}$; 
here the value of $\theta_{\mathrm {ISCO}}$ is computed by numerically 
solving $F^{\mathrm {T3}} (\theta_{\mathrm {ISCO}}) 
=  (v_{\mathrm {ISCO}})^3 / (\pi m)$, 
and we perform the similar calculation to obtain $\theta_{\mathrm {pole}}$ 
using $v_{\mathrm {pole}}$.

Similar to~\eqref{T2-H}, the BH absorption part of the solutions 
$\phi^{\mathrm {T3}}_{\mathrm H}(\theta)$ and 
$F^{\mathrm {T3}}_{\mathrm H}(\theta)$ may be expressed as 
\begin{align}\label{T3-H}
\phi^{\mathrm {T3}}_{\mathrm {H}} 
&= 
-\frac{1}{\nu}
\left\{
\ln \left( \frac{\theta}{\theta_{\mathrm {reg}}} \right) 
\phi^{\mathrm {T3}}_{\mathrm {Flux},5}
+
{\theta^2} 
\left(
\phi^{\mathrm {T3}}_{\mathrm {Flux},7}
+
\nu \, \phi^{\mathrm {T3}}_{\mathrm {BH},7}
\right)
+
O(\theta^3)
\right\}\,, \cr
F^{\mathrm {T3}}_{\mathrm {H}}
&= 
\frac{\theta^8}{8 \pi m}
\left\{
F^{\mathrm {T3}}_{\mathrm {Flux},5}
+
{\theta^2} 
\left(
F^{\mathrm {T3}}_{\mathrm {Flux},7}
+
\nu \, F^{\mathrm {T3}}_{\mathrm {BH},7}
\right)
+
O(\theta^3)
\right\}\,,
\end{align}
where $\phi^{\mathrm {T3}}_{\mathrm {Flux},5}$ and 
$F^{\mathrm {T3}}_{\mathrm {Flux},5}$ as well as  
$\phi^{\mathrm {T3}}_{\mathrm {Flux},7}$ and 
$F^{\mathrm {T3}}_{\mathrm {Flux},7}$ are 
the LO ($2.5$PN) and NLO ($3.5$PN) BH absorption parts of the solutions 
that only account for the contribution of the horizon energy flux 
with the substitution 
$\delta m = \delta \nu = \delta \chi_s = \delta \chi_a 
= \Gamma^{i}_{\mathrm {H}} = 0$, respectively, 
while $\phi^{\mathrm {T3}}_{\mathrm {H},7}$ and 
$F^{\mathrm {T3}}_{\mathrm {H},7}$ denote 
the corrections due to the LO secular change 
in the BH mass and spin. 
Their explicit expressions read 
\begin{align}\label{T3-phiH}
\phi^{\mathrm {T3}}_{\mathrm {Flux},5}
&= 
\left(  
-{\frac {5}{128}} + {\frac {15}{128}} \nu 
\right) \chi_s (1 + 3 \chi_s^2)
+ 
\left( 
-{\frac {45}{128}} + {\frac {45}{128}} \nu 
\right) \Delta \chi_a \chi_s^2 \cr 
& \quad 
+ \left(   
-{\frac {45}{128}} +{\frac {135}{128}} \nu 
\right) \chi_a^2 \chi_s
+ \left(  
-{\frac {5}{128}} + {\frac {5}{128}} {\nu}
\right) 
\Delta \chi_a (1 + 3 \chi_a^2) \,, \cr
\phi^{\mathrm {T3}}_{\mathrm {Flux},7}
&= 
\left( 
-{\frac {134845}{344064}}
+{\frac {129945}{114688}} \nu 
+{\frac {975}{4096}} \nu^2 
\right) \chi_s^3
\cr & \quad
+ \left( 
-{\frac {134845}{114688}}
+{\frac {120145}{114688}} \nu 
+{\frac {1875}{4096}} \nu^2 
 \right) \Delta \chi_a \chi_s^2 \cr 
& \quad 
+ \left\{  
\left( 
-{\frac {134845}{114688}} 
+{\frac {389835}{114688}} \nu 
+{\frac {2925}{4096}} \nu^2 
\right) \chi_a^2 
\right. \cr & \quad \left.
-{\frac {153745}{1032192}} 
+{\frac {49615}{114688}} \nu 
+{\frac {325}{4096}} \nu^2 
\right\} \chi_s \cr 
& \quad 
+ \left( 
-{\frac {134845}{344064}} 
+{\frac {120145}{344064}} \nu 
+{\frac {625}{4096}} \nu^2 
\right) \Delta \chi_a^3 
\cr & \quad
+ \left( 
-{\frac {153745}{1032192}} 
+{\frac {139045}{1032192}} \nu 
+{\frac {625}{12288}} \nu^2 
\right) \Delta \chi_a \,,\cr
\phi^{\mathrm {T3}}_{\mathrm {BH},7}
&= 
\left(
{\frac {8835}{114688}}
-{\frac {2385}{14336}} \nu 
\right) \chi_s (1 + 3 \chi_s^2) 
+ \left( 
{\frac {79515}{114688}} - {\frac {6345}{57344}} \nu 
\right) \Delta \chi_a \chi_s^2 \cr 
& \quad 
+ \left( 
{\frac {79515}{114688}}
-{\frac {21465}{14336}} \nu 
\right) \chi_a^2 \chi_s 
+ \left( 
{\frac {8835}{114688}} 
-{\frac {705}{57344}} \nu 
\right) \Delta \chi_a (1 + 3 \chi_a^2 )\,, 
\cr & \quad 
\end{align} 
and 
\begin{align}\label{T3-fH}
F^{\mathrm {T3}}_{\mathrm {Flux},5} 
&=  
 \left( \frac{1}{128} -  \frac{3}{128} \nu \right) 
\chi_s (1 + 3 \chi_s^2 ) 
+ \left( \frac{9}{128} - \frac{9}{128} \nu \right) 
\Delta \chi_a \chi_s^2 \cr 
& \quad 
+ \left( \frac{9}{128} - \frac{27}{128} \nu \right) 
\chi_a^2 \chi_s 
+ \left( \frac{1}{128} - \frac{1}{128} \nu \right) 
\Delta \chi_a (1 + 3 \chi_a^2)
\,, \cr
F^{\mathrm {T3}}_{\mathrm {Flux},7}
&= 
\left( 
{\frac {26969}{172032}} 
- {\frac {25989}{57344}} \nu 
-{\frac {195}{2048}} \nu^2 
\right) \chi_s^3 
+ 
\left(
{\frac {26969}{57344}}
-{\frac {24029}{57344}} \nu 
-{\frac {375}{2048}}  \nu^2 
\right) \Delta \chi_a \chi_s^2 \cr 
& \quad 
+ \left\{  
\left( 
{\frac {26969}{57344}} 
- {\frac {77967}{57344}} \nu 
- {\frac {585}{2048}} \nu^2 
\right) \chi_a^2 
+
{\frac {30749}{516096}} - {\frac {9923}{57344}} \nu - {\frac {65}{2048}} \nu^2
\right\} \chi_s  \cr 
& \quad
+
\left( 
{\frac {26969}{172032}}
-{\frac {24029}{172032}} \nu 
-{\frac {125}{2048}} \nu^2 
\right) \Delta \chi_a ^3  
\cr & \quad
+ 
\left( 
{\frac {30749}{516096}} 
- {\frac {27809}{516096}} \nu 
- {\frac {125}{6144}} \nu^2 
\right) \Delta \chi_a\,,  \cr 
F^{\mathrm {T3}}_{\mathrm {BH},7}
&= 
\left(
-{\frac {1767}{57344}}+{\frac {477}{7168}} \nu 
\right) \chi_s ( 1 + 3 \chi_s^2 )
+
\left( 
- {\frac {15903}{57344}} + {\frac {1269}{28672}} \nu 
\right) \Delta \chi_a \chi_s^2 \cr 
& \quad 
+ 
\left( 
- {\frac {15903}{57344}} + {\frac {4293}{7168}} \nu 
\right) \chi_a^2 \chi_s 
+ \left( 
-{\frac {1767}{57344}} + {\frac {141}{28672}} \nu 
\right) \Delta \chi_a (1 + 3 \chi_a^2) \,.
\end{align} 
The initial and terminating conditions for TaylorT3 are slightly complicated 
as the dimensionless time variable $\theta$ implicitly involves 
a reference time $t^{\mathrm {T2}}_{\mathrm {ref}}$ in~\eqref{t-T2}. 
At a given initial frequency $F_0 = (v_0)^3 / (\pi m)$, 
the value of $t^{\mathrm {T2}}_{\mathrm {ref}}$ 
has to be tuned so that $t = 0$ by numerically solving~\eqref{f-T3} 
with $F^{\mathrm {T3}} = F_0$ in terms of $\theta$. 
Same as TaylorT2, we also recall that 
our solution in~\eqref{phi-T3} and~\eqref{f-T3} 
are valid only when $\theta < \theta_{\mathrm {pole}}$.

Furthermore, we note that the evolution of $F^{\mathrm {T3}}(\theta)$ is 
\textit{not monotonic}. 
In fact, $F^{\mathrm {T3}}(v(\theta))$ begins to decrease before 
$v$ reaches $v_{\mathrm {ISCO}}$ (or $v_{\mathrm {pole}}$) 
and even less than zero between $v_{\mathrm {ISCO}}$ 
and $v_{\mathrm {pole}}$. 
This unphysical behavior is reported in~\cite{Buonanno:2009zt} 
for the non-spinning case, 
and we find the same appears for the spinning cases in general. 
Therefore, TaylorT3 evolution must be terminated before either at 
$\theta_{\mathrm {fin}}$ such that $(d F^{\mathrm {T3}}/ d \theta) = 0$ 
or $\theta_{\mathrm {pole}}$ if they are smaller than 
the nominal value such as $\theta_{\mathrm {ISCO}}$.

\subsection{TaylorF2}
\label{subsec:F2}

TaylorF2 is an approximation for waveforms in the frequency domain, 
which is the most commonly used for the purpose of GW data analysis 
and other application. 
Using the stationary phase approximation, the frequency-domain waveform 
can be computed from the Fourier representation of the time-domain waveform,
which may be written as~\cite{Mishra:2016whh,Damour:2002kr,Arun:2008kb,
Ajith:2011ec}
\begin{equation}\label{h-F}
{\tilde h}(f)
=
A(f) \, e^{-i [ {\Psi_{\mathrm {SPA}}}(f) - \pi / 4] }\,,
\quad 
\Psi_{\mathrm {SPA}}(f) \equiv 2 \pi f t(f) - \Psi(f)\,,
\end{equation}
with the frequency-domain amplitude 
[recall~\eqref{t-polarization} in the above]
\begin{equation}\label{def-A}
A(f) \equiv 
C \frac{2 \nu(v_f) m(v_f)}{D_L} (\pi m(v_f) f)^{2/3} 
\left(\left. \frac{d F(v)}{dt}\right|_{v = v_f} \right)^{-1/2}\,.
\end{equation}
Here, $C$ is a numerical constant 
that depends on the relative position and inclination of 
the inspiraling BBH system with respect to the detector, 
and $F(v_f) = v_f^3 / (\pi m)$ is the GW frequency of the dominant 
$(2,2)$ spin-weighted-spherical harmonic mode evaluated 
at the saddle point $v_f$ [recall~\eqref{def-v2}].

In the adiabatic approximation, the time derivative of $F(v)$ 
can be written as 
\begin{equation}\label{dFdt-F2}
\left.
\frac{d F(v)}{dt} 
\right|_{v = v_f}
= 
\frac{3 v_f^2}{\pi m}
\left. 
\frac{d v}{dt} 
\right|_{v = v_f} \,,
\end{equation}
and its PN expansion is simply obtained by the corresponding 
TaylorT4 expression of ${d v^{\mathrm {T4}}}/{d t}$ in~\eqref{dvdt-T4}. 
Then, the substitution of this back into~\eqref{def-A} gives 
a closed analytic expression of the amplitude 
$A^{\mathrm {F2}}(f)$ up to $3.5$PN order. 

However, we note that the higher PN corrections in $A^{\mathrm {F2}}(f)$ 
\textit{do not} come from that to the (time-domain) amplitude 
of the waveform, which is truncated at the Newtonian order 
in~\eqref{h-F}. 
The time-domain amplitude is currently only available 
to the $3$PN accuracy for the non-spinning terms~\cite{Blanchet:2008je}, 
and the $2$PN accuracy for the SO and SS 
terms~\cite{Buonanno:2012rv} beyond the Newtonian order
\footnote{
The partial results for the higher PN corrections to the amplitude 
(of the dominant harmonic) are also
known~\cite{Blanchet:2011zv,Faye:2012we,Faye:2014fra,Marchand:2016vox}.
}.
This means that the frequency-domain amplitude $A^{\mathrm {F2}}(f)$ 
is incomplete beyond $2$PN order unless we include 
appropriate higher PN contributions to the time-domain amplitude, 
but, unfortunately, they are beyond the present state-of-the-art. 
We therefore do not list the explicit PN expressions of 
$A^{\mathrm {F2}}(f)$ here, 
but it can be straightforwardly obtained using the result in this paper. 
See (5.7) in~\cite{Ajith:2011ec} for the explicit $3.5$PN expression 
for $A^{\mathrm {F2}}(f)$, but including 
the SO and SS terms only up to $2.5$PN order.
The complete expressions for the frequency-domain amplitude 
to $2$PN order, 
which are calculated from the corresponding $2$PN time-domain amplitude 
with all possible spin effects, 
are also listed in~(8a) -- (8c) in~\cite{Mishra:2016whh} 
and Appendix.~D of~\cite{Arun:2008kb}. 

The frequency-domain phase $\Psi_{\mathrm {SPA}}(f)$ 
in~\eqref{h-F} is obtained by solving 
the following set of equations:
\begin{equation}\label{dpsidt-F2}
\frac{d \Psi_{\mathrm {SPA}}}{df} - 2 \pi t = 0\,, 
\quad
\frac{dt}{df} 
+ 
\frac{\pi m}{3 v^2} 
\frac{( {\partial {\cal E}} / {\partial v} )_{m,\,S}}
{{\cal F}_{\mathrm {eff}}(v)}
= 0\,.
\end{equation}
In these equations, the expression 
$( {\partial {\cal E}} / {\partial v} )_{m,\,S}$ and 
${\cal F}_{\mathrm {eff}}(v)$ can be obtained from~\eqref{def-vE} 
and~\eqref{def-Eff}, respectively. 
If we leave the PN expression of 
${( {\partial {\cal E}} / {\partial v} )_{m,\,S}}/
{{\cal F}_{\mathrm {eff}}(v)}$ as a ratio of polynomials, 
as is done for TaylorT1 in section~\ref{subsec:T1}, 
the numerical integration of~\eqref{dpsidt-F2} 
gives TaylorF1 approximant of the phase 
$\Psi_{\mathrm {SPA}}^{\mathrm {F1}}(f)$. 
On the other hand, if we re-expand 
${( {\partial {\cal E}} / {\partial v} )_{m,\,S}}/
{{\cal F}_{\mathrm {eff}}(v)}$ 
as a single Taylor expansion in $v$ and truncated 
at the appropriate PN order, which is $3.5$PN order 
in our calculation, 
the solution produces the closed form TaylorF2 expression 
of the phase $\Psi_{\mathrm {SPA}}^{\mathrm {F2}}(f)$.

Similar to~\eqref{phi-T2}, we write for the full solution, 
\begin{equation}\label{phi-F2}
\Psi^{\mathrm {F2}}_{\mathrm {SPA}}(f)
= 
2 \pi f t_c - \Psi_c 
+
\Psi^{\mathrm {F2}}_{\infty}(v(f))
+
\Psi^{\mathrm {F2}}_{\mathrm {H}}(v(f))\,,
\end{equation}
where the constants $t_c$ and $\Phi_c$ can be chosen arbitrary. 
The non-absorption part of the phase $\Psi^{\mathrm {F2}}_{\infty}$ 
is obtained by solving the differential equations 
\begin{equation}\label{dpsidt-F2-inf}
\frac{d \Psi_{\infty}}{df} - 2 \pi t = 0\,, 
\quad
\frac{dt}{df} 
+ 
\frac{\pi m}{3 v^2} 
\frac{( {\partial {E}} / {\partial v} )_{m,\,S}}
{{F}_{\infty}}
= 0\,,
\end{equation}
in which we expand the ratio of polynomials 
$( {\partial {E}} / {\partial v} )_{m,\,S}/ {F}_{\infty}$
to $3.5$PN order. 
The solution may have the structure of 
\begin{equation}\label{F2-inf}
\Psi^{\mathrm {F2}}_{\infty} (f)
= 
\frac{3}{128 \nu v^5}
\left(
\Psi^{\mathrm {F2}}_{\mathrm {NS}}
+ 
{v^3} \Psi^{\mathrm {F2}}_{\mathrm {SO}} 
+
{v^4} \Psi^{\mathrm {F2}}_{\mathrm {SS}} 
+
{v^7} \Psi^{\mathrm {F2}}_{\mathrm {SSS}} 
+ 
O(v^8)
\right)\,,
\end{equation}
[recall that $v = (\pi m^{\mathrm {I}} f)^{1/3}$]. 
The explicit $3.5$PN expression for $\Psi^{\mathrm {F2}}_{\mathrm {NS}}$ 
is given in~(3.18) of~\cite{Buonanno:2009zt}. 
$\Psi^{\mathrm {F2}}_{\infty}$ with all possible spin-dependent 
contributions up to $3.5$PN order can be obtained from
~(6a) -- (6c) in~\cite{Mishra:2016whh} together with~(6.22) 
in~\cite{Arun:2008kb}
(see also~\cite{Khan:2015jqa,Wade:2013hoa}),
but we repeat it here for completeness adopting our notation: 
\begin{align}\label{F2-Psi0}
\Psi^{\mathrm {F2}}_{\mathrm {SO}}
&= 
 \left\{  
 \left( 
 -{\frac {25150083775}{3048192}}
 +{\frac {10566655595}{762048}} \nu 
 -{\frac {1042165}{3024}} \nu^2 
 +{\frac {5345}{36}} \nu^3
  \right) \chi_s \right. \cr 
& \quad \left. 
+ \left( 
 -{\frac {25150083775}{3048192}}
 +{\frac {26804935}{6048}} \nu 
 -{\frac {1985}{48}} \nu^2 
 \right) \Delta \chi_a \right\} {v}^{4} \cr 
& \quad 
+ \pi \left\{
 \left( 
 {\frac {2270}{3}} -520 \nu   \right) \chi_s +{\frac {2270}{3}} 
 \Delta \chi_a 
 \right\} {v}^{3} \cr 
& \quad 
+ \left\{  
\left( 
-{\frac {732985}{2268}}
+{\frac {24260}{81}} \nu 
+{\frac {340}{9}} \nu^2 
 \right) \chi_s \right. \cr 
& \quad \left. 
+ \left( 
-{\frac {732985}{2268}} 
-{\frac {140}{9}} \nu 
\right) \Delta \chi_a \right\} 
\left\{
1 + 3 \ln \left( \frac{v}{v_{\mathrm {reg}}} \right) 
\right\} {v}^{2} \cr 
& \quad 
+ \left( {\frac {113}{3}}-{\frac {76}{3}} \nu \right) \chi_s 
+{\frac {113}{3}} \Delta \chi_a 
\,,\cr
\Psi^{\mathrm {F2}}_{\mathrm {SS}}
&=  
\pi 
\left\{
\left( - 570  + 40 \nu  \right) \chi_s^2 
-1140 \Delta \chi_a \chi_s
+ \left(  -570 + 2240\,{\nu}  \right) \chi_a^2 
\right\} {v}^{3} \cr 
& \quad 
+ \left\{ 
 \left( 
 {\frac {75515}{288}}
-{\frac {232415}{504}} \nu 
+{\frac {1255}{9}} \nu^2
  \right) \chi_s^2 \right. \cr 
& \quad \left.   
+ \left( 
{\frac {75515}{144}} -{\frac {8225}{18}} \nu 
\right) \Delta \chi_a \chi_s 
+ \left( 
{\frac {75515}{288}} -{\frac {263245}{252}} \nu -480 \nu^2 
\right) \chi_a^2 
 \right\} {v}^{2} \cr 
& \quad  
+ \left( -{\frac {405}{8}} + \frac{5}{2}\,{\nu} \right) \chi_s^2 
-{\frac {405}{4}} \Delta \chi_a \chi_s 
+ \left( -{\frac {405}{8}} + 200\,\nu \right) \chi_a^2 
\,, \cr
\Psi^{\mathrm {F2}}_{\mathrm {SSS}}
&= 
\left( 
{\frac {14585}{24}} - {\frac {475}{6}} \nu + {\frac {100}{3}} \nu^2 
\right) \chi_s^3
+ 
\left( 
{\frac {14585}{8}} - {\frac {215}{2}} \nu 
\right) \Delta \chi_a \chi_s^2 \cr 
& \quad 
+ \left( 
{\frac {14585}{8}}  - 7270\,{\nu} + 80\,{{\nu}}^{2}
\right) \chi_a^2 \chi_s 
+ \left( 
{\frac {14585}{24}}-2380\,{\nu} \right) \Delta \chi_a^3\,. 
\end{align} 

Similar to~\eqref{T2-phi0}, 
$v_{\mathrm {reg}}$ in $\Psi^{\mathrm {F2}}_{\mathrm {SO}}$ 
may be chosen either the value at ISCO $v_{\mathrm {ISCO}}$ 
of the Kerr metric with the final total mass $m(v_{\mathrm {ISCO}})$ 
and effective spin $\chi_{\mathrm {eff}}(v_{\mathrm {ISCO}})$ 
[recall~\eqref{def-chi-eff}],
or that of the pole $v_{\mathrm {pole}}$ in the PN energy flux $F_{\infty}$.
We also recall that the $3.5$PN (relative $1.5$PN) term 
in $\Psi^{\mathrm {F2}}_{\mathrm {SS}}$ is still \textit{incomplete} 
unless we include unknown SS tail contributions. 

Like the expression in~\eqref{T2-H}, 
the BH absorption part of the phase $\Psi^{\mathrm {F2}}_{\mathrm {H}}$ may 
have the form 
\begin{equation}\label{F2-H}
\Psi^{\mathrm {F2}}_{\mathrm {H}} 
= 
\frac{3}{128 \nu}
\left[
\left\{
1 + 3 \ln \left( \frac{v}{v_{\mathrm {reg}}} \right) 
\right\}
\Psi^{\mathrm {F2}}_{\mathrm {Flux},5}
+
{v^2} \left(
\Psi^{\mathrm {F2}}_{\mathrm {Flux},7}
+
\nu \, \Psi^{\mathrm {F2}}_{\mathrm {BH},7}
\right)
+
O(v^3)
\right]\,,
\end{equation}
where $\Psi^{\mathrm {F2}}_{\mathrm {Flux},5}$ 
and $\Psi^{\mathrm {F2}}_{\mathrm {Flux},7}$
solely denote the contributions from the LO ($2.5$PN) 
and the NLO ($3.5$PN) horizon energy flux 
with the substitution 
$\delta m = \delta \nu = \delta \chi_s = \delta \chi_a 
= \Gamma^{i}_{\mathrm {H}} = 0$, respectively, 
while $\Psi^{\mathrm {F2}}_{\mathrm {BH},7}$ accounts for  
the correction due to the LO change in the BH mass and spin. 
They read
\footnote{
The expression for $\Psi^{\mathrm {F2}}_{\mathrm {Flux},5}$ 
in previous versions was off by the factor of $3$ 
due to omitting the prefactor of $3$ 
in front of $\ln ({v}/{v_{\mathrm {reg}}})$ in Eq.~\eqref{F2-H}. 
We thank Zihan Zhou and Horng Sheng Chia for bringing this typo 
to our attention. Note that the Maple code used for this work 
had this expression implemented correctly, and hence the results 
reported in this paper remain unchanged.
}
\begin{align}\label{F2-PsiH}
\Psi^{\mathrm {F2}}_{\mathrm {Flux},5} 
&=  
\left( -\frac{10}{9} + \frac{10}{3}\,{\nu} \right) \chi_s (1 + 3 \chi_s^2) 
+ \left( -10 + 10 \nu  \right) \Delta \chi_a \chi_s^2 \cr 
& \quad 
+ \left( -10 + 30\,{\nu}  \right) \chi_a^2 \chi_s 
+ \left( -\frac{10}{9} + \frac{10}{9} \nu \right) 
\Delta \chi_a (1 + 3 \chi_a^2) \,, \cr
\Psi^{\mathrm {F2}}_{\mathrm{Flux},7}
&=  
 \left( 
- {\frac {7285}{56}} + {\frac {21295}{56}} \nu + {\frac {135}{2}} \nu^2 
 \right) \chi_s^3
+ \left( 
-{\frac {21855}{56}} + {\frac {20175}{56}} \nu + {\frac {285}{2}} \nu^2
\right) \Delta \chi_a \chi_s^2 \cr
& \quad 
+ \left\{  
\left( 
-{\frac {21855}{56}} + {\frac {63885}{56}} \nu + {\frac {405}{2}} \nu^2 
\right) \chi_a^2 
-{\frac {8335}{168}} + {\frac {24445}{168}} \nu + {\frac {45}{2}} \nu^2
\right\} \chi_s \cr 
& \quad 
+ \left( 
- {\frac {7285}{56}} + {\frac {6725}{56}} \nu + {\frac {95}{2}} \nu^2
\right) \Delta \chi_a^3 
+ \left( 
- {\frac {8335}{168}} + {\frac {7775}{168}} \nu + {\frac {95}{6}} \nu^2 
\right) \Delta \chi_a \,, \cr
\Psi^{\mathrm {F2}}_{\mathrm{BH},7}
&=  
\left(  {\frac {8825}{336}}  - {\frac {3175}{56}} \nu
\right) \chi_s (1 + 3 \chi_s^3)
+ 
\left( 
{\frac {26475}{112}} - {\frac {75}{2}} \nu 
\right) \Delta \chi_a \chi_s^2 \cr 
& \quad 
+ 
\left( 
{\frac {26475}{112}} - {\frac {28575}{56}} \nu 
\right) \chi_a^2 \chi_s 
+
\left( 
{\frac {8825}{336}} - {\frac {25}{6}} \nu 
\right) \Delta \chi_a (1 + 3 \chi_a^2)\,.  
\end{align} 
Taking into account the definition in~\eqref{phi-F2}, 
this combined with~\eqref{F2-H} is the same as 
what was given in~\eqref{phi-F2-0} as the BH-absorption phase term.

\section{The match between waveforms with and without black-hole absorption }
\label{sec:match}
The inspiral PN templates (Taylor template families) 
constructed in the preceding section~\ref{sec:Taylor} allow 
to have a more quantitative estimate of the importance of the BH absorption 
in the context of GW data analysis. In this section, 
after a brief introduction of the matched filtering
we compute the \textit{match}~\cite{Owen:1995tm,Owen:1998dk} 
between the frequency-domain PN template TaylorF2 
[see section~\ref{subsec:F2}] with and without each effect 
of BH absorption, namely, the horizon flux and the secular change 
in the BH mass and spin accumulated in the inspiral phase. 
The macth allows to quantify the difference between two waveforms 
with the mindset of GW data analysis, 
and measures the ``faithfulness''~\cite{Damour:1997ub} 
of TaylorF2 templates with BH absorption 
in detecting GW signals of BBHs by Advanced LIGO and LISA.

\subsection{Matched filtering}
\label{subsec:MF}

We first flesh out the basic of matched filtering 
in the GW data analysis.
The material covered in this subsection is fairly standard 
for the literature and our presentation is largely patterned after 
Ajith~\cite{Ajith:2011ec}.

Suppose that $h(t; {\bm {\lambda}})$ is the GW signal 
observed in a detector, 
depending on the set of physical parameters of the source ${\bm {\lambda}}$; 
e.g., they are initial masses and spins of each BH in BBHs in our case. 
We assume that the detector noise $n(t)$ follows 
a stationary, zero-mean Gaussian distribution, 
characterized by its (one-sided) power spectral density (PSD) $S_{h}(f)$
\footnote{In this paper, we do not consider the time-dependence 
of the noise property and non-Gaussian noise, 
both of which are observed 
in the real instrumental data~\cite{Bose:2016sqv}. 
This is another limitations of our work.
In that case, more advanced methods to better discriminate
signals from noise is required 
(see, e.g.,~\cite{Allen:2005fk,Allen:2004gu}). }.
Then, the  (frequency) \textit{overlap} between two GW signals 
$h_{1,2}(t)$ in terms of the noise-weighted inner product 
is defined by~\cite{Cutler:1994ys}
\begin{equation}\label{overlap}
\left\langle h_1,\, h_2 \right \rangle 
\equiv
2 \int_{f_{\mathrm{min}}}^{f_{\mathrm{max}}}
\frac{ {\tilde h}_1 (f) {\tilde h}_2^{*} (f)  
+ {\tilde h}_2 (f) {\tilde h}_1^{*} (f)} {S_{h}(f)} df\,,
\end{equation}
where ${\tilde h}_{1,2}(f)$ is the Fourier transforms of 
the real functions $h_{1,2}(t)$, 
the asterisk denotes its complex conjugate 
and $\{f_{\mathrm{min}},\,f_{\mathrm{max}}\}$ are certain cutoff frequencies 
determined by the setup that we consider; 
the frequency ranges used in our analysis for Advanced LIGO and LISA 
will be described in section~\ref{subsec:match}.

In short, the GW data analysis problem is extracting a specific GW signal 
$h(t; {\bm {\lambda}})$ buried in noisy detector data, say, 
$d(t) \equiv h(t; {\bm {\lambda}}) + n(t)$ 
[assuming that the noise is additive].  
Under above assumption, 
it is known that the optimal filter for detecting 
$h(t; {\bm {\lambda}})$ in a date stream $d(t)$ 
is the \textit{matched filter}~\cite{Allen:2005fk}; 
$h(t; {\bm {\lambda}})$ is cross-correlated with $d(t)$.
The matched-filter output is the overlap~\eqref{overlap} 
between the given normalized (filter) template 
${\hat h} (f) \equiv 
{\tilde h} (f) / \sqrt{\langle h,\,h\rangle}$ 
and the data $d(t)$
\begin{equation}\label{SNR}
\rho \equiv \langle {\hat h}({\bm {\lambda}}) ,\, d \rangle\,.
\end{equation}
This defines the signal-to-noise ratio (SNR) for the filter 
$h(t; {\bm {\lambda}})$, and it is known that 
this is maximized (optimized) when the template parameters $\lambda$ 
match those of the actual GW signals. 
Namely, the optimal SNR for $h(t; {\bm {\lambda}})$ is given by 
$\rho_{\mathrm {opt}} \equiv 
\langle {h}({\bm {\lambda}}), {h}({\bm {\lambda}}) \rangle^{1/2}$.

The templates and GW signals in data stream 
depend on the set of intrinsic physical parameters 
of the BBH ${\bm {\lambda}}$ and its orientation relative to the detector 
as well as two extrinsic parameters $\Psi_{\mathrm c}$ and 
$t_{\mathrm c}$~\cite{Owen:1995tm,Sathyaprakash:1994nj}, 
which are the time and GW phase when the BBH is coalescence. 
In general, a matched-filtering search for GW signals 
can be computationally expensive 
because of the large variety of possible waveforms to be filtered. 
Because none of these parameters for GW signals 
are basically known \textit{a priori}, 
we can afford to tolerate the systematic errors 
in unknown parameters $\Psi_{\mathrm c},t_{\mathrm c}$ 
and/or ${\bm {\lambda}}$ of the templates  
and we are at liberty to maximize the amplitude of the SNR $\rho$ 
over them.  
This reduces the computational cost of the matched-filtering search, 
but still we have to investigate how much the SNR is lost 
by using templates that have ``wrong'' parameter values 
of the target GW signals.

A good measure for such loss of SNR is the
\textit{match}~\cite{Owen:1995tm,Owen:1998dk}
\footnote{Recall that the match is different from the fitting factor; 
the fitting factor is defined by the optimized match over all 
the template parameters ${\bm {\lambda}}$~\cite{Apostolatos:1995pj}. 
}.
Consider the template $x (\Psi_c,\,t_c,\,{\bm {\lambda}})$ 
with intrinsic physical parameters ${\bm {\lambda}}$ 
and the extrinsic parameters $\Psi_{\mathrm c}$ and $t_{\mathrm c}$ 
as well as the target GW signal $h({\bm {\lambda'}})$ 
with (another) intrinsic physical parameters ${\bm {\lambda'}}$ 
in the data stream; the value of time-of-coalescence $t_c'$ 
and the corresponding coalescence GW phase $\Psi_c'$ for the target GW signal 
are assumed to be zero. 
Then, the match is defined by the overlap~\eqref{overlap} 
between the normalized template ${\hat x} (\Psi_c,\,t_c,\,{\bm {\lambda}})$ 
and the normalized GW signal ${\hat h}({\bm {\lambda'}})$ 
maximized over $\Psi_c$ and $t_c$: 
\begin{equation}\label{match0}
{\mathrm {match}}
\equiv
\max_{\Psi_c,\,t_c} 
{\langle {\hat x} (\Psi_c,\,t_c,\,{\bm {\lambda}})
,\, {\hat h} ({\bm {\lambda'}})
\rangle } \,.
\end{equation}
This measures the fraction of the ``optimal SNR'' 
for $h({\bm {\lambda'}})$ using the template $x ({\bm {\lambda}})$. 
We generally say that $x ({\bm {\lambda}})$ with high values 
of match (match $ \gtrsim 0.97$) 
are ``effectual'' in detection and faithful 
in estimating the intrinsic parameters 
of $h ({\bm {\lambda'}})$~\cite{Damour:1997ub}. 
This is the criteria for the template imperfection 
that we adopted in section~\ref{subsec:result2}.

\subsection{Computing the match}
\label{subsec:match}

To compute the match~\eqref{match0}, 
one must first provide a reference GW signal $h({\bm {\lambda'}})$ 
and a set of templates ${x} (\Psi_c,\,t_c,\,{\bm {\lambda}})$.
Because the match is most efficiently computed in the frequency domain, 
we model $h({\bm {\lambda'}})$ and ${x} (\Psi_c,\,t_c,\,{\bm {\lambda}})$ 
using the the so-called ``restricted-Newtonian'', 
frequency-domain, $3.5$PN TaylorF2 waveforms constructed 
in section~\ref{subsec:F2} [recall~\eqref{h-F2} and~\eqref{h-F}]: 
\begin{equation}\label{F2-full}
{\tilde h}(f;\Psi_c,\,t_c,\,m,\,\nu,\,\chi_i) 
\equiv 
{\cal A}\, f^{-7/6} 
e^{i \Psi^{\mathrm {F2}}_{3.5{\mathrm {PN}}}(f\,;\Psi_c,t_c,m,\nu,\chi_i)}\,.
\end{equation}
In the non-precessing case such as ours, 
the intrinsic parameters ${\bm {\lambda}}$ for the BBH are 
the initial values of the total mass $m$, the symmetric mass ratio $\nu$, 
and the dimensionless spin parameter $\chi_i$ of each BH. 
Here, the ``Newtonian'' amplitude ${\cal A}$ is computed 
from the PN expansion of the frequency-domain 
amplitude $A(f) /  f^{-7/6}$ in~\eqref{def-A} by truncating it 
in the leading PN (``Newtonian'') order.  
This is a constant depending on the masses, spins, 
distance and orientation of the BBH 
relative to the observer, thus does not affect the match. 
The explicit expressions for the $3.5$PN phase 
$\Psi^{\mathrm {F2}}_{3.5{\mathrm {PN}}}$ was derived in~\eqref{phi-F2} 
and this is a sum of the standard non-absorption part 
$\Psi^{\mathrm {F2}}_{\infty}$ in~\eqref{F2-inf} 
for the spinning point-particle binary and the BH-absorption part 
$\Psi^{\mathrm {F2}}_{\mathrm {H}}$ in~\eqref{F2-H}. 
In this analysis, the two arbitrary constants $t_c$ and $\Psi_c$ 
in $\Psi^{\mathrm {F2}}_{3.5{\mathrm {PN}}}$ can be 
specified as the time and GW phase when a BBH is coalescence, 
and we can set $v_{\mathrm {reg}} = 1$ without loss of generality.

We are interested in the (loss of) match 
between GW waveforms for BBHs with and without effects of BH absorption. 
For the match calculation, we therefore chose the TaylorF2 waveforms 
${\tilde h}_{\mathrm {H}}(f\,;{\bm {\lambda}})$ 
with the complete $3.5$PN phase $\Psi^{\mathrm {F2}}_{3.5{\mathrm {PN}}}$ 
in~\eqref{phi-F2} as our reference GW signal. 
Meanwhile, the templates are chosen to be TaylorF2 waveforms 
${\tilde x}_{\mathrm {T}}(f\,;{\bm {\lambda}})$ 
with the same amplitude ${\cal A}$ 
and intrinsic parameters of the BBH 
${\bm {\lambda}} = \{m,\nu,\chi_i\}$ 
as ${\tilde h}_{\mathrm {H}}(f\,;{\bm {\lambda}})$, 
but its phase $\Psi^{\mathrm {F2}}_{\mathrm {T}}$ differs from 
$\Psi^{\mathrm {F2}}_{3.5{\mathrm {PN}}}$, neglecting any of 
contributions due to BH absorption; 
namely the LO ($2.5$PN) and NLO ($3.5$PN) horizon-flux contributions 
$\Psi^{\mathrm {F2}}_{\mathrm {Flux},5}$ 
and $\Psi^{\mathrm {F2}}_{\mathrm {Flux},7}$, respectively, 
as well as the LO ($3.5$PN) contribution 
$\Psi^{\mathrm {F2}}_{\mathrm {BH},7}$ 
due to the secular change in the BH masses and spins 
during the inspiral phase. 
Specifically, in section~\ref{subsec:result2} 
the following five $\Psi^{\mathrm {F2}}_{\mathrm {T}}$ 
for ${\tilde x}_{\mathrm {T}}$ were considered 
[recall~\eqref{phi-F2-0}]: 
\begin{enumerate}
\item the neglect of LO horizon-flux term: 
$ \Psi^{\mathrm {F2}}_{\mathrm {T}} 
\equiv
\Psi^{\mathrm {F2}}_{3.5{\mathrm {PN}}}
- 
(3 (1 + 3 \ln(v)) \Psi^{\mathrm {F2}}_{\mathrm {Flux},5}) / (128 \nu)$; 
\item the neglect of NLO horizon-flux term: 
$\Psi^{\mathrm {F2}}_{\mathrm {T}} 
\equiv
\Psi^{\mathrm {F2}}_{3.5{\mathrm {PN}}}
- 
(3 v^2 \Psi^{\mathrm {F2}}_{\mathrm {Flux},7} )/ (128 \nu) $; 
\item the neglect of the LO term due to the secular change 
in BH mass and spins:  
$\Psi^{\mathrm {F2}}_{\mathrm {T}} 
\equiv
\Psi^{\mathrm {F2}}_{3.5{\mathrm {PN}}}
- 
(3 v^2 \Psi^{\mathrm {F2}}_{\mathrm {BH},7} ) / 128$; 
\item the neglect of all phase terms due to BH absorption:  
$\Psi^{\mathrm {F2}}_{\mathrm {T}} 
\equiv
2 \pi f t_c - \Psi_c +
\Psi^{\mathrm {F2}}_{\infty}$; 
\item the neglect of LO cubic-in-spin term 
in the non-absorption phase term $\Psi^{\mathrm {F2}}_{\infty}$: 
$\Psi^{\mathrm {F2}}_{\mathrm {T}} 
\equiv
\Psi^{\mathrm {F2}}_{3.5{\mathrm {PN}}}
- 
(3 v^2 \Psi^{\mathrm {F2}}_{\mathrm {SSS}}) / (128 \nu) $.
\end{enumerate}

The remaining input for the match~\eqref{match0} is 
the model for the PSD of the detector noise $S_{h}(f)$. 
The analytical fits to the PSDs of existing and planned GW detectors 
are conveniently summarized in~\cite{Samajdar:2017mka}.
For Advanced LIGO, we use the fit to the PSD 
of its ``zero-detuning, high-power'' configuration~\cite{Harry:2010zz}
and we import this expression from (4.7) of~\cite{Ajith:2011ec}:
\begin{align}\label{S-LIGO}
&S_{h}(f) \equiv \cr
& \quad 10^{-48} \left(
0.0152 x^{-4} + 0.2935 x^{9/4} + 2.7951 x^{3/2} 
- 6.5080 x^{3/4} + 17.7622
\right)\, {\mathrm{[Hz^{-1}]}}\,,
\end{align}
where $x \equiv (f / 245.4)  \,{\mathrm{Hz^{-1}}}$. 
For LISA, we use the latest fit to its (sky-averaged) PSD 
introduced by (1) of Babak et al~\cite{Babak:2017tow}:
[recall that we use $ G = c = 1$] 
\footnote{
In the spirit of proof-of-principle, 
we here ignore the orbital motion of LISA 
and consider only the single detector configuration.
Such orbital motion generates the modulations 
to the waveforms. 
While the modulation in GW is irrelevant to our analysis, 
this is more important for 
the sky localization of the BBHs~\cite{Cutler:1997ta}. 
We also do not include the confusion noise components 
due to galactic and extragalactic
binaries~\cite{Audley:2017drz,Bender:1997hs}.
}
\begin{equation}\label{S-LISA}
S_{h}(f) 
\equiv 
\frac{20}{3}
\frac{
4 S_h^{\mathrm {acc}}(f) + 2 S_h^{\mathrm {loc}} 
+ S_h^{\mathrm {sn}} + S_h^{\mathrm {omn}}}
{L^2}
\left\{
1 + 
\left(
\frac{2 L f }{0.41}
\right)^2
\right\}
\, {\mathrm{[Hz^{-1}]}}\,,
\end{equation}
where $ L = 2.5 \times 10^9 \,{\mathrm {m}}$ is the arm length. 
Noise contributions $S_h^{\mathrm {acc}}(f), S_h^{\mathrm {loc}}, 
S_h^{\mathrm {sn}}$ and $S_h^{\mathrm {omn}}$ come from 
low-frequency acceleration, local interferometer noise, 
shot noise and other measurement noise, respectively. 
Notably, the analytic form for $S_h^{\mathrm {acc}}(f)$
accounts for the level of improvement successfully demonstrated 
by the LISA Pathfinder~\cite{Armano:2016bkm}: 
\begin{align}\label{accS-LISA}
& S_{h}^{\mathrm {acc}}(f) \equiv  \cr
& \quad  
\left[
9.00 \times 10^{-30}
+ 
3.24 \times 10^{-28}
\left\{
\left( \frac{3.00 \times 10^{-5} {\mathrm{[Hz]}} }{f} \right)^{10}
+
\left( \frac{1.00 \times 10^{-4} {\mathrm{[Hz]}} }{f} \right)^{2}
\right\}
\right] \cr 
& \times 
\left(
\frac{1.00 {\mathrm{[Hz]}}}{2 \pi f}
\right)^4
\, {\mathrm{[m^2 \, Hz^{-1}]}}\,.
\end{align}
The other noise components $S_h^{\mathrm {loc}}, S_h^{\mathrm {sn}}$ 
and $S_h^{\mathrm {omn}}$ are all constants and they are given by 
\begin{align}\label{othS-LISA}
S_{h}^{\mathrm {loc}}
& \equiv
2.89 \times 10^{-24} \, {\mathrm{[m^2 \, Hz^{-1}]}}\,, \cr 
S_{h}^{\mathrm {sn}}
& \equiv
7.92 \times 10^{-23} \, {\mathrm{[m^2 \, Hz^{-1}]}}\,, \cr 
S_{h}^{\mathrm {omn}}
& \equiv
4.00 \times 10^{-24} \, {\mathrm{[m^2 \, Hz^{-1}]}}\,.
\end{align}

The optimization with respect to $\Psi_c$ and $t_c$ 
in the match~\eqref{match0} becomes trivial in the frequency domain 
when using only the dominant $(2,2)$ mode such as ours; 
in general, however, more sophisticated methods for computing the match are 
required when the GW signals include the higher harmonics 
as well~\cite{Damour:1997ub,McWilliams:2010eq}. 
We compute the match by first maximizing it 
over unknown phase $\Psi_c$ analytically, 
making use of the complex matched-filter output~\cite{Allen:2005fk}: 
\begin{align}\label{match1}
{\mathrm {match}}
=
4\, 
\max_{t_c} \int_{f_{\mathrm{min}}}^{f_{\mathrm{max}}}
\frac{ {\hat x}_{\mathrm {T}} (f\,;{\bm {\lambda}}) 
{\hat h}_{\mathrm {H}}^{*} (f\,;{\bm {\lambda}})} {S_{h}(f)}\, 
e^{2 \pi i f t_c}df\,,
\end{align}
and subsequently search its maximum value over $t_c$. 
In this paper, we consider the frequencies in the interval 
$m f \in [0.0035,\,0.018]$ for Advanced LIGO and 
$m f \in [2.0 \times 10^{-4} \, \nu^{-3/8},\, m f_{\mathrm {ISCO}}]$ for LISA, 
where $f_{\mathrm {ISCO}}$ is twice of the ISCO frequency of Kerr metric 
defined in terms of the initial total mass and 
initial effective spin of the BBH 
[recall~\eqref{def-ISCO}and~\eqref{def-chi-eff}]; 
this choice is a rudimentary one that serves as a proof-of-principle.

We find that the computation~\eqref{match1} would be slightly challenging 
(particularly for supermassive BBHs observed by LISA), 
since its integrand becomes highly-oscillating functions 
with the high-mass ratio and high spins, 
and an increasingly high resolution is required to achieve convergence. 
We overcome this technical issue by computing~\eqref{match1} 
to use \textit{Maple}'s function \textsf{NLPSolve} 
with the appropriate numerical integration controls offered 
by \textit{Maple}, in which a higher numerical precision can be easily 
achieved.

\subsection{Results and discussion}
\label{subsec:match-mas}

Our main results are displayed in figure~\ref{fig:LIGOcs} 
for Advanced LIGO and figure~\ref{fig:LISAcs} for LISA. 
In these figures, we considered the BBH configurations 
with initial aligned-spins ${\chi}_{1} = {\chi}_2$ 
that range from $0.30$ to $0.998$, 
assuming that the initial total mass of BBHs are $m = 60.0 M_{\odot}$ 
and $m = 10^{6} M_{\odot}$, respectively.  
For completeness, we here show two more groups of results 
for different BBH configurations. 

\begin{figure}[tbp]
\begin{tabular}{cc}  
\begin{minipage}[t]{.45\hsize}
  \centering
  \includegraphics[clip, width=\columnwidth]{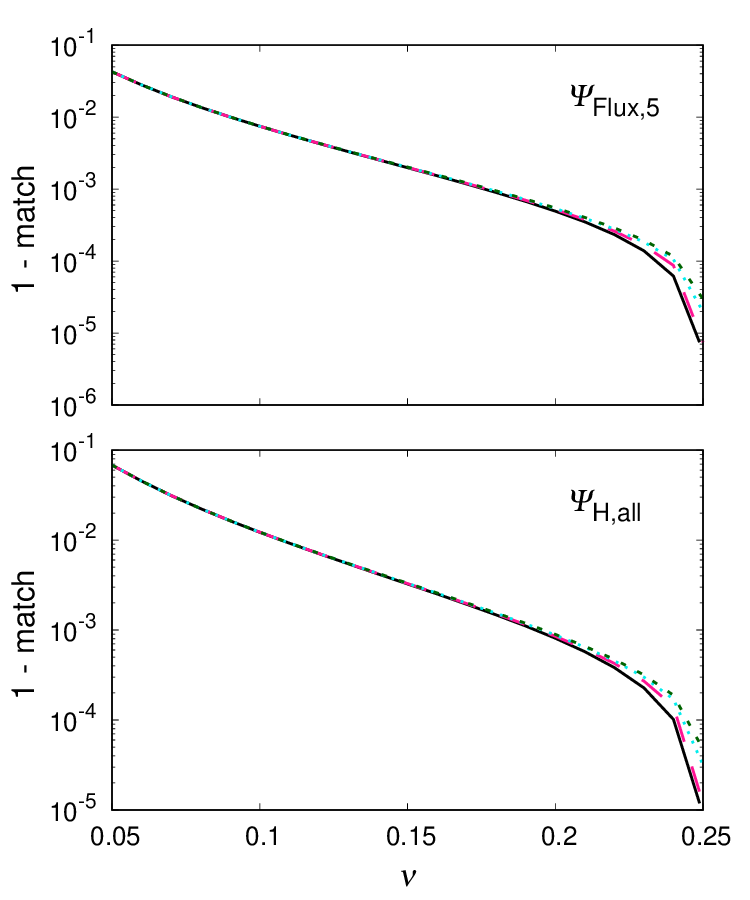}
\end{minipage}
\qquad 
\begin{minipage}[t]{.45\hsize}
  \centering
  \includegraphics[clip, width=\columnwidth]{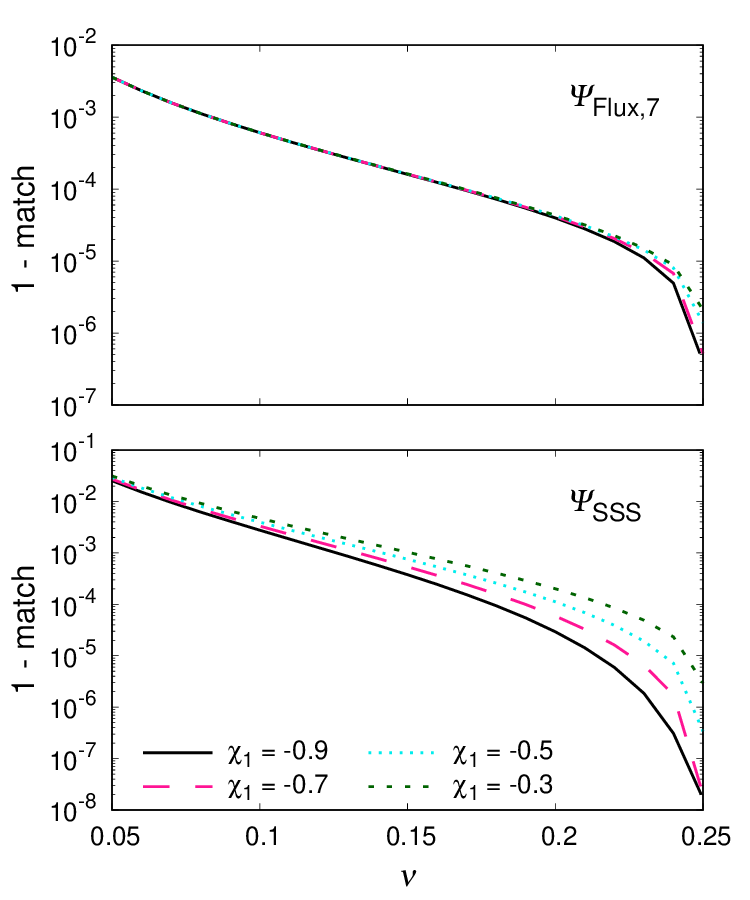}
  \end{minipage}
\end{tabular}
\caption{The mismatch ($ \equiv 1 - {\mathrm {match}}$) 
between two TaylorF2 templates with and without 
each phase correction due to BH absorption 
accumulated in the Advanced LIGO frequency band $m f \in [0.0035,\,0.018]$, 
where the initial total mass and spin of the large BH 
(labeled by `$2$';\,$m_2 \geq m_1$) are chosen to be $m = 60.0 M_\odot$  
and ${\chi}_{2} = +0.90$, respectively. 
The results are plotted as a function of 
the symmetric mass ratio $\nu$ for different values 
of the initial anti-aligned spin of 
the small BH (labeled by `$1$') $\chi_1$, 
and they are grouped into four panels according to 
what is neglected in the GW phase~\eqref{phi-F2-0}; 
\textit{Top left:} the neglect of the LO horizon-flux term 
$\Psi^{\mathrm {F2}}_{\mathrm {Flux},5}$. 
\textit{Top right:} the neglect of the NLO horizon-flux term 
$\Psi^{\mathrm {F2}}_{\mathrm {Flux},7}$. 
\textit{Bottom left:} the neglect of all phase terms due to BH absorption 
$\Psi^{\mathrm {F2}}_{\mathrm {H,all}}$, 
including all horizon-flux terms 
$\Psi^{\mathrm {F2}}_{\mathrm {Flux},5}$ and
$\Psi^{\mathrm {F2}}_{\mathrm {Flux},7}$ 
as well as the LO term due to the secular change in BH intrinsic parameters 
$\Psi^{\mathrm {F2}}_{\mathrm {BH},7}$. 
\textit{Bottom right:} (for comparison) 
the neglect of the LO cubic-in-spin term  
$\Psi^{\mathrm {F2}}_{\mathrm {SSS}}$ in the non-absorption, 
point-particle phase term $\Psi^{\mathrm {F2}}_{\infty}$. 
}
\label{fig:LIGOas}
\end{figure}
%
%
\begin{figure}[tbp]
\begin{tabular}{cc}  
\begin{minipage}[t]{.45\hsize}
  \centering
  \includegraphics[clip, width=\columnwidth]{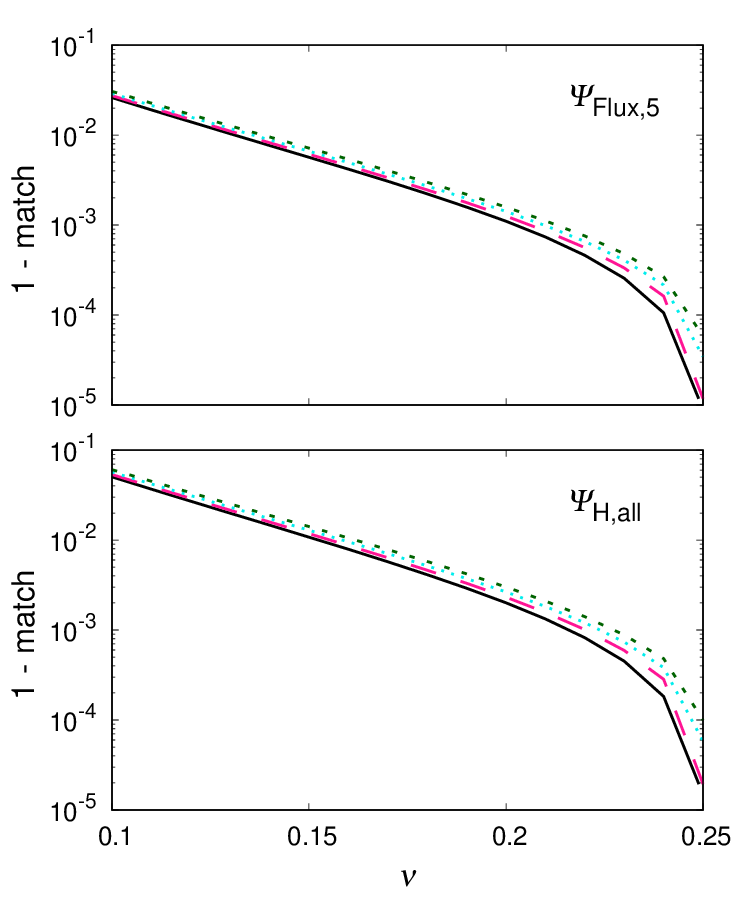}
\end{minipage}
\qquad 
\begin{minipage}[t]{.45\hsize}
  \centering
  \includegraphics[clip, width=\columnwidth]{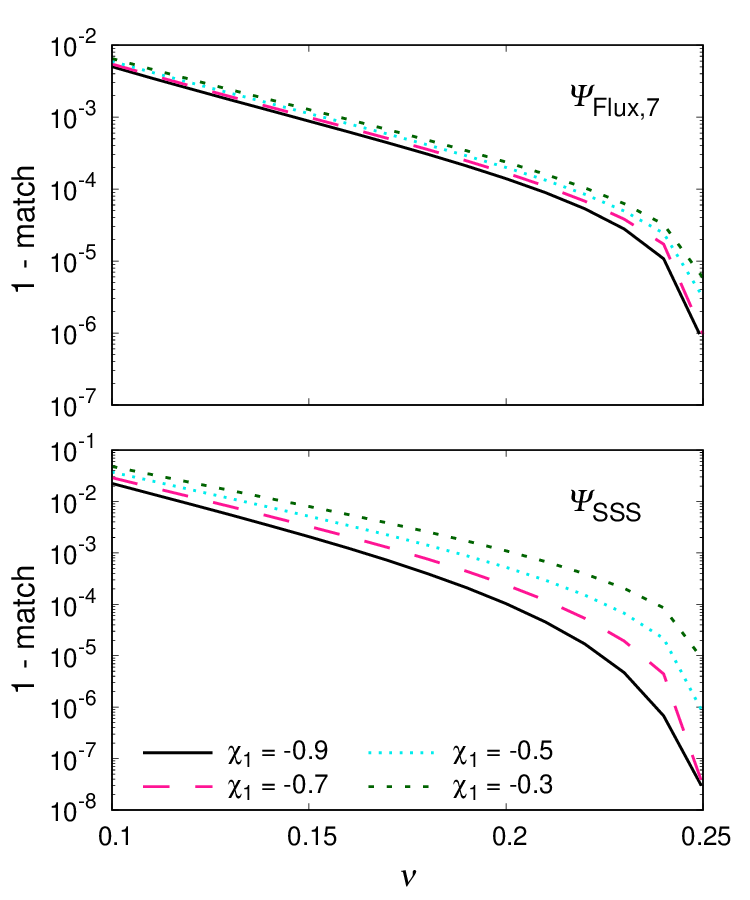}
  \end{minipage}
\end{tabular}
\caption{
The mismatch ($ \equiv 1 - {\mathrm {match}}$) 
between the two TaylorF2 templates with and without 
each phase correction due to BH absorption 
accumulated in a space based detector, 
LISA frequency band 
$m f \in [2.0 \times 10^{-4} \, \nu^{-3/8},\,  m f_{\mathrm {ISCO}}]$, 
where the initial total mass and the spin of the large BH
(labeled by `$2$') are chosen to be $m = 10^{6} M_\odot$ and 
${\chi}_{2} = +0.90$, respectively. 
The results are plotted as a function of 
the initial symmetric mass ratio $\nu$ for different values 
of the initial anti-aligned spin of the small BH $\chi_1$.  
The label and grouping of the panels are the same as 
for figure~\ref{fig:LIGOas}.}
\label{fig:LISAas}
\end{figure}

The recent observation GW170104~\cite{Abbott:2017vtc} disfavors 
the aligned-spin configuration. 
Motivated by this measurement, the first group is concerned 
with BBHs with anti-aligned spins. 
Suppose BBH systems that have the same total-mass as those 
in figures~\ref{fig:LIGOcs} and~\ref{fig:LISAcs} 
but have both anti aligned-spins 
$\chi_{1,2} = \{-0.30,\,-0.50,\,-0.70,\,-0.998\}$. 
We point out that mismatch for each template ${\hat x}_{\mathrm {T}}$ 
with the reference GW signal ${\tilde h}_{\mathrm {H}}$ 
for such BBHs with anti-aligned spins are smaller 
than that for corresponding aligned-spin BBHs 
with $\chi_{1,2} = \{+0.30,\,+0.50,\,+0.70,\,+0.998\}$ 
by the factor of $O(10)$. 
In fact, we find that all mismatch is below the $10^{-3}$ mark 
for such anti-aligned-spin BBHs. It is therefore not significant even when 
the BBH is in the nearly anti-extremal limit $\chi_{1,2} = -0.998$. 

We also consider the mismatch between  
${\tilde x}_{\mathrm {T}}$ and ${\tilde h}_{\mathrm {H}}$ 
for different values of initial anti-aligned spin of the small BH $\chi_1$ 
(labeled by `$1$';\,$m_1 \geq m_2$), 
while the spin of the large BH (labeled by `$2$') 
is assumed to be $\chi_{2} = +0.90$, in order to explore 
the asymmetric, anti-aligned spin configurations.
The results for such BBHs with initial total masses $m = 60.0 M_{\odot}$ 
that are observable by Advanced LIGO 
and for those with initial total masses $m = 10^{6} M_{\odot}$ detected 
by LISA are summarized in figures~\ref{fig:LIGOas} 
and~\ref{fig:LISAas}, respectively. 
Notice that neglecting $\Psi^{\mathrm {F2}}_{\mathrm {BH},7}$ always produces 
the mismatch below the $10^{-5}$ mark, which are not significant 
for all BBHs that are considered here. 
For this reason, we did not plot the corresponding mismatch in these figures.

Figures~\ref{fig:LIGOas} and~\ref{fig:LISAas} show that 
all mismatch except that from $\Psi^{\mathrm {F2}}_{\mathrm {SSS}}$ 
are dominated by the spin of the large BH $\chi_2$. 
The spin of the small BH $\chi_1$ is largely irrelevant 
unless BBHs are almost equal-mass configurations $\nu \gtrsim 0.22$, 
where none of mismatch are significant.
This is an expected feature because the spin of the small BH is 
likely to be unimportant in the high mass-ratio regime. 
The mismatch from the neglect of $\Psi^{\mathrm {F2}}_{\mathrm {H,all}}$ 
(the cubic-in-spin phase term  $\Psi^{\mathrm {F2}}_{\mathrm {SSS}}$) 
is therefore significant only in the high mass-ratio regime 
$\nu \lesssim 0.07$ ($\nu \lesssim 0.06$) for Advanced LIGO and 
$\nu \lesssim 0.15$ ($\nu \lesssim 0.25$) for LISA; 
refer back to figures~\ref{fig:LIGOcs} and~\ref{fig:LISAcs}.

The $\chi_1$-dependence of the match in the almost equal-mass regime, 
which is particularly pronounced for that 
from $\Psi^{\mathrm {F2}}_{\mathrm {SSS}}$, 
is rooted in the fact that coefficients of 
the term in $\Psi^{\mathrm {F2}}_{\mathrm {Flux},5},\,
\Psi^{\mathrm {F2}}_{\mathrm {Flux},7}$ 
and $\Psi^{\mathrm {F2}}_{\mathrm {SSS}}$ 
that involves the anti-symmetric spin parameter $\chi_a$ 
[recall~\eqref{chi-as}] 
are not so small compared to those only proportional 
to the symmetric spin parameter $\chi_s^3$. 
In fact, the coefficient of $\chi_a^2 \chi_s$ in 
$\Psi^{\mathrm {F2}}_{\mathrm {SSS}}$ are quite large $\sim O(10^{3})$ 
for almost equal-mass BBHs; recall~\eqref{F2-Psi0}. 
This explains why the dependence of $\chi_1$ is the most visible 
for the asymmetric spin configuration 
$\chi_1 = -0.30,\,\chi_2 = + 0.90$ in the almost equal-mass regime 
of these figures.

\begin{figure}[tbp]
\begin{tabular}{cc}  
\begin{minipage}[t]{.45\hsize}
  \centering
  \includegraphics[clip, width=\columnwidth]{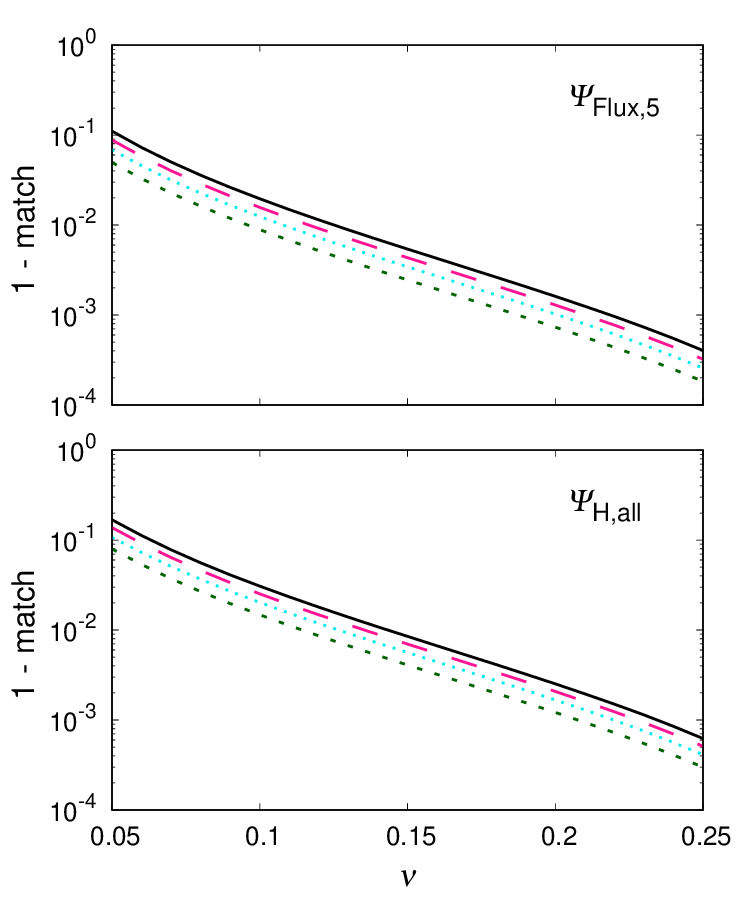}
\end{minipage}
\qquad 
\begin{minipage}[t]{.45\hsize}
  \centering
  \includegraphics[clip, width=\columnwidth]{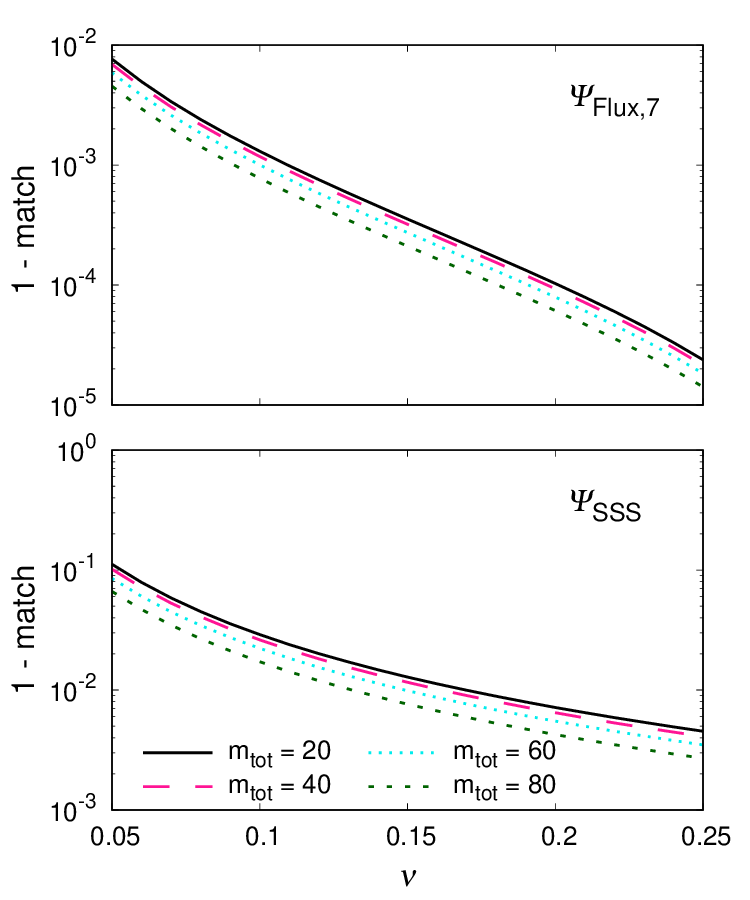}
  \end{minipage}
\end{tabular}
\caption{The mismatch ($ \equiv 1 - {\mathrm {match}}$) 
between two TaylorF2 templates with and without 
each phase correction due to BH absorption 
accumulated in the Advanced LIGO frequency band $m f \in [0.0035,\,0.018]$, 
where the initial aligned-spins are chosen to be near extremal values 
${\chi}_{1,2} = 0.998$. 
For $m = 80.0 M_{\odot}$, we set the lower cutoff frequency 
at $f = 10.0 {\mathrm {Hz}}$. 
The results are plotted as a function of 
the initial symmetric mass ratio $\nu$ for different values 
of the initial total mass $m_{\mathrm {tot}} \equiv m / M_\odot$.
The label and grouping of the panels are the same as 
for figures~\ref{fig:LIGOas}.}
\label{fig:LIGOm}
\end{figure}
%

\begin{figure}[tbp]
\begin{tabular}{cc}  
\begin{minipage}[t]{.45\hsize}
  \centering
  \includegraphics[clip, width=\columnwidth]{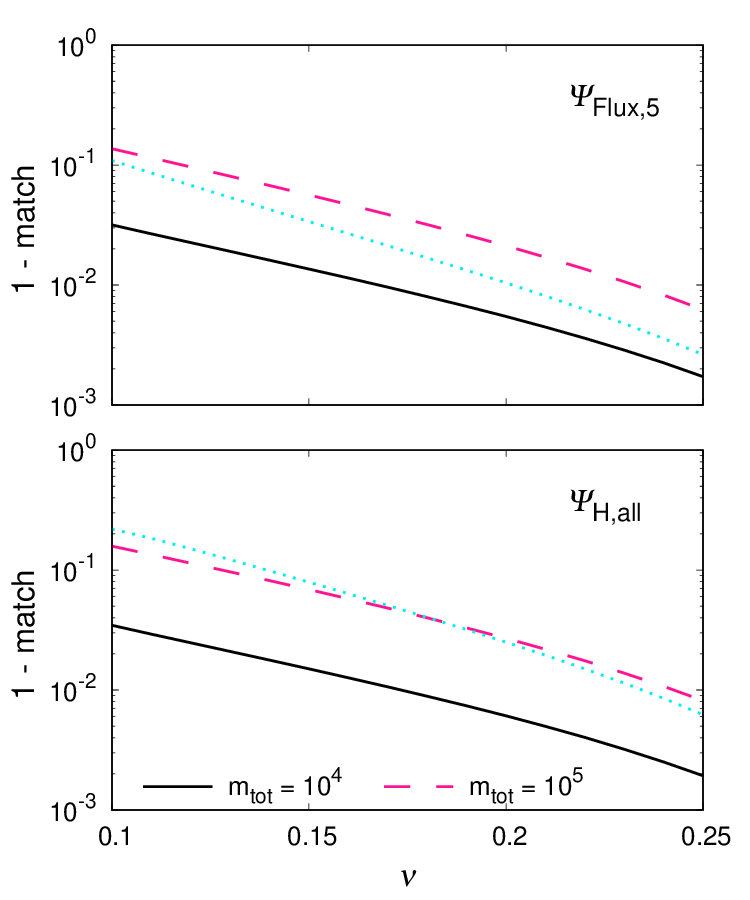}
\end{minipage}
\qquad 
\begin{minipage}[t]{.45\hsize}
  \centering
  \includegraphics[clip, width=\columnwidth]{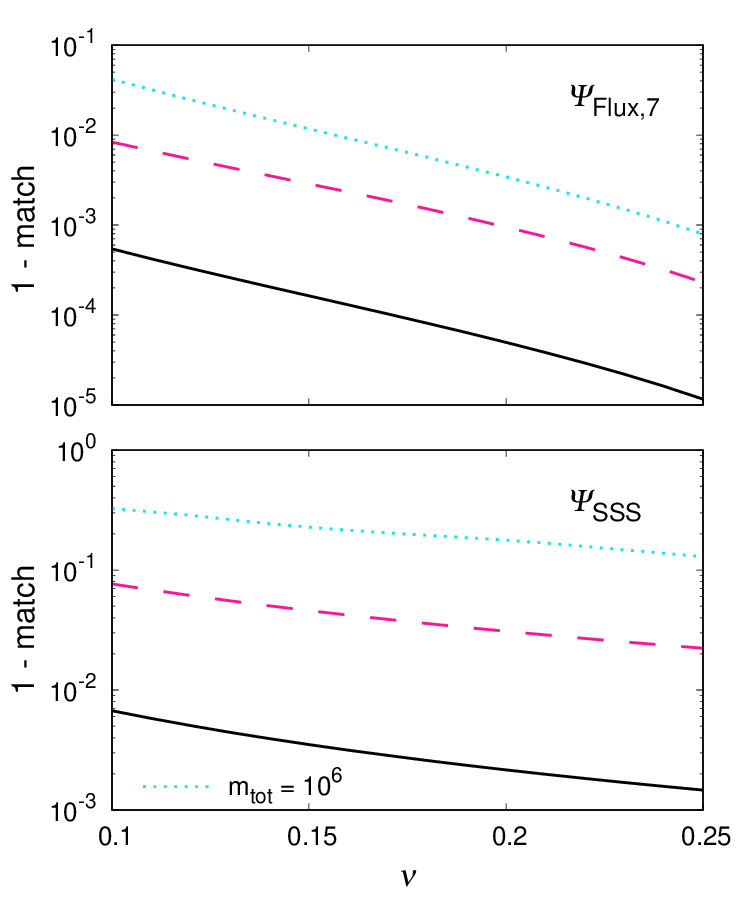}
  \end{minipage}
\end{tabular}
\caption{The mismatch ($ \equiv 1 - {\mathrm {match}}$) 
between the two TaylorF2 templates with and without 
each phase correction due to BH absorption accumulated 
in a LISA frequency band 
$m f \in [2.0 \times 10^{-4} \, \nu^{-3/8},\, m f_{\mathrm {pole}}]$, 
where the initial aligned-spins are chosen to be ${\chi}_{1,2} = 0.998$. 
Notice that the upper cutoff of the frequency is 
$m f_{\mathrm {pole}}$ rather than $m f_{\mathrm {ISCO}}$ 
to validate TaylorF2. We also set the upper cutoff frequency 
at $f = 1.0 {\mathrm {Hz}}$ for $m = 10^{4} M_{\odot}$ case 
($f_{\mathrm {pole}} \sim 2.2 {\mathrm {Hz}}$ in this case).
The label and grouping of the panels are the same as 
for figure~\ref{fig:LIGOm}.}
\label{fig:LISAm}
\end{figure}

The second group considers the mismatch for each template 
${\tilde x}_{\mathrm {T}}$ with the reference signal 
${\tilde h}_{\mathrm {H}}$ 
for different values of initial total masses $m$ of BBHs, 
while initial aligned-spins of BBHs are 
assumed to be $\chi_{1,2} = 0.998$. 
The results for BBHs that are observable Advanced LIGO 
and LISA are summarized in figures~\ref{fig:LIGOm} 
and~\ref{fig:LISAm}, respectively. 
Once again, the mismatch due to the neglect of 
$\Psi^{\mathrm {F2}}_{\mathrm {BH},7}$ is not plotted here; 
the resulting mismatch is always below the $10^{-5}$ mark in both cases, 
and does not become significant. 

Figure~\ref{fig:LIGOm} shows that all mismatch are largely 
independent of the total mass; 
the lower total-mass system produces larger mismatch, 
and the plotted mismatch are different from each other only 
by the factor of $O(1)$. 
This is simply because the frequency range that we use for Advanced LIGO 
covers only the early inspiral phase in our analysis; 
recall that the upper cutoff frequency is chosen to be 
the relatively low frequency $m f_{\mathrm {max}} = 0.018$, 
largely motivated by the inspiral portion of 
the phenomenological ``PhenomD'' model~\cite{Khan:2015jqa} 
as well as that of the NINJA project~\cite{Ajith:2012az,Aasi:2014tra}. 
For BBHs with the total mass 
$m / M_{\odot} = \{ 20.0,\,40.0,\,60.0,\,80.0 \}$ 
considered in figure~\ref{fig:LIGOm}, 
the cutoff frequencies 
$m f_{\mathrm {min}} = 0.0035$ and 
$m f_{\mathrm {max}} = 0.018$ are translated to 
$f_{\mathrm {min}} \sim \{36.0,\,18.0,\,12.0,\,10.0\} $ Hz 
and
$f_{\mathrm {max}} \sim \{183.0,\,91.0,\,61.0,\,46.0\} $ Hz, respectively; 
the lower cutoff frequency for $m = 80.0 M_{\odot}$ configuration
is instead selected at $f_{\mathrm {min}} = 10.0$ Hz 
because the noise PSD below this frequency is not well characterized. 
We see that none of them reaches the minimum of the Advanced LIGO's noise 
PSD in~\eqref{S-LIGO}, locating at $f_{\mathrm {LIGO}} \sim 250.0$ Hz. 
Hence, the mismatch for $m = 20.0 M_{\odot}$ configuration becomes largest 
as the frequency overlaps between ${\tilde x}_{\mathrm {T}}$ and 
${\tilde h}_{\mathrm {H}}$ are in the Advanced LIGO's wider sensitivity band 
(and thus have more time to accumulate a phase difference).  
It should be noted that our conclusion 
of figure~\ref{fig:LIGOm} is therefore valid only 
for this specific choice of the Advanced LIGO's frequency range. 

Indeed, we see that the mismatch for the LISA case plotted 
in figure~\ref{fig:LISAm} shows the different dependence 
on the total mass $m$. 
For example, the mismatch from $\Psi^{\mathrm {F2}}_{\mathrm {H,all}}$ 
for two configurations $m = \{10^5, 10^6 \} M_{\odot}$ 
are almost identical to each other (within the order-of-magnitude) 
and is significant in the almost equal-mass regime $\nu \lesssim 0.19$, 
while that for the configuration  $m = 10^4 M_{\odot}$ is smaller 
by the factor of $O(10)$. 
Recall that the templates used in the LISA case is terminated near 
the ISCO of Kerr spacetime, including the late inspiral phase. 
The cutoff frequencies for the case of LISA 
$m f_{\mathrm {min}} = 2.0 \times 10^{-4}$ and 
$m f_{\mathrm {max}} = m f_{\mathrm {pole}}$ are approximately 
translated to 
$m f_{\mathrm {min}} \sim 6.8 \times \{10^{-3},\,10^{-4},\,10^{-5}\}$ Hz 
and 
$m f_{\mathrm {max}} \sim \{1.0,\,2.2 \times 10^{-1},\,2.2 \times 10^{-2}\}$ 
Hz for the BBH with $m = \{10^4, 10^5, 10^6 \} M_{\odot}$, respectively; 
the upper cutoff frequency for the $m = 10^4 M_{\odot}$ configuration 
is instead selected at $f_{\mathrm {max}} = 1.0$ Hz 
because the noise PSD above this frequency may not be well characterized. 
Since the minimum of LISA's noise PSD in~\eqref{S-LISA} 
is at $f_{\mathrm {LISA}} \sim 8.3 \times 10^{-3}$ Hz, 
where LISA is most sensitive to the GW signals, 
each configuration covers quite different frequency range of the noise PSD. 
The fact that the $m = 10^{4} M_{\odot}$ configuration always produces 
the smallest mismatch is a direct consequence of this; 
$f_{\mathrm {LISA}}$ is only marginally covered for this configuration.

\section*{Acknowledgments}
We thank Katerina Chatziioannou, Alexandre Le Tiec, Eric Poisson, 
Riccardo Sturani and an anonymous referee 
for useful discussion and for constructive comments on this manuscript. 
SI would like to thank the financial support 
from Ministry of Education - MEC during his stay 
at IIP-Natal-Brazil, where parts of this project were completed.
HN acknowledges support from MEXT Grant-in-Aid for Scientific Research 
on Innovative Areas, ``New developments in astrophysics
through multi-messenger observations of gravitational wave sources'',  
No.~24103006, JSPS KAKENHI Grant, No.~JP17H06358,
and JSPS Grant-in-Aid for Scientific Research (C), No.~16K05347.

\section*{References}

\end{document}